\documentclass[iop]{emulateapj-rtx4}
\shortauthors{Sekanina \& Chodas}
\shorttitle{Comet C/2011 W3: Physical and Orbital Study}
\slugcomment{Version \today }
\newcommand{\Rsun}{$R_{\mbox{\scriptsize \boldmath $\odot$}}$}
\newcommand{\Lsun}{${\cal L}_{\mbox{\scriptsize \boldmath $\odot$}}$}
\newcommand{\Msun}{${\cal M}_{\mbox{\scriptsize \boldmath $\odot$}}$}
\newcommand{\gsun}{$\gamma_{\mbox{\scriptsize \boldmath $\odot$}}$}
\newcommand{\gapeq}{$\;$\raisebox{0.3ex}{$>$}\hspace{-0.28cm}\raisebox{-0.75ex}{$\sim$}$\;$}
\begin{document}
\title{Comet C/2011 W3 (Lovejoy):\ Orbit Determination,
 Outbursts,\\Disintegration of Nucleus, Dust-Tail Morphology,
 and\\Relationship to New Cluster of Bright Sungrazers}
\author{Zdenek Sekanina \& Paul W. Chodas}
\affil{Jet Propulsion Laboratory, California Institute of Technology \\
  4800 Oak Grove Drive, Pasadena, CA 91109}
\email{Zdenek.Sekanina@jpl.nasa.gov}

\begin{abstract}
We describe the physical and orbital properties of C/2011 W3.  After
surviving the perihelion passage, the comet was observed to undergo major
physical changes.  The permanent loss of the nuclear condensation and the
formation of a narrow spine tail was observed first at Malargue, Argentina,
on December 20 and then systematically at Siding Spring, Australia.  The
process of disintegration culminated with an outburst (terminal fragmentation
event) on December 17.6 UT.  The postperihelion tail, observed for $\sim$3
months, was the product of activity over $<$2 days.  Because of the delayed
response to the hostile environment in the immediate proximity of the Sun,
the nucleus' breakup and crumbling was probably caused by thermal stress due
to the penetration of intense heat pulse deep into the nucleus' interior after
perihelion.  The same mechanism may be responsible for cascading fragmentation
of sungrazers at large heliocentric distances.  The observed behavior is at
odds with the rubble-pile model, since the residual mass of the nucleus after
perihelion, estimated at $\sim$10$^{12}$ g (a sphere $\sim$150--200 m across),
still possessed significant cohesive strength.  The spine tail --- the product
of the terminal outburst --- was a synchronic feature, whose brightest part
contained submillimeter-sized dust particles, released at velocities not
exceeding 30 m\,s$^{-1}$.  The loss of the nuclear condensation prevented an
accurate orbital-period determination by traditional techniques.  Since the
missing nucleus must have been located on the synchrone, whose orientation and
sunward tip have been measured, we compute the astrometric positions of this
missing nucleus as the coordinates of the points of intersection of the spine
tail's axis with the lines of forced orbital-period variation, derived from
the orbital solutions based on high-quality preperihelion astrometry from the
ground.  The resulting orbit gives \mbox{$698 \pm 2$} years for the osculating
orbital period, which proves that C/2011 W3 is the first major member of the
expected new, 21st-century cluster of bright Kreutz-system sungrazers, whose
existence was predicted by these authors in 2007.  From the spine tail's
evolution, we determine that its measured tip, populated by dust particles
1-2 mm in diameter, receded antisunward from the computed position of the
missing nucleus.  The bizarre appearance of the comet's dust tail in images
taken only hours after perihelion with the coronagraphs on board the SOHO and
STEREO spacecraft is readily understood.  The disconnection of the comet's
head from the tail released before perihelion and an apparent activity
attenuation near perihelion have a common cause --- sublimation of all dust
at heliocentric distances smaller than about 1.8 solar radii.  The tail's
brightness is strongly affected by forward scattering of sunlight by dust.
From an initially broad range of particle sizes, the grains that were imaged
the longest had a radiation-pressure parameter \mbox{$\beta \simeq 0.6$},
diagnostic of submicron-sized silicate grains and consistent with the
existence of the dust-free zone around the Sun.  The role and place of C/2011
W3 in the hierarchy of the Kreutz system and its genealogy via a 14th century
parent suggest that it is indirectly related to the celebrated sungrazer
X/1106 C1, which, just as the first-generation parent of C/2011 W3, split
from a common predecessor during the previous return to perihelion.
\end{abstract}

\keywords{comets: general --- comets: individual (comet of A.D.\ 467,
 X/1106 C1, comet of 1314, X/1381 V1, C/1843 D1, C/1880 C1, C/1882 R1,
 C/1887 B1, C/1945 X1, C/1963 R1, C/1965 S1, C/1970 K1, D/1993 F2,
 C/2011 W3) --- methods: data analysis}

\section{Introduction}

The Kreutz system of sungrazing comets (e.g., Kreutz 1888, 1891, 1901,
Marsden 1967, 1989, 2005, Sekanina 2002a, 2003, Sekanina \& Chodas 2004,
2007, 2008) offers the ultimate example of an advanced phase of cascading
fragmentation (Seka\-nina \& Chodas 2007), a process that was shown to occur
throughout the orbit, including the aphelion region (150--200 AU from the
Sun).  Separation velocities acquired by fragments at each fragmentation event
generate orbit variations that depend strongly on the breakup's heliocentric
distance (Sekanina 2002a).  At a single tidally-triggered or tidally-assisted
event in the immediate proximity of perihelion (which takes place in the
Sun's inner corona), a separation velocity of 1 m s$^{-1}$ causes the two
fragments to return to perihelion at times that are typically some 80
years apart.  The same separation velocity at a single nontidal splitting
event far from the Sun, including the aphelion region, affects the orbital
period hardly at all, but, depending on the velocity's direction, can
introduce material changes in the other orbital elements: up to $\sim \!
\frac{1}{7}$ {\Rsun} (1 {\Rsun} = Sun's radius or 0.0046548 AU) in the
perihelion distance (thus allowing some fragments to collide with the Sun's
photosphere at the next perihelion passage), up to $\sim$5$^\circ$ in the
longitude of the ascending node and the argument of perihelion, and up to
$\sim$1$^\circ$ in the orbit inclination.  Given that the separation velocity
can easily reach a few meters per second, the resulting effects on the orbital
elements substantially exceed those by the indirect planetary perturbations
considered by Marsden (1967).

The observed long-term temporal distribution of the bright sungrazers
does indeed show a tendency toward clumping, with the interval between
two consecutive clusters averaging about 80 years (e.g., Marsden 1967,
Seka\-nina \& Chodas 2007).  In addition, the four major fragments of
comet C/1882 R1, the brightest known member of the Kreutz system since
the beginning of the 18th century (Kreutz 1888, 1891) were found to have
{\it actual} orbital periods that increase also with an average step of
80 years, from $\sim$600 to $\sim$840 years (Seka\-nina \& Chodas 2007).

Considering this recurrence cycle, given that the two most recent clusters
of bright Kreutz sungrazers peaked in the 1880s and 1960s, noting that the
clusters were preceded, two to more than three decades earlier, by precursor
sungrazers, and also recognizing that the arrival rate of SOHO Kreutz system
comets has been climbing ever since the launch of the spacecraft (Sekanina
\& Chodas 2007, Knight et al.\ 2010), the authors predicted in 2007 that
{\it ``another cluster of bright sungrazers is expected to arrive in the
coming decades, the earliest member possibly just several years from now''}
(Sekanina \& Chodas 2007).

With T. Lovejoy's discovery of C/2011 W3 (Green 2011), the question
has arisen whether this is indeed the first major member of the predicted
21st-century cluster.  The answer depends critically on the accurate
determination of the comet's orbital period $P$.  If $P$ is about 400 years
or shorter, the comet should be a fragment of one of the sungrazers reported
in the course of the 17th-century or even more recently, and has nothing in
common with a new cluster.  On the other hand, if the orbital period is
substantially longer than 400 years, then C/2011 W3 should indeed belong to
the new cluster.

\begin{table}[ht]
\noindent
\begin{center}
{\footnotesize {\bf Table 1}\\[0.1cm]
{\sc Temporal Coverage of the Head of Comet C/2011 W3\\by the SOHO and
STEREO Imaging Instruments$^{\rm a}$}\\[0.2cm]
\begin{tabular}{c@{\hspace{0.2cm}}c@{\hspace{0.2cm}}c@{\hspace{0.3cm}}c@{\hspace{0.05cm}}c}
\hline\hline\\[-0.15cm]
& Imaging & Spatial &
\multicolumn{2}{@{\hspace{-0.15cm}}c}{Coverage:\ 2011 December (UT)}\\[-0.02cm]
& instru- & resolu- &
\multicolumn{2}{@{\hspace{-0.15cm}}c}{\rule[0.6ex]{4.16cm}{0.4pt}}\\[-0.02cm]
Spacecraft & ment & tion$^{\rm b}$ & preperihelion
& postperihelion\\[0.1cm]
\hline\\[-0.15cm]
SOHO     & C2   & 11$^{\prime\prime}\!$.4 & 15.75--15.96 & 16.07--16.22 \\
& C3   & 56$^{\prime\prime}$     & 14.09--15.90 & 16.15--18.36 \\[0.1cm]
STEREO-A & COR1 & 7$^{\prime\prime}\!$.5  & 15.87--15.96 & 16.09--16.16 \\
 & COR2 & 14$^{\prime\prime}\!$.7 & 15.35--15.92 & 16.13--16.66 \\
& HI1  & 70$^{\prime\prime}$     & 12.01--14.92 & 16.65--22.37 \\[0.1cm]
STEREO-B & COR1 & 7$^{\prime\prime}\!$.5  & 15.88--15.96 & 16.23--16.45 \\
 & COR2 & 14$^{\prime\prime}\!$.7 & 15.35--15.90 & 16.38--17.35 \\
& HI1 & 70$^{\prime\prime}$     & 10.81--15.01 & 17.87--26.95 \\[0.05cm]
\hline\\[-0.25cm]
All & \ldots\ldots & \dots\ldots & 10.81--15.96 & 16.07--26.95 \\[0.05cm]
\hline\\[-0.2cm]
\multicolumn{5}{l}{\parbox{8.2cm}{$^{\rm a}$\,{\scriptsize Based
on the authors' inspection{\vspace{-0.03cm}} of the images available at
these websites:\ http://sohodata.nascom.nasa.gov/cgi-bin/data\_query
{\vspace*{-0.03cm}}for SOHO and{\vspace{-0.03cm}}
http://stereo-ssc.nascom.nasa.gov/cgi-bin/images for STEREO.}}}\\[0.15cm]
\multicolumn{5}{l}{\parbox{8.2cm}{$^{\rm b}$\,{\scriptsize Per pixel or
per two binned pixels, as used.}}}\\[0.3cm]
\end{tabular}}
\end{center}
\end{table}

\section{Preperihelion Ground-Based Observations}

When discovered on November 27, C/2011 W3 had only $\sim$18 days to reach
perihelion (which occurred on 2011 December 16.0 UT), and there was little
hope for an accurate determination of the orbital period from preperihelion
data, regardless of their quality.  Observing circumstances were unfavorable
for ground-based imaging, since the comet's elongation from the Sun was
rapidly decreasing, from 50$^\circ$ at discovery to merely 17$^\circ\!$.6 on
December 10, when its position was measured from the ground for the last time
before perihelion.  Still, more than 100 astrometric positions were obtained
during this two-week period (Spahr et al.\ 2011, 2012), the great majority of
which was sufficiently accurate and mutually consistent to be used for
deriving a high-quality set of elements, except for the orbital period.
For example, two sets of elliptical elements computed by Williams (2011a,
2011b) from 91 observations between November 27 and December 8 and from
94 observations between November 27 and December 10, gave osculating orbital
periods of, respectively, \mbox{$376 \pm 51$} years (leaving a mean residual
of $\pm$0$^{\prime\prime}\!.7$) and \mbox{$680 \pm 64$} years (leaving
$\pm$0$^{\prime\prime}\!.8$).

\section{Spaceborne Observations Near Perihelion Passage}

Around perihelion, from December 11 to 22, the comet was too close to
the Sun to obtain any astrometric positions from the ground.  The comet
was, however, extensively observed with instruments on board several
satellites and deep-space probes.  Astrometric data were in fact extracted
from the comet's images seen in the coronagraphs and other imaging devices
on the SOHO and both STEREO spacecraft (Kracht 2011, 2012, Spahr et al.\
2012).  The coverage of the comet's motion by the three spacecraft
was excellent over this period of time, as is apparent from Table 1.
We use the standard abbreviations for the relevant instruments:\ C2 and C3
for the two coronagraphs on the SOHO spacecraft and COR1 for the inner
coronagraph, COR2 for the outer coronagraph, and HI1 for the first
heliospheric imager on either of the two STEREO spacecraft.  Included in
Table 1 is the spatial resolution of the instruments, based on the
information from Brueckner et al. (1995) for SOHO and from Howard
et al. (2008) for STEREO.  The astrometric data from the measured
images of the comet obtained by these instruments turned out to be so
poor that they could not be combined with the ground-based positions,
often of a subarcsec accuracy, to derive high-quality orbital solutions.
The best among the spaceborne data are the six positions obtained by
Kracht (2011) from the COR2 images of STEREO-B between December 16.49
and 16.57 UT, shortly after perihelion, but even they are not accurate
enough to be included in the final iteration of the orbit.

\begin{figure*}[ht]
 \vspace*{-4.85cm}
 \hspace*{-0.7cm}
 \centerline{
 \scalebox{1.1}{
 \includegraphics{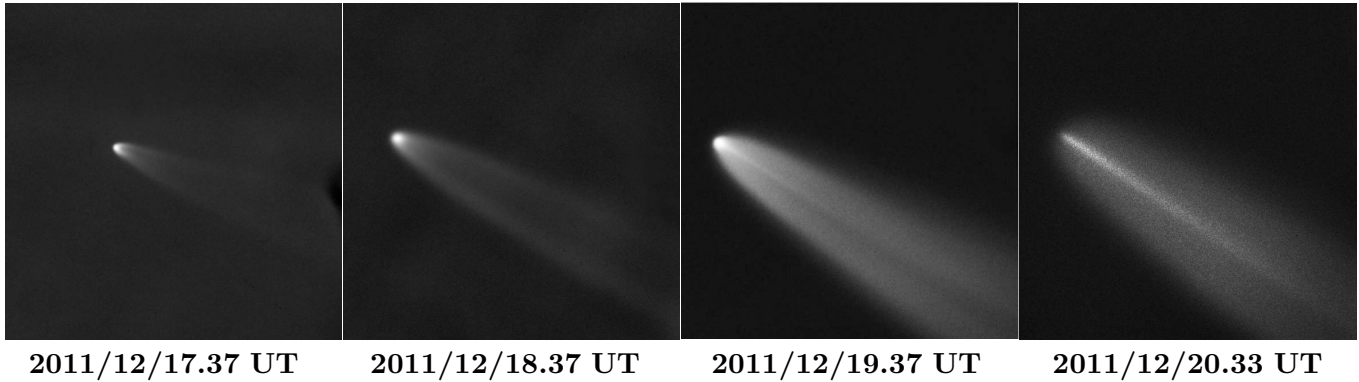}}}
 \vspace*{-21.95cm}
 \caption{Some of the earliest postperihelion images of Comet C/2011 W3
taken --- at the request of J.\ \v{C}ern\'y --- by J.\ Ebr, M.\ Prouza, P.\
Kub\'anek, and M.\ Jel\'{\i}nek with the FRAM 30-cm f/10 Schmidt-Cassegrain
reflector, a robotic, remotely controled telescope at the Pierre Auger
Observatory, Malargue, Argentina.  Each frame is approximately 11$^\prime$ on
a side, corresponding to some 430,000 km in the first image and 375,000 km in
the last image.  North is up, east to the left.  The comet was 5$^\circ\!$.8
from the Sun during the first exposure, 8$^\circ\!$.5 during the second,
11$^\circ\!$.0 during the third, and 13$^\circ\!$.3 during the last one.
These images provided the first evidence of the major physical changes in the
comet's morphological appearance, which culminated with the sudden, complete
loss of the nuclear condensation on December 20. (Image credit: J.\
\v{C}ern\'y, Czech Astronomical Society.)}
\vspace*{0.6cm}
\end{figure*}

The extensive sets of images taken by the various instruments on board 
the SOHO and STEREO spacecraft prove more useful for examining the comet's
dust-tail morphology (Sec.\ 10) and may also be useful for studying the light
curve in the general proximity of perihelion, even though many of the CCD
frames show the comet's head saturated.  We will not discuss the complex
issues of the SOHO and STEREO photometry, but would like to call attention
to a few very preliminary findings from our cursory inspection of the images
taken with the C2 and C3 coronagraphs on board SOHO.  Most of the inspected
images display the saturation artifact known as ``blooming,'' which has for
similar integration times been assumed to measure approximately the amount
of excess brightness by the number of affected pixels and therefore by the
length of the overflow streak.  Implementation of this admittedly
oversimplified rule does, however, lead to conclusions that are consistent
with the results based on firmer, independent evidence addressed later in
this paper.  It appears that the comet's activity became significantly lower
very close to perihelion over a period of several hours.  This apparent
attenuation is almost certainly a product of intensive sublimation of dust,
as examined in some detail in Secs.\ 10 and 11.

In any case, it seems that the comet's brightness reached a maximum about 0.3
day before perihelion, followed by rapid fading.  Up to three subsequent
outbursts (some perhaps multiple) may have occurred, peaking at, respectively,
0.4, 0.8, and about1$\frac{1}{2}$ days after perihelion, or on December 16.4,
16.8, and $\sim$17.5 UT.  A rigorous study of the light curve should verify
these preliminary results. \\[-0.05cm]

\section{Early Postperihelion Observations from Ground, and Sudden
Transformation of Comet's Appearance}
\vspace{0.3cm}

Although, astonishingly, the comet was imaged in daylight from the ground
as early as 1--2 days after perihelion by several observers (Kronk 2011),
the frames contained no reference stars to derive the comet's astrometric
positions.  Among these early postperihelion observations was a set of
images taken by a group of Czech observers between December 17.4 and 20.3
UT, who used a robotic, remotely controled telescope, a FRAM 30-cm f/10
Schmidt Cassegrain reflector of the Pierre Auger Observatory at Malargue,
Argentina.  While no astrometry was possible, comparison of the images
obtained on the four days (Fig.\ 1) shows major changes in the appearance
of the comet's head (\v{C}ern\'y 2011).  From December 17 to 18 the nuclear
condensation seemed to have grown and brightened a little, extending in
a broad fan in the tailward direction, but otherwise the morphology
remained essentially the same.  From December 18 to 19 the change was more
pronounced; even though the nuclear condensation remained clearly visible,
the quasi-parabolic contours of the tail became filled with much more
material and one can discern a streamer that extended for a few arcminutes
nearly along the tail axis.  Should it persist and change its orientation
predictably with time, such a feature is diagnostic of a sudden, brief
outburst of dust from the nucleus (Sec. 5).  By itself, an event of this
kind may or may not be part of a cataclysmic process that results in the
demise of the nucleus.  But the stunning change in the comet's appearance
between December 19 and 20 strongly suggests that during this episode,
portended by the morphological changes during the previous days, the nucleus
entirely disintegrated.  On December 20 the nuclear condensation {\it
completely disappeared\/} (Fig.\ 1) and the streamer, usually referred to as
a spine tail and much longer and more prominent now than the previous day,
dominated the comet's head and near-tail region.

On December 20 the comet was imaged at Malargue for the last time.  Although
the spaceborne observations with the HI1 imagers on both the STEREO-A and
STEREO-B spacecraft continued past this date (Table 1), their spatial
resolution was not sufficient to show the changes detected with clarity in
the Malargue images.  Fortunately, the observing conditions on the ground
were generally improving as the comet's elongation from the Sun steadily
continued to increase, and after a three-day gap earth-based monitoring
of the comet's head region resumed.

\begin{figure*}[ht]
 \vspace*{-4.9cm}
 \hspace*{-0.6cm}
 \centerline{
 \scalebox{1.095}{
 \includegraphics{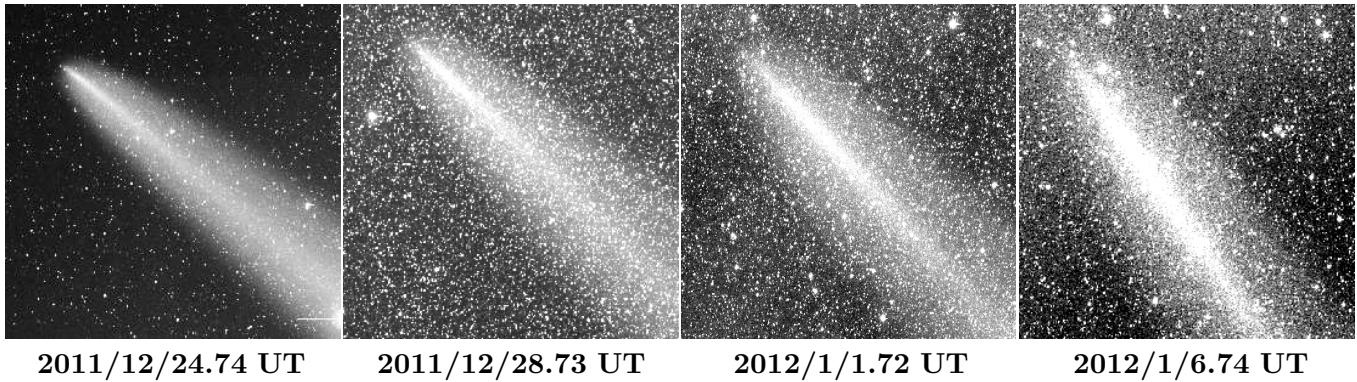}}}
 \vspace*{-21.9cm}
 \caption{ The head and adjoining portion of the spine tail of comet
C/2011 W3 taken between 2011 December 24 and 2012 January 6 by R.\ H.\
McNaught with the Uppsala 50-cm f/3.5 Schmidt Telescope, Siding Spring
Survey.  The fields of view are 48$^\prime$ on a side, corresponding to
1.39 million km on the first image and 1.04 million km on the last image.
North is up and the east to the left.  The Sun is to the upper left, but
at an angle to the direction of the tail.  Clearly seen is the clockwise
rotation and a gradual widening of the spine tail, which is encompassed by a
much broader quasi-parabolic envelope.  The surface brightness of the spine
tail reaches a flat maximum at some distance from the sunward tip; this
distance steadily increases with time.  The estimated location of the peak
surface brightness is populated by dust particles subjected to solar radiation
pressure of \mbox{$\beta_{\rm peak} \simeq 0.006$--0.007}, whose diameter is
\mbox{400--500} $\mu$m at an assumed bulk density of 0.4 g cm$^{-3}$.  [Image
credit:\ R.\ H.\ McNaught, Siding Spring Survey (UA/NASA/ANU).]}
\vspace*{0.5cm}
\end{figure*}

\section{Follow-up Postperihelion Ground-Based Observations, and Origin of
 Spine Tail}

The continuation of the high-resolution monitoring of the comet was very
important for finding out whether the loss of the nuclear condensation, so
unambiguously documented by the December 20 imaging, became indeed permanent
and irreversible. If the nucleus did indeed disappear, there would be no
definite point to bisect, and this condition would seem to thwart any
attempt at getting accurate postperihelion astrometric data and computing
a high-quality set of orbital elements, including a well-determined orbital
period.

Starting on December 23, a systematic series of observations of what remained
of the comet was begun by McNaught (2012) with the Uppsala 50-cm f/3.5 Schmidt
telescope of the Siding Spring Survey in Australia.  As seen from the examples
displayed in Fig.\ 2, the comet had nearly the same appearance in all images
taken between December 23 and January 18, fully confirming the morphology
in the Malargue images of December 20.  There was no trace of a nuclear
condensation, only a gradually vanishing ``hood'' at the sunward end of a
faint feature with quasi-parabolic contours, on which a much brighter,
rectilinear, spine-like tail was superimposed.  In the absence of a better
choice, McNaught measured what he perceived as possible candidate positions
for the tip of the tail at its sunward end, arguably the least objectionable
substitute for the missing nucleus.

In spite of this handicap, McNaught's imaging observations provide useful
information by allowing one to measure rather precisely (with a $\pm$1$^\circ$
precision) the systematic clockwise rotation of the spine tail, clearly
apparent from Fig.\ 2.  Combined with the measurements of the streamer on
the Malargue exposures from December 19--20, the position angle data measured
by the authors are listed in Table 2.  Each of the measured orientations in
column 2 was compared with a set of calculated position angles of {\it
synchronic\/} features, loci of dust particles of different sizes subjected
to a range of accelerations by solar radiation pressure but released from the
nucleus at the same time (e.g., Finson \& Probstein 1968).  The distribution
of the times of particle release, for which the synchrones' position angles
match exactly the measured orientations, have shown a sharp concentration in
time, with a random scatter of less than $\pm$0.4 day, providing evidence that
the spine tail was indeed a product of a major, fairly brief outburst (or a
rapid sequence of outbursts) that peaked on December $17.6 \pm 0.2$ UT, about
1.6 days after perihelion; the comet was then \mbox{$0.144 \pm 0.012$} AU
from the Sun. (For a preliminary report, see Seka\-nina 2012.)  The predicted
variations of the spine tail until mid-February 2012 are displayed in Fig.\ 3.

The event of December 17.6 (be it single or multiple), which is possibly
identical with the last of the three postperihelion outbursts mentioned at the
end of Sec.\ 3, must have begun only a fraction of a day earlier, probably
about December 17.2 UT, and led to the disappearance of the nuclear
condensation some 2--2$\frac{1}{2}$ days later and to the termination of
activity (Sec.\ 12).  Although details of this process are unknown at present,
it exhibits characteristics very similar to those of the cataclysmic
fragmentation of another sungrazer, comet C/1887 B1 (e.g., Kreutz 1901;
Marsden 1967, 2005; Sekanina 1984, 2002a; Sekanina \& Chodas 2004).  This
general conclusion is supported by the absence, from December 20 on, of any
traces of a second tail that would have contained freshly ejected dust; such
a tail would have preceded the spine tail some 5$^\circ$ to 7$^\circ$ in the
clockwise direction.

\begin{table}
\noindent
\begin{center}
{\footnotesize {\bf Table 2}\\[0.1cm]
{\sc Orientation of Spine Tail of Comet C/2011 W3}\\[0.2cm]
\begin{tabular}{l@{\hspace{0.08cm}}l@{\hspace{0.08cm}}r@{\hspace{0.75cm}}c@{\hspace{0.5cm}}c@{\hspace{0.95cm}}l}
\hline\hline\\[-0.15cm]
& & &
\multicolumn{2}{@{\hspace{-0.65cm}}c}{Position angle of spine tail}\\[-0.06cm]
& & &
\multicolumn{2}{@{\hspace{-0.65cm}}c}{\rule[0.6ex]{3.6cm}{0.4pt}}\\[-0.06cm]
\multicolumn{3}{@{\hspace{-0.3cm}}c}{Date (UT)} & measured & residual$^{\rm a}$
& Observer(s)\\[0.05cm]
\hline\\[-0.15cm]
2011 & Dec. & 19.37 & 239\rlap{$^\circ$} & +0$^\circ\!\!$.15 & Ebr et
al.$^{\rm b}$ \\
&      & 20.33 & 237 & $-$0.34 & $\;\;\;\;$" \\
&      & 23.75 & 233 &   +0.29 & McNaught \\
&      & 24.74 & 231 & $-$0.48 & $\;\;\;\;$" \\
&      & 26.74 & 229 & $-$0.12 & $\;\;\;\;$" \\
&      & 27.74 & 228 &   +0.02 & $\;\;\;\;$" \\
&      & 28.73 & 227 &   +0.11 & $\;\;\;\;$" \\
&      & 29.73 & 225 & $-$0.82 & $\;\;\;\;$" \\
&      & 30.73 & 225 &   +0.23 & $\;\;\;\;$" \\
&      & 31.73 & 224 &   +0.27 & $\;\;\;\;$" \\
2012 & Jan. &  1.72 & 223 &   +0.29 & $\;\;\;\;$" \\
&      &  2.73 & 222 &   +0.37 & $\;\;\;\;$" \\
&      &  3.73 & 221 &   +0.53 & $\;\;\;\;$" \\
&      &  6.74 & 214 & $-$0.17 & $\;\;\;\;$" \\
&      & 12.45 & $\;\:$52 & $-$0.34 & $\;\;\;\;$" \\
&      & 16.56 & $\;\:$51 & $-$0.01 & $\;\;\;\;$" \\
&      & 18.58 & $\;\:$52 & +0.06 & $\;\;\;\;$" \\[0.05cm]
\hline\\[-0.2cm]
\multicolumn{6}{l}{\parbox{8.1cm}{$^{\rm a}$\,{\scriptsize The difference
is:\ measured minus calculated{\vspace{-0.04cm}} from a synchronic feature
generated by dust outburst on 2011 December 17.6 UT.}}} \\[0.09cm]
\multicolumn{6}{l}{\parbox{8.1cm}{$^{\rm b}$\,{\scriptsize Images reported
by J. \v{C}ern\'y.}}}
%
%
\end{tabular}}
\end{center}
\end{table}

\section{Development of a Novel Technique for Orbit Determination}

There have been attempts to determine the orbital elements of C/2011 W3
using some of McNaught's measurements of the tip of the spine tail.  The
results of one such effort have been published by Williams (2011c).
Although he did not list the residuals, he quoted McNaught's remark that
``the tip of the `spine' \ldots lies pretty much on the front edge of the
parabolic hood.  It is rather less well defined [on Dec.\ 24] than on
Dec.\ 23.  For this reason I cannot be sure just how closely I am measuring
the same point as on Dec.\ 23'' (Williams 2011c).  Even though McNaught
continued his observations and reductions of the positions of the tip well
into January, his results from the post-Dec.\ 24 images have not been
published, nor have they been used by Williams to further update the orbit.

\begin{figure}[hb]
 \vspace*{-4.9cm}
 \hspace*{-0.25cm}
 \centerline{
 \scalebox{0.85}{
 \includegraphics{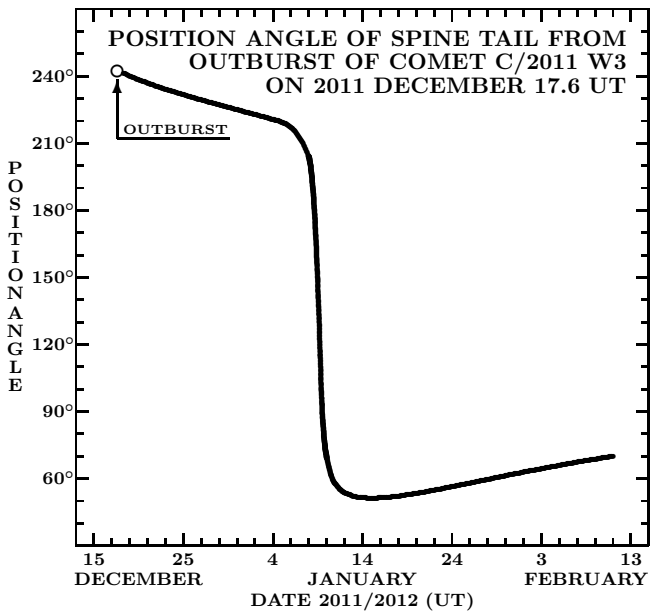}}}
 \vspace*{-11.5cm}
 \caption{ Temporal variations in the position angle (equinox J2000.0)
of the spine tail, described as a synchronic feature originating from an
outburst on 2011 December 17.6 UT.  The steep rate of change between 2012
January 8 and 11 is due to the comet's approach to within 1$^\circ\!$.5 of
the south celestial pole on January 9.}
\end{figure}

Fortunately, McNaught communicated the results of his continuing observations
to one of our colleagues, who provided us with the information conveyed
(Chesley 2012).  The outcome is 46 astrometric positions of the tip of the
spine tail, in addition to the three from December 23--24 already published
(Spahr et al.\ 2012).  In the following we describe a novel technique that we
devised to exploit this set of McNaught's astrometric observations, together
with his images of the spine tail.  The goal is to extract accurate positions
of the missing nucleus and to employ them in our orbit-determination efforts.

Every dust particle released from a comet's nucleus pursues its own orbit in
space, which differs from, and is independent of, the comet's subsequent orbit.
The orbital deviations are determined by the release time and circumstances as
well as the magnitude of solar radiation pressure that the particle has been
subjected to after release.  Radiation pressure forces the particle to move in
a field of reduced effective gravity compared with that acting on the comet.
The field intensity is described by the ratio $\beta$ between the acceleration
$\gamma_{\rm pr}$ due to radiation pressure and the Sun's gravitational
acceleration {\gsun}.  If $G$ is the gravitational constant and {\Msun} the
mass of the Sun, then, at a heliocentric distance $r$, \mbox{{\gsun}$(r) =
G\!\!$ {\Msun}/$r^2$}, whereas the acceleration $\gamma_{\rm pr}(r)$ is given
as a product of radiation pressure, $\wp(r)$, and the particle's effective
cross sectional area for radiation pressure, $A_{\rm pr}$, per its mass, $m$.
Here \mbox{$\wp(r)$ = {\Lsun}/$(4 \pi c r^2)$}, with {\Lsun} being the Sun's
total radiation energy emitted per unit time and $c$ the speed of light, while
$A_{\rm pr}$ is the product of the particle's geometrical cross sectional area,
$A$, and the efficiency for radiation pressure, $Q_{\rm pr}$.  The ratio
$\beta$ can thus be written in the form (e.g., Seka\-nina et al.\ 2001)
\begin{equation}
\beta = \frac{\gamma_{\rm pr}(r)}{\mbox{\gsun}(r)} = \frac{{\mbox{\Lsun}}
 Q_{\rm pr} A}{4 \pi c G \mbox{{\Msun}} m} = \frac{1.15 \, Q_{\rm pr}}{\rho x},
\end{equation}
where the last expression on the right applies specifically to a spherical
grain of a bulk density $\rho$ (in g cm$^{-3}$) and diameter $x$ (in $\mu$m).
For large particles (tens of microns across and larger) \mbox{$Q_{\rm pr}
\simeq 1$} (the effective cross section for radiation pressure is nearly
identical to the geometrical cross section) and $\beta$ is generally smaller
than 0.1.  For particles released from a comet orbiting the Sun in a
nearly parabolic path, even this small value of $\beta$ suffices to force
the dust move in distinctly hyperbolic orbits {\it concave\/} to the Sun.
For strongly absorbing submicron-sized grains $\beta$ can easily exceed unity
(a field with acceleration by radiation pressure exceeding gravitational
acceleration), and such particles move in hyperbolic orbits {\it convex\/}
to the Sun.  In the special case of \mbox{$\beta = 1$}, particles are
subjected to no force and therefore move along straight lines.

As follows from Eq.\ (1), the ratio $\beta$ is generally independent of
heliocentric distance.  However, the situation gets a little more complicated
when the comet is extremely close to the Sun, as briefly discussed in Sec.\
11.

It is because of the effects of solar radiation pressure that the dust
particles of different sizes released during a brief outburst line up in
the tail along a synchrone, on which the nucleus is located at \mbox{$\beta
= 0$}.  And because the motions of these particles through the tail are
independent of any changes in the comet's motion and behavior {\it after\/}
the outburst, including any misfortunes that the nucleus may then incur,
the undisturbed positions of the missing nucleus can be recovered from the
motion of the synchrone, if one can determine the point at which \mbox{$\beta
= 0$}.  Without an additional constraint the projected position of a
disintegrated nucleus is increasingly uncertain, but primarily in one
dimension only, {\it along\/} the synchrone.

In practice, a synchrone has a finite breadth for a variety of reasons:\
a finite duration of the outburst; being the product of a sequence
of outbursts; and/or dust particles acquiring lateral velocities upon their
release.  The general tendency for the synchronic features in the dust
tails of comets is to get broader with time, which gradually increases the
positional uncertainty {\it across\/} the synchrone.

Over a limited range of fairly low $\beta$ values, the synchrone is a straight
line with very high precision (often to better than $\pm$0$^\circ\!$.1 in
the position angle).  If the tip of the spine tail of C/2011 W3 measured by
McNaught is located on the synchrone's axis, the coordinates of the tip and
the spine tail's orientation define the synchrone's equation that also fits
the undisturbed position of the missing nucleus.  If \mbox{[$\alpha_{\rm
obs}(t)$, $\delta_{\rm obs}(t)$]} are the measured equatorial coordinates of
the tip of the spine tail and $p(t)$ is the tail's position angle at time $t$,
the equatorial coordinates of the missing nucleus, \mbox{[$\alpha(t)$,
$\delta(t)$]}, must satisfy a condition
\begin{equation}
15 \; \frac{\alpha(t) - \alpha_{\rm obs}(t)}{\delta(t) - \delta_{\rm obs}(t)}
  \cos{\delta_{\rm obs}(t)} = \tan{p(t)},
\end{equation}
where $\alpha(t)$ and $\alpha_{\rm obs}(t)$ are in hours and $\delta(t)$ and
$\delta_{\rm obs}(t)$ in degrees.

To further constrain the equatorial coordinates $\alpha(t)$ and $\delta(t)$,
we recall from Sec.\ 2 that the quality and congruence of the astrometric
positions from the preperihelion ground-based observations allowed us to
determine accurate sets of orbital elements, except for the osculating period
$P$.  If these early astrometric data are used to fit orbits with a number
of different {\it forced\/} values of the orbital period, the missing nucleus
must at any given time be located on this {\it line of orbital-period
variation\/} computed for that time.  This condition provides a second
constraint on the equatorial coordinates \mbox{[$\alpha$, $\delta$]} of
the missing nucleus, which can readily be found as those of the {\it point
of intersection\/} of the line of orbital-period variation with the line
that satisfies the equation of synchrone in Eq.\ (2).  This result also
provides useful information on the orbital period $P$ and determines the
separation distance of the measured tip of the spine tail from the missing
nucleus and thereby the critical value of $\beta$ for the largest dust
grains that are detected at the tail's tip.

Because of the lack of any reliable information on the orbital period at
the beginning of this exercise, it is necessary to search for the solution
iteratively.  In the first approximation, we choose three widely different
values of the orbital period, $P_{-1}$, $P_0$, and $P_1$, such that
\mbox{$P_{-1} < P_0 < P_1$} and \mbox{$P_0 - P_{-1} = P_1 - P_0 = \Delta
P_0$}, derive sets of orbital elements from the early astrometry, and
for each $t$ compute the corresponding topocentric coordinates
$[\alpha_{-1}(t), \delta_{-1}(t)]$, $[\alpha_0(t), \delta_0(t)]$, and
$[\alpha_1(t),\delta_1(t)]$.  We assume that in the given range of
orbital periods $P(t)$ both coordinates can be interpolated by fitting,
separately for each coordinate, a quadratic law
\begin{eqnarray}
\alpha(t) &  = & A(t) \! + \! B(t) [P(t)
 \! - \! P_0] \! + \! C(t) [P(t) \! - \! P_0]^2, \nonumber \\[-0.05cm]
 & & \\[-0.38cm]
\delta(t) & = & F(t) \! + \! G(t) [P(t) \! - \! P_0] \! + \! H(t)
[P(t) \! - \! P_0]^2, \nonumber
\end{eqnarray}
where
\begin{eqnarray}
A(t) &  =  & \alpha_0(t),  \nonumber \\[0.1cm]
B(t) &  =  & \frac{\alpha_1(t) - \alpha_{-1}(t)}{2\,\Delta P_0}, \\[0.1cm]
C(t) &  =  & \frac{\alpha_{-1}(t) + \alpha_1(t) - 2 \,
\alpha_0(t)}{2 \, (\Delta P_0)^2}, \nonumber
\end{eqnarray}
and similarly for $F(t)$, $G(t)$, and $H(t)$.  After inserting for $\alpha(t)$
and $\delta(t)$ from Eqs.\ (3) into Eq.\ (2) we obtain for \mbox{$\Delta P(t)
= P(t) \! - \! P_0$}:
\begin{equation}
 \begin{array}{l}
[H(t) \! - \! \zeta(t) C(t)] \, [\Delta P(t)]^2 \! + \! [G(t) \! - \! \zeta(t)
  B(t)] \, \Delta P(t) \\[0.1cm]
  + F(t) \! - \! \delta_{\rm obs}(t) - \zeta(t) [A(t) \! - \! \alpha_{\rm
  obs}(t)] = 0,
 \end{array}
\end{equation}
where
\begin{equation}
\zeta(t) = 15 \cos{\delta_{\rm obs}(t)} \cot{p(t)}.
\end{equation}
One of the $\Delta P(t)$ roots of Eq.\ (5), which must give $P(t)$ outside
the chosen range of \mbox{$\langle P_{-1}, P_1 \rangle$}, is ignored.  The
other root is used to determine the desired topocentric coordinates
[$\alpha(t)$, $\delta(t)$] from Eq.\ (3).  At the end of each iteration
the values of $P(t)$ are averaged:
\begin{equation}
\langle P \rangle = \frac{1}{N} \, \sum_{i=1}^{N} P(t),
\end{equation}
where $N$ is the number of images included in the exercise.  It is also
verified that scatter of the $P$ values is not excessive (not more than
a few percent of $\langle P \rangle$) and that their distribution is
essentially random.  If $\langle P \rangle$ differs substantially from
the starting value of $P_0$, one should contemplate another iteration by
replacing this value of $P_0$ with $\langle P \rangle$, tightening the
interval $\Delta P_0$, and repeating the entire procedure, including the
determination of the time of outburst that controls the position angles
of the spine tail.  The ultimate test is provided by the residuals of
the corrected equatorial coordinates of the missing nucleus from the  new
orbital solution.

\begin{figure*}[ht]
 \vspace*{-4cm}
 \hspace*{0.26cm}
 \centerline{
 \scalebox{0.85}{
 \includegraphics{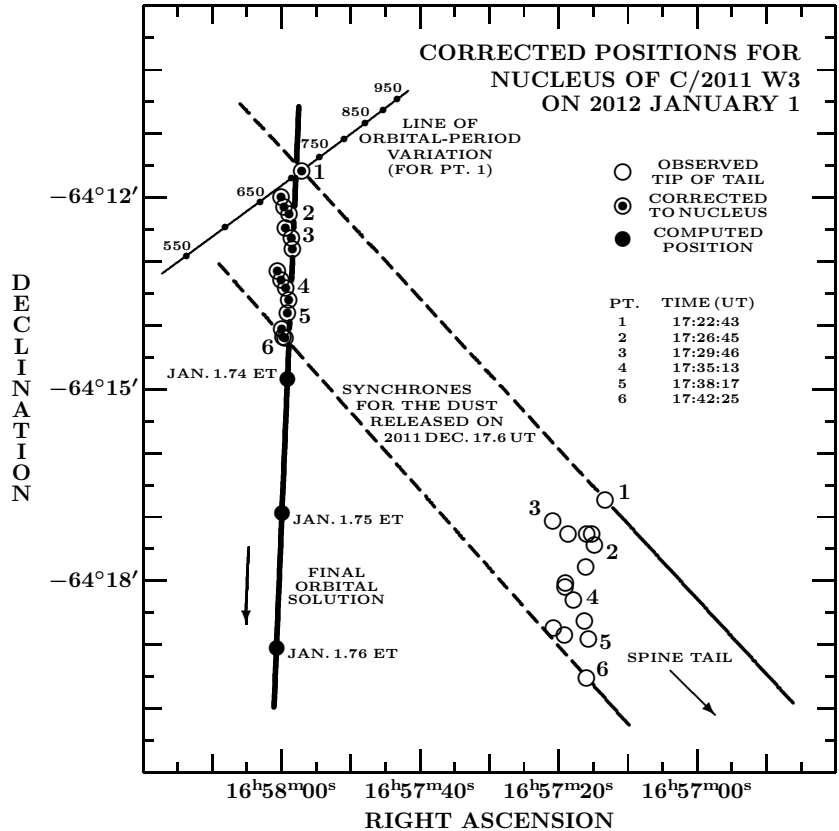}}}
 \vspace*{-10.25cm}
 \caption{Schematic illustration of the novel technique developed for
determining the corrected topocentric positions of the missing nucleus of
Comet C/2011 W3 (equinox J2000.0).  This example refers to the 15 astrometric
positions of the tip of the spine tail measured by R. H.  McNaught on 2012
January 1 between 17:22:43 and 17:42:25 UT (open circles).  Plotted for
the first (Point 1) and the last (Point~6) positions is the spine tail's
orientation (medium-thick lines) as a synchrone that refers to the dust
outburst on December 17.6 UT.  Projected in the direction opposite the
direction of the spine tail (broken lines shown again only for Points 1 and
6), each of these lines intersects the corresponding line of orbital-period
variation, which consists of the predicted topocentric positions of the missing
nucleus calculated from the preperihelion high-quality observations as a
function of the {\it forced\/} orbital period $P$.  Drawn only for the time
of Point~1, this line of orbital-period variation is shown as a thin line
calibrated with a number of $P$ values (in years).  The points of intersection,
plotted as circled dots, are the reconstructed positions of the missing
nucleus.  We note that the line of orbital-period variation for Point~1
intersects the synchrone at a value of $P$ slightly exceeding 700 years.  The
heavy line is the resulting orbital solution with the ephemeris positions
at three times, plotted by filled circles. The positions of the spine tail's
measured tips are seen to be several arcminutes away from the comet's orbit;
these distances illustrate the very large magnitude of the introduced errors,
if the tip measurements are not corrected.}
\vspace*{0.4cm}
\end{figure*}

\section{Computations and the Final Orbit}

To keep the number of necessary iterations of the proposed technique to
a minimum, we needed the best possible first-approximation set of
orbital elements, {\it including\/} $P$.  On the one hand, we learnt
that the preperihelion observations alone were inadequate to provide such
a set.  On the other hand, we found ourselves in an unenviable
situation in regard to the sources of postperihelion ground-based data,
with no astrometry possible either because of the absence of
reference stars in images in which the comet still possessed a nuclear
condensation (to bisect in a measuring machine) or because of the loss
of the condensation (when reference stars were plentiful).  Under these
circumstances, we decided to use some of the six positions measured
by Kracht (2011) in the STEREO-B COR2 images taken about $\frac{1}{2}$ day
after perihelion (Sec. 3).  Combined with the high-quality ground-based
positions obtained before perihelion, they offered a solution
acceptable for our purpose, with a maximum residual of 9$^{\prime\prime}$,
and yielded an osculating period of \mbox{$785 \pm 14$} years.

Based on this finding, we used the same sample of observations to compute
three sets of orbital elements by successively forcing the orbital period
to 600, 800, and 1000 years.  At this point we applied the
technique described in Sec.\ 6 to McNaught's 46 astrometric observations
made between 2011 December 23 and 2012 January 6, and from Eq.\ (3) we
obtained the first set of predicted positions for the missing nucleus.  The
gist of the procedure is depicted in Fig.\ 4 on an example of 15 positional
measurements obtained by McNaught on 2012 January 1.  The average orbital
period resulting from this exercise was equal to \mbox{$\langle P \rangle
= 709 \pm 6$} years.  We got ready for a second iteration by inspecting
the used positional data in the orbital run.  Having found a solution that is
much closer to the true orbital period of the comet, we were no longer
critically dependent on the postperihelion data points from STEREO-B.
They were not used in the second iteration, because the preperihelion 
astrometric observations alone allowed computer runs with new {\it
forced\/} values of the orbital period.  For the nominal run, we chose,
in accordance with the developed procedure, \mbox{$P_0 = 709$} years, and the
resulting new set of orbital elements was used to correct the derived
position angles of the spine tail.  We narrowed down the interval of
$\Delta P$ to $\sim$5$\sigma$ of \mbox{$\langle P \rangle$}, or 30 years
(i.e., $P_{-1} = 679$ years and $P_1 = 739$ years), and proceeded with the
second iteration of the missing nucleus' positions.  Comparing the results
of the two iteration cycles, we noticed that in 38 out of the 46
observations the agreement in the derived positions of the missing nucleus
was 6$^{\prime\prime}$ or better in either coordinate, and that in the
two worst matches the differences amounted to 20$^{\prime\prime}$ and
10$^{\prime\prime}$, respectively.  We felt that another iteration was
unnecessary because it would only lead to changes in the subarcsecond range.
Consequently, the iterative process was at this point terminated.
\begin{table*}
\begin{center}
{\footnotesize {\bf Table 3} \\[0.08cm]
{\sc Derived Topocentric Positions of the Missing Nucleus of Comet C/2011 W3
and Related Spine-Tail Data}\\[0.08cm]
\begin{tabular}{l@{\hspace{0.08cm}}l@{\hspace{0.08cm}}r@{\hspace{0.9cm}}c@{\hspace{0.8cm}}c@{\hspace{0.5cm}}c@{\hspace{0.18cm}}c@{\hspace{0.35cm}}c@{\hspace{0.75cm}}c@{\hspace{0.5cm}}c}
\hline\hline\\[-0.25cm]
& & & \multicolumn{2}{@{\hspace{-0.25cm}}c}{Equatorial coordinates}
& Distance of & Acceleration & Particle
& \multicolumn{2}{@{\hspace{-0.17cm}}c}{Residuals$^{\rm a}$}\\[-0.02cm]
\multicolumn{3}{@{\hspace{-0.08cm}}c}{Observation time}
& \multicolumn{2}{@{\hspace{-0.25cm}}c}{\rule[0.7ex]{3.85cm}{0.4pt}}
& tail's tip from & parameter, & diameter,
& \multicolumn{2}{@{\hspace{-0.17cm}}c}{\rule[0.7ex]{2.25cm}{0.4pt}}\\[-0.02cm]
\multicolumn{3}{@{\hspace{-0.08cm}}c}{$t$ (UT)}
& R.A.(2000) & Decl.(2000) & nucleus, $\ell_{\rm tip}$ & $\beta_{\rm tip}$ 
& $x_{\rm tip}$\,(mm) & R.A. & Decl.\\[0.06cm]
\hline \\[-0.25cm]
& & & $^{{\rm h} \;\;\; {\rm m} \;\;\; {\rm s}}$
& $\;\;\;\;\;^{\circ \;\;\;\:\prime \;\;\;\:\,\prime\:\!\!\prime}$
& $^\prime\;$ & & & $\;\;\;\:^{\prime\:\!\!\prime}$
& $\;\;\;\:^{\prime\:\!\!\prime}$\\[-0.25cm]
2011 & Dec. & 23.75224 & 16 59 21.69 & $-$39 25 32.8 & 1.03 & 0.00198
& 1.45 &   +3.5 & $-$4.8 \\
&      & 24.74267 & 16 58 18.93 & $-$41 44 14.7 & 1.40 & 0.00202
& 1.42 &   \llap{(}+4.4 & $-$6.1\rlap{)} \\
&      & 24.74785 & 16 58 18.54 & $-$41 44 57.0 & 1.07 & 0.00154
& 1.87 &   +3.2 & $-$3.8 \\
&      & 26.73669 & 16 56 50.13 & $-$46 40 58.2 & 1.20 & 0.00107
& 2.69 &   +0.7 &   +1.6 \\
&      & 26.73853 & 16 56 50.14 & $-$46 41 16.6 & 0.73 & 0.00065
& 4.42 &   +1.3 & +0.4 \\
&      & 26.74041 & 16 56 50.32 & $-$46 41 38.6 & 1.19 & 0.00106
& 2.71 &   +3.8 & $-$4.0 \\
&      & 27.74339 & 16 56 23.95 & $-$49 20 34.6 & 3.04 & 0.00221
& 1.30 &   +0.6 &   +1.6 \\
&      & 27.74468 & 16 56 24.05 & $-$49 20 49.1 & 2.54 & 0.00185
& 1.55 &   +1.8 & $-$0.3 \\
&      & 28.73164 & 16 56 10.10 & $-$52 03 48.9 & 2.75 & 0.00166
& 1.73 & \llap{(}$-$4.2 &   +8.6\rlap{)} \\
&      & 29.73101 & 16 56 09.99 & $-$54 55 52.1 & 2.89 & 0.00147
& 1.96 & \llap{(}$-$5.7 &   +9.4\rlap{)} \\
&      & 30.72663 & 16 56 26.64 & $-$57 54 19.8 & 2.79 & 0.00122
& 2.36 &   +3.8 & $-$2.5 \\
&      & 31.73210 & 16 57 00.27 & $-$61 01 01.5 & 4.89 & 0.00184
& 1.56 & $-$3.1 &   +4.1 \\
&      & 31.73270 & 16 56 59.89 & $-$61 01 05.6 & 5.15 & 0.00194
& 1.48 & \llap{(}$-$6.0 &   +6.8\rlap{)} \\ 
&      & 31.73329 & 16 57 00.37 & $-$61 01 15.4 & 5.40 & 0.00203
& 1.42 & $-$2.7 &   +3.7 \\
&      & 31.73387 & 16 57 00.26 & $-$61 01 21.0 & 4.55 & 0.00171
& 1.68 & $-$3.7 &   +4.8 \\
&      & 31.73447 & 16 57 01.36 & $-$61 01 35.1 & 4.70 & 0.00177
& 1.62 &   +4.0 & $-$2.5 \\
2012 & Jan. &  1.72411 & 16 57 57.11 & $-$64 11 34.9 & 7.01 & 0.00231
& 1.24 & \llap{(}$-$8.8 &   +7.0\rlap{)} \\
&      &  1.72505 & 16 58 00.08 & $-$64 11 59.9 & 7.18 & 0.00236
& 1.22 & \llap{(+}10.2$\!\!\!\:$ & $-$6.9\rlap{)} \\
&      &  1.72599 & 16 57 59.62 & $-$64 12 08.4 & 6.99 & 0.00230
& 1.25 & \llap{(}+6.7 & $-$4.3\rlap{)} \\
&      &  1.72691 & 16 57 58.90 & $-$64 12 15.5 & 7.05 & 0.00232
& 1.24 &   +1.5 & $-$0.6 \\
&      &  1.72786 & 16 57 59.43 & $-$64 12 28.8 & 6.52 & 0.00215
& 1.34 &   +4.5 & $-$2.7 \\
&      &  1.72901 & 16 57 58.57 & $-$64 12 37.8 & 6.04 & 0.00199
& 1.44 & $-$1.7 &   +1.9 \\
&      &  1.72995 & 16 57 58.44 & $-$64 12 47.9 & 6.78 & 0.00223
& 1.29 & $-$3.0 &   +2.8 \\
&      &  1.73089 & 16 58 00.61 & $-$64 13 09.1 & 6.65 & 0.00219
& 1.31 & \llap{(+}10.7$\!\!\!\:$ & $-$7.3\rlap{)} \\
&      &  1.73184 & 16 58 00.05 & $-$64 13 17.3 & 6.56 & 0.00216
& 1.33 & \llap{(}+6.6 & $-$4.3\rlap{)} \\
&      &  1.73279 & 16 57 59.39 & $-$64 13 25.0 & 6.64 & 0.00218
& 1.32 &   +1.8 & $-$0.8 \\ 
&      &  1.73395 & 16 57 58.98 & $-$64 13 36.3 & 6.83 & 0.00225
& 1.28 & $-$1.4 &   +1.6 \\
&      &  1.73492 & 16 57 59.18 & $-$64 13 48.4 & 6.95 & 0.00229
& 1.26 & $-$0.6 &   +0.9 \\
&      &  1.73587 & 16 58 00.01 & $-$64 14 03.2 & 6.53 & 0.00215
& 1.34 &   +4.4 & $-$2.7 \\
&      &  1.73682 & 16 57 59.46 & $-$64 14 11.3 & 6.19 & 0.00204
& 1.41 &   +0.3 &  +0.4 \\
&      &  1.73779 & 16 57 59.75 & $-$64 14 22.9 & 7.00 & 0.00230
& 1.25 &   +1.7 &   +0.3 \\
     &      &  2.72541 & 16 59 28.62 & $-$67 29 52.8 & 6.39 & 0.00186
     & 1.55 & $-$4.9 &   +2.7 \\
     &      &  2.72634 & 16 59 28.49 & $-$67 30 03.3 & 5.40 & 0.00157
     & 1.83 & \llap{(}$-$6.2 &   +3.5\rlap{)} \\
     &      &  2.72731 & 16 59 29.53 & $-$67 30 18.0 & 6.16 & 0.00179
     & 1.61 & $-$0.9 &   +0.6 \\
     &      &  2.72826 & 16 59 27.73 & $-$67 30 23.7 & 7.11 & 0.00207
     & 1.39 & \llap{($-$}11.8$\!\!\!\:$ &   +6.4\rlap{)} \\
     &      &  2.72920 & 16 59 27.78 & $-$67 30 34.8 & 6.75 & 0.00196
     & 1.47 & \llap{($-$}12.2$\!\!\!\:$ &   +6.7\rlap{)} \\
     &      &  3.73303 & 17 01 45.81 & $-$70 54 27.0 & 7.45 & 0.00194
     & 1.48 & \llap{($-$}20.7$\!\!$ & +6.5\rlap{)} \\
     &      &  3.73398 & 17 01 45.84 & $-$70 54 38.6 & 7.76 & 0.00202
     & 1.42 & \llap{($-$}21.3$\!\!\!\:$ & +6.7\rlap{)} \\
     &      &  3.73493 & 17 01 45.82 & $-$70 54 50.1 & 7.94 & 0.00206
     & 1.40 & \llap{($-$}22.2$\!\!\!\:$ & +6.9\rlap{)} \\
     &      &  3.73587 & 17 01 44.23 & $-$70 54 58.8 & 7.81 & 0.00203
     & 1.42 & \llap{($-$}30.8$\!\!\!\:$ & +9.9\rlap{)} \\
     &      &  3.73682 & 17 01 49.66 & $-$70 55 19.4 & 7.59 & 0.00197
     & 1.46 & $-$5.0 &   +1.1 \\
     &      &  6.73909 & 17 22 41.59 & $-$81 24 11.4 & 8.60 & 0.00169
     & 1.70 & \llap{($-$}28.4$\!\!\!\:$ & $-$5.0\rlap{)} \\
     &      &  6.74001 & 17 22 54.11 & $-$81 24 21.1 & 8.65 & 0.00170
     & 1.69 & $-$2.3 & $-$3.0 \\
     &      &  6.74095 & 17 22 39.36 & $-$81 24 45.4 & 8.04 & 0.00158
     & 1.82 & \llap{($-$}37.4$\!\!\!\:$ & $\:$\llap{$-$}15.3\rlap{)} \\
     &      &  6.74189 & 17 22 46.40 & $-$81 24 52.5 & 8.51 & 0.00168
     & 1.71 & \llap{($-$}23.7$\!\!\!\:$ & $\:$\llap{$-$}10.4\rlap{)} \\
     &      &  6.74282 & 17 22 58.62 & $-$81 24 55.5 & 8.94 & 0.00176
     & 1.63 &    +1.6 & $-$1.6 \\
     &      & 12.44813 & $\;\:$4 10 23.94 & $-$78 38 18.4 & \llap{2}3.29
     & 0.00340 & 0.85 & \llap{($-$6}26.6$\!\!\!\:$
     & $\:$\llap{$-$5}31.2\rlap{)$^{\rm b}$} \\
     &      & 16.56300 & $\;\:$4 29 30.00 & $-$65 58 51.9 & \llap{1}2.67
     & 0.00171 & 1.68 & \llap{($-$}98.0$\!\!\!\:$
     & $\:$\llap{$-$1}25.8\rlap{)} \\
     &      & 18.58108 & $\;\:$4 32 31.89 & $-$60 35 49.1 & \llap{1}0.75
     & 0.00143 & 2.01 & \llap{($-$1}04.9$\!\!\!\:$
     & $\:$\llap{$-$1}53.3\rlap{)} \\[0.05cm]
\hline\\[-0.2cm]
\multicolumn{10}{l}{\parbox{15.7cm}{$^{\rm a\,}${\scriptsize The difference
 is:\ the position derived from measurement minus the{\vspace*{-0.06cm}}
 position computed from the final orbital solution; the residual in R.A.\
 includes the factor cos(Decl.); the positions whose residuals are
 parenthesized have not been used in the solution.}}} \\
\multicolumn{10}{l}{\parbox{15.7cm}{$^{\rm b\,}${\scriptsize The starting
 position of the tail's tip appears to be grossly in error.}}}\\[-0.05cm]
\end{tabular}}
\end{center}
\end{table*}

For the times of McNaught's exposures, the resulting topocentric positions
of the missing nucleus from the final iteration are listed in columns 2 and
3 of Table 3.  Included also are the derived positions from the last three
observations that McNaught made on January 12, 16, and 18, although they are
too inaccurate to be used in the orbital solution.  In column 4 the table
lists the angular distance $\ell_{\rm tip}$ of the measured tip of the spine
tail from the nucleus' predicted position.  From an ephemeris of the December
17.6 synchrone this angular distance is converted in column 5 into the
radiation-pressure parameter $\beta_{\rm tip}$, which in the given range
varies practically linearly with the angular distance and which, according
to Eq.\ (1), is diagnostic of the diameter $x_{\rm tip}$ of the dust particles
that were detected by McNaught at the tip of the spine tail.  These are
the largest particles in the observed tail.  For an assumed particle bulk
density of 0.4 g cm$^{-3}$ (e.g., Richardson et al.\ 2007), this particle
diameter is given in column 6.  Remarkably, Table 3 and Fig.\ 5 show that the
values of $\beta_{\rm tip}$ and $x_{\rm tip}$ did not vary systematically
during the nearly four weeks of McNaught's observation, averaging,
respectively, \mbox{$0.00191 \pm 0.00042$} and \mbox{$1.6 \pm 0.5$} mm.
On the other hand, the angular distance $\ell_{\rm tip}$ was increasing
nonuniformly, but steadily, with time, resembling the rate of recession of
a separated companion fragment from the nucleus of its parent comet.
%

The tabulated topocentric positions of the missing nucleus were at this point
incorporated into the input file for the final orbit determination run.
It should be remembered that McNaught's intention was to measure positions
of the candidates of the tail's tip, not positions of extreme points on the
tail's {\it axis\/}, which would be ideal for our orbit determination efforts.
On the days on which he took more than one image, it is likely that the
tail's tip candidates differed from image to image, as is abundantly clear
from his comments to Williams (2011c).  And as the spine tail steadily
broadened, the chance that any measured tip candidate happened to be in
fact located very close to the axis became increasingly remote.  One should
therefore expect that for only some of the derived astrometric positions of
the missing nucleus will the residuals from the final orbit be acceptable for
the orbit determination run.  Also expected is that the quality of these
positions should deteriorate with the spine tail's growing breadth and
therefore with time.

\begin{figure}[hb]
 \vspace*{-3.8cm}
 \hspace*{-0.45cm}
 \centerline{
 \scalebox{0.935}{
 \includegraphics{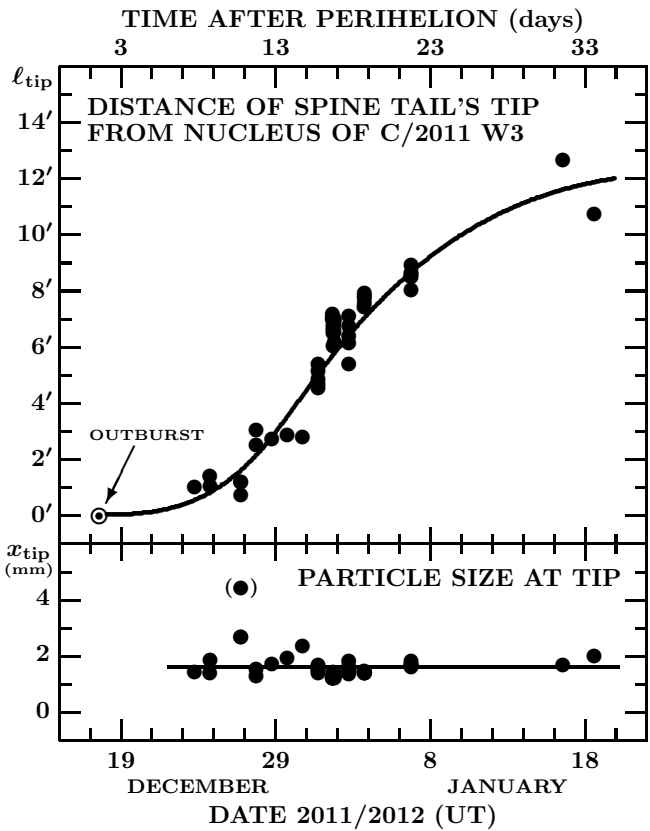}}}
 \vspace*{-12.7cm}
 \caption{Angular distance $\ell_{\rm tip}$ of the spine tail's tip, measured
by McNaught in his images of C/2011 W3, taken with the Uppsala Schmidt
Telescope between December 23 and January 18, as a function of time (top);
and the derived diameter of dust particles situated at the tip (bottom).
The distance increases with time, whereas the particle dimensions remain
the same.}
\vspace*{0.2cm}
\end{figure}

The final orbital solution, which includes the relativistic effect, is
presented in Table 4.  It is based on a total of 123 observations between
2011 November 27 and 2012 January 6, all ground-based, which left in
either coordinate a residual not exceeding 5$^{\prime\prime}$.  Of these,
96 are preperihelion (from November 27 to December 10) and 27 are
postperihelion.  All postperihelion entries come from Table 3.  This means
that the position of the missing nucleus was successfully recovered
from nearly 60 percent of McNaught's positions of the tail's tip from
images he took between Decmeber 23 and January 6.  Given the odds our
approach was facing, we consider this number to be quite satisfactory.  As
expected, the three positions between January 12 and 18 turned out
to offer unacceptable residuals, and the starting position from
January 12 was clearly incorrect.  The residuals of all the positions
of the missing nucleus are listed in the last two columns of Table 3, with
the rejected positions parenthesized.  The residuals from the preperihelion
observations are excellent, showing no systematic trends whatsoever, and
largely in the subarcsecond range.  The two most significant contributions,
with images in both cases obtained robotically, were from the Pierre
Auger Observatory, Malargue, Argentina (code I47, 42 accepted positions)
and from the Remote Astronomical Society Observatory, Mayhill, New Mexico
(code E03, 21 accepted observations); combined they account for nearly
two-thirds of all used preperihelion positional data.
\begin{table*}
\begin{center}
{\footnotesize {\bf Table 4}\\[0.1cm]
{\sc Final Set of Orbital Elements Adopted for Comet C/2011 W3
  and\\Integration One Revolution Back in Time (Eq. J2000.0)}\\[0.2cm]
\begin{tabular}{l@{\hspace{0.6cm}}r@{\hspace{0.07cm}}c@{\hspace{0.07cm}}l@{\hspace{0.5cm}}c}
\hline\hline\\[-0.15cm]
Orbital element & \multicolumn{3}{@{\hspace{0.6cm}}c}{Current
  apparition}  & Previous return$^{\rm a}$ \\[0.1cm]
\hline\\[-0.15cm]
Epoch of osculation (ET) & & \llap{2011 D}e\rlap{c.\,25.0} &
 & 1329 Jan.\,6.0 \\[0.05cm]
Time of perihelion passage, $t_\pi$ (ET) & 2011 Dec.\,16.011810 & $\pm$
        & 0.000040 & 1329 Jan.\,4.9336 \\
Argument of perihelion, $\omega$ & 53$^\circ\!$.5103 & $\pm$
        & 0$^\circ\!$.0020 & $\;\:$52$^\circ\!$.815 \\
Longitude of ascending node, $\Omega$ & 326$^\circ\!$.3694 & $\pm$
        & 0$^\circ\!$.0027 & 325$^\circ\!$.276 \\
Orbital inclination, $i$  & 134$^\circ\!$.3559 & $\pm$ & 0$^\circ\!$.0012
 & 133$^\circ\!$.811 \\
Perihelion distance, $q$ (AU) & 0.00555381 & $\pm$ & 0.00000007 & 0.0059198 \\
Orbital eccentricity, $e$ & 0.99992942 & $\pm$ & 0.00000014
 & 0.9999237 \\ [0.05cm]
Osculating orbital period, $P$ (yr) & 698 & $\pm$ & 2 & 684 \\ [0.1cm]
\hline\\[-0.3cm]
\multicolumn{5}{l}{\parbox{12cm}{$^{\rm a\,}${\scriptsize Dates in the
Julian calendar.}}}
\end{tabular}}
\end{center}
\vspace*{-0.05cm}
\end{table*}

Out of a total of 46 positions available to us from astrometric measurements
of the images taken on board the SOHO and STEREO spacecraft, none satisfied
our adopted cutoff of $\pm$5$^{\prime\prime}$ for the residuals.  As expected
from our discussion in Sec.\ 3, the least discordant were again the measurements
by Kracht (2011) from the six COR2-B images of Dec.\ 16, which left residuals
of up to 27$^{\prime\prime}$ in right ascension and up to 15$^{\prime\prime}$
in declination.  The 24 positions from the STEREO-B images between Dec.\ 10
and 14 showed surprisingly small residuals in declination, of up to
12$^{\prime\prime}$, but strongly systematic, entirely unacceptable residuals
of up to 60$^{\prime\prime}$ in right ascension.  The 8 data points from the
SOHO images on Dec.\ 14 had the opposite problem; 5 of them had residuals
less than 10$^{\prime\prime}$ in right ascension, but all were off by
40$^{\prime\prime}$ to 87$^{\prime\prime}$ in declination.  The least
satisfactory was a set of 8 positions from STEREO-A (listed by Spahr et al.\
2012), which had residuals scattered wildly from $-$94$^{\prime\prime}$
to +97$^{\prime\prime}$ in right ascension and from $-$78$^{\prime\prime}$ to
+4$^{\prime\prime}$ in declination.

From the orbital period in Table 4 we conclude that comet C/2011 W3 is the {\it
first major member of the new, 21st-century cluster of bright Kreutz
sungrazers\/} whose existence was predicted by the authors of this paper in
2007 (Sec.\ 1).  The comet cannot be the return of a fragment of any of the
sungrazing comets observed since the 17th century,
contrary to speculations based on preliminary
sets of orbital elements.  The perihelion distance agrees closely with the
perihelion distances of the bright sungrazers C/1843 D1, C/1880 C1, and to
a lesser degree C/1963 R1, but all three angular elements have values unlike
any of the known bright members of the system.\footnote{As a matter of
historic curiosity, it should be pointed out that the first orbit for the
sungrazer C/1945 X1, calculated by Cunningham (1946a, 1946b), was in fact
somewhat similar to the orbit of C/2011 W3 ($\omega = 50^\circ\!$.9, $\Omega
= 322^\circ\!$.3, $i = 137^\circ\!$.0, $q = 0.0063$ AU), but it was derived
from only crudely determined positions of the comet.  The photographic plates
with the comet's images were properly measured and reduced only in 1952
(Marsden 1967) and none of Marsden's (1989) resulting orbits shows any obvious
resemblance to that of C/2011 W3.}  This suggests that there may exist yet
another subcategory of bright sungrazers that has never been considered in
the evolutionary models of the Kreutz system.  Table 4 also shows the results
of integrating the orbit back one revolution about the Sun, to the early 14th
century, suggesting the presence of fairly minor perturbations.  The related
issues and implications of the comet's orbit are addressed in Sec.\ 13.

\section{Particle Velocities in Spine Tail and Its Anomalous Brightness
 Profile}

Besides the clearly apparent clockwise rotation of the spine tail, both the
Malargue images in Fig.\ 1 and the Siding Spring images in Fig.\ 2 show
a quasi-parabolic envelope, in which the spine tail is immersed and which,
based on its approximate axial orientation, may have originated from the
previous outburst that peaked on about December 16.8 UT or 0.8 day after
perihelion (Sec.\ 3).  While the spine tail dominates the images
brightness-wise, the previous outburst was apparently more violent, as
suggested by a much greater breadth of the envelope and, consequently, much
higher ejection velocities of the released dust.  Indeed, if the width of
the spine tail and the envelope are both due entirely to lateral velocities
of the particles at the time of release, these velocities come out to be
typically up to \mbox{20--30} m s$^{-1}$ for the spine tail, but up to at
least 150 m s$^{-1}$ for the envelope.  Whereas the latter velocities are
fairly typical of microscopic dust ejected due to an interaction with
outflowing gas, the former are so unusually low that they suggest that
the entire residual mass of the nucleus simply collapsed.

One may question whether the episode that led to the formation of the spine
tail can at all be called an outburst.  It may be more appropriate to refer
to it as a {\it cataclysmic fragmentation event\/} or {\it terminal
collapse\/} or {\it complete disintegration\/}, but we continue to use all
these terms interchangeably even after we address the physical nature of the
process in Sec.\ 12.

The distribution of surface brightness along the spine tail, which displays a
broad maximum at some distance from the sunward tip in all McNaught's images,
is positively peculiar.  The gradual rise in the near-head brightness
with increasing distance from the nucleus is anomalous in that it contradicts
a typical power law for the particle-size distribution of dust comets.  In a
low-ejection-velocity approximation, the Finson-Probstein (1968) approach
provides the following expression for the variation of the surface brightness
$\cal I$ as a function of the radiation-pressure acceleration parameter
$\beta$ along a synchronic feature observed at a time $t_{\rm obs}$ and made
up of dust particles released from the nucleus at a time $t_{\rm obs}\!- \tau$:
\begin{equation}
{\cal I}(\beta) \; {\textstyle \propto} \; \frac{p \, \phi(\alpha)}{r^2} \,
 \frac{\beta^{\,k-4}}{|\Im(\xi,\eta;\beta,\tau)|},
\end{equation}
where $p$ is the geometric albedo of the particles, $\phi(\alpha)$ their
phase function at a phase angle $\alpha$ [normalized to $\phi(0^\circ) = 1$],
$x^{-k} dx$ is the differential distribution function of particle diameters
$x$, $r$ is the heliocentric distance, and $\Im$ is a Jacobian that converts
the radial and transverse rectangular coordinates, $\xi$, $\eta$, into
$\beta$ and $\tau$
\begin{equation}
\Im(\xi,\eta;\beta,\tau) = \left| \begin{array}{cc}
{\displaystyle \frac{\partial \xi}{\partial \beta}}
 & {\displaystyle \frac{\partial \xi}{\partial \tau}} \\[0.25cm]
{\displaystyle \frac{\partial \eta}{\partial \beta}}
 & {\displaystyle \frac{\partial \eta}{\partial \tau}}
\end{array} \right| .
\end{equation}
At a geocentric distance $\Delta$, the coordinates $\xi$ and $\eta$ are
projected onto the tangent plane of the sky at the comet's nucleus and
reckoned from it, $\xi$ pointing away from the Sun and $\eta$ being normal
to it in either direction, as only the absolute value of $\Im$ is used in
Eq.\ (8).

\begin{figure*}[ht]
 \vspace*{-4.75cm}
 \hspace*{-0.4cm}
 \centerline{
 \scalebox{1}{
 \includegraphics{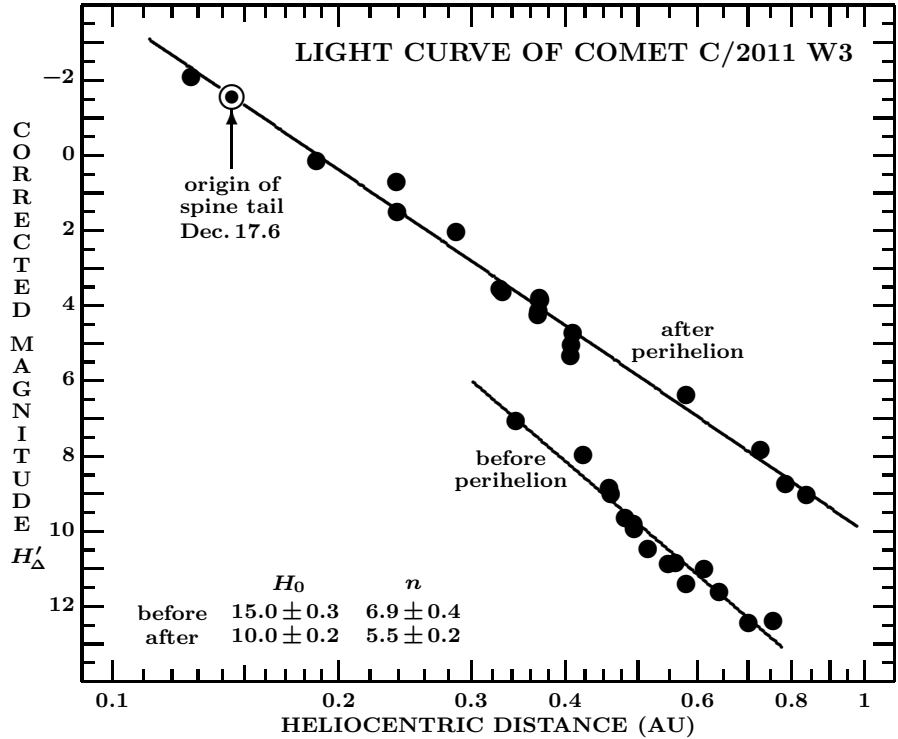}}}
 \vspace*{-14.95cm}
 \caption{Preliminary light curve of comet C/2011 W3 based on visual
and CCD total brightness estimates made from the ground.  The magnitudes
plotted have been normalized to 1 AU from earth with the inverse square power
law and to a zero phase angle with the ``compound'' Henyey-Greenstein law, as
modified by Marcus.  Personal and instrumental effects have been corrected to
the limited degree possible.  The comet is shown to have been much brighter
after perihelion than before it.}
\vspace*{0.5cm}
\end{figure*}

We have already mentioned that at small values of $\beta$ (say, \mbox{$\beta
< 0.1$}), the population of particles released at a time $t_{\rm obs}\!- \tau$
and accelerated at a rate $\beta$ is located at an angular distance $\ell$
from the nucleus, which is proportional to $\beta$,
\begin{equation}
{\ell}(\beta) = K \beta,
\end{equation}
where $K$ is a constant.  Similarly, for small values of
$\beta$, \mbox{$\partial \xi/\partial \beta \: {\textstyle \propto} \: \beta
\!\cdot\!  \Delta$} and \mbox{$\partial \eta/\partial \beta \: {\textstyle
\propto} \: \beta \!\cdot\! \Delta$}, while \mbox{$\partial \xi/\partial
\tau \: {\textstyle \propto} \: \Delta$} and \mbox{$\partial \eta/ \partial
\tau \: {\textstyle \propto} \: \Delta$}, so that for such large particles
\begin{equation}
\Im(\xi,\eta;\beta,\tau) \; {\textstyle \propto} \; \beta \!\cdot\! \Delta^2
\end{equation}
and
\begin{equation}
{\cal I}({\ell}) \; {\textstyle \propto} \; \frac{p \, \phi(\alpha)}{r^2
 \Delta^2} \, {\ell}^{k-5}.
\end{equation}
Since the exponent $k$ for the size-distribution function of cometary dust is
typically in the range of 3.5 to 4.0 (e.g., Sitko et al.\ 2011), the surface
brightness along the spine tail of C/2011 W3 is expected to drop with
increasing distance $\ell$ from the nucleus with some power of $\ell$,
usually between ${\ell}^{-1.5}$ and ${\ell}^{-1}$.  From Eq.\ (12) it is
obvious that the condition $d(\log {\cal I})/d(\log {\ell}) \gg 0$ in the
part of the spine tail between the tip and the peak surface brightness
implies a condition
\begin{equation}
k \gg 5.
\end{equation}
Such an extremely steep size distribution function can only be explained by
crumbling of the largest particles (say, $x > 1$ cm) into progressively
smaller ones.  Crude visual inspection of McNaught's spine-tail images
suggests that the surface brightness attains the peak near \mbox{$\beta_{\rm
peak} \simeq 0.006$--0.007}, or at particle diameters of $\sim$400 to 500
$\mu$m, independent of the observation time.  It is therefore possible that
the fragmentation process essentially terminated once fragments of large
debris were reduced to sizes of $\sim$100 $\mu$m.

\section{Light Curve, Tail Brightness, and Total Mass Estimate for Released
Dust}

It is hoped that eventually a detailed light curve will be available for
comet C/2011 W3.  At this time, however, only fragmentary information has
been published.  With the brightness data collected primarily from Spahr et
al.\ (2011, 2012) and Green (2012a, 2012b), Figure 6 presents a preliminary
light curve, a plot against the heliocentric distance of the magnitude,
$H_\Delta^\prime$, normalized to 1 AU from Earth with the inverse square
power law of the geocentric distance and to a zero phase angle with the
``compound'' Henyey-Greenstein law, as modified for cometary dust by Marcus
(2007).  The plotted magnitudes were all identified by the observers as total,
either visual or CCD, and were corrected by us for personal and
instrumental effects to the limited degree possible and referred to the visual
photometric system of an average unaided eye.  The phase-effect correction is
important especially for the postperihelion observations, for which the phase
angle was varying between 90$^\circ$ and 130$^\circ$ and the effects of forward
scattering of sunlight were significant.

The comet was much brighter after perihelion than before, unquestionably due to
its cataclysmic fragmentation (Secs.\ 5 and 8).  Figure 6 does not explicitly
show the outbursts that are mentioned in Sec.\ 3, but this is hardly
significant given the poor coverage of the light curve near the Sun and low
accuracy of the data.  From Fig.\ 6 it appears that to a first approximation,
one can fit both branches of the light curve with the traditional power law
$r^{-n}$, with $n$ equal to \mbox{$6.9 \pm 0.4$} before perihelion and
\mbox{$5.5 \pm 0.2$} after perihelion, and with the intrinsic magnitude (at
1 AU from both the Sun and earth), $H_0$, equal \mbox{$15.0 \pm 0.3$} before
perihelion and \mbox{$10.0 \pm 0.2$} after perihelion.  It is possible that
without the outbursts the postperihelion light curve would have gotten
flattened at heliocentric distances below $\sim$0.2 AU.

\begin{table*}
\begin{center}
{\footnotesize {\bf Table 5}\\[0.1cm]
{\sc Upper Limit on Total Amount of Dust from Fragmentation of Comet
C/2011 W3}\\[0.2cm]
\begin{tabular}{c@{\hspace{0.3cm}}c@{\hspace{0.6cm}}c@{\hspace{0.6cm}}c@{\hspace{0.7cm}}c@{\hspace{0.4cm}}c@{\hspace{0.2cm}}c}
\hline\hline\\[-0.15cm]
Upper limit on & \multicolumn{3}{@{\hspace{-0.5cm}}c}{Total mass of dust,
 $({\cal M}_{\rm dust})_{\rm max}$\,(g)}
 & \multicolumn{3}{@{\hspace{-0.2cm}}c}{Effective diameter, ${\cal D_{\rm
 max}}$\,(km)} \\[-0.06cm]
particle size$^{\rm a}$,
 & \multicolumn{3}{@{\hspace{-0.5cm}}c}{\rule[0.6ex]{5.2cm}{0.4pt}}
 & \multicolumn{3}{@{\hspace{-0.2cm}}c}{\rule[0.6ex]{4.0cm}{0.4pt}} \\[-0.06cm]
$x_{\rm max}$ (cm) & $k = 3.5$ & $k = 3.8$ & $k = 4.1$ & $k = 3.5$ & $k = 3.8$
 & $k = 4.1$ \\[0.05cm]
\hline\\[-0.15cm]
10 & $0.9 \times 10^{14}$ & $5.5 \times 10^{12}$ & $7.7 \times 10^{11}$ & 0.76
   & 0.30 & 0.15 \\
10\rlap{$^2$} & $3.0 \times 10^{14}$ & $9.0 \times 10^{12}$ & $8.2 \times
   10^{11}$ & 1.12 & 0.35 & 0.16 \\
10\rlap{$^3$} & $9.3 \times 10^{14}$ & $\llap{1}4.5 \times 10^{12}$ & $8.6
   \times 10^{11}$ & 1.65 & 0.41 & 0.16 \\[0.1cm]
\hline\\[-0.25cm]
\multicolumn{7}{l}{\parbox{11.78cm}{$^{\rm a}$\,{\scriptsize Lower limit to
 particle diameters is always $x_{\rm min} = 0$.1 $\mu$m, the bulk density
 $\rho_{\rm dust} = 0.4$ g cm$^{-3}$.}}}
\end{tabular}}
\end{center}
\end{table*}

The primary purpose for incorporating the light curve into the scope of this
paper has been our intention to use it for estimating the total mass of dust
involved in the fragmentation process culminating on December 17.6 UT and
therefore the residual mass of the nucleus at breakup.  If the visual
brightness at the time were due entirely to scattering of sunlight by
dust released from the disintegrating nucleus into the atmosphere, one could
readily compute the total cross sectional area of the particulate material,
once a value for the geometric albedo of the optically thin cloud is adopted.
In practice, unfortunately, sodium atoms radiating in the doublet near 5900
{\AA} are a strong contributor to the visual brightness of SOHO sungrazing
comets (Biesecker et al.\ 2002), so that the light curve provides us with only
an upper limit to the cross sectional area of the dust particles and indirectly
to the mass of the disintegrated nucleus.

An independent approach to determining the amount of released dust is to
estimate the total visual brightness of the comet's tail at a time sufficiently
long after the fragmentation process has been completed.  Because the tail
includes all dust released by the comet since its perihelion passage, the
result in this case is an estimate of the mass of the nucleus at, or very
shortly after, perihelion.  This result would represent a lower limit:\ not
only would all sodium have sublimated away and ionized long before such an
observation, but by this time the surface brightness of some of the released
debris would have already dropped below the detection threshold of visual
observers.

The immediate goal of either of the two approaches is the determination of
$H_{\rm eff}$, the effective visual magnitude of the estimated light scattered
by the debris, normalized to the heliocentric and geocentric distances of 1 AU
with an inverse square power law, to a zero phase angle with the Marcus (2007)
formula, and to the photometric system of an average unaided eye.  Expressed
in km$^2$, the effective cross sectional area of the dust in the cloud,
$X_{\rm dust}$, is given by an equation
\begin{equation}
X_{\rm dust} = \frac{1.54}{p} \times 10^{6-0.4 H_{\rm eff}},
\end{equation}
where $p$ is the geometric albedo of the dust, for which we use a value of
0.04 (e.g., Lamy et al.\ 2009).  For the Sun's visual magnitude we adopt
in Eq.\ (14) a value of $-$26.65 (corresponding to the $V$ magnitude of
$-$26.75).  There is no explicit phase function involved because $H_{\rm
eff}$ has already been corrected for this effect.

To be able to estimate the total mass of dust that corresponds to the derived
cross sectional area, one has to assume the following: (i) an upper limit on
diameters of the particles, $x_{\rm max}$, (ii) the lower limit, $x_{\rm min}$,
(iii) their size distribution function between $x_{\rm min}$ and $x_{\rm max}$,
which, as in the preceding section, is given by a power law $x^{-k} dx$, and
(iv) their average bulk density, $\rho_{\rm dust}$.  The relation between the
total mass of dust, ${\cal M}_{\rm dust}$, and the cross sectional area,
$X_{\rm dust}$, is given by
\begin{equation}
{\cal M}_{\rm dust} = \frac{2 (k - 3)}{3 (4 - k)} \, \Theta \, \rho_{\rm dust}
 \, X_{\rm dust} \, x_{\rm min}^{k-3} x_{\rm max}^{4-k}
\end{equation}
when $3 < k < 4$, and by
\begin{equation}
{\cal M}_{\rm dust} = \frac{2 (k -3)}{3 (k - 4)} \, \Theta \rho_{\rm dust} \,
 X_{\rm dust} \, x_{\rm min}
\end{equation}
when $k > 4$.  The coefficient $\Theta$, usually near unity, is given by an
expression
\begin{equation}
\Theta = \frac{1 - (x_{\rm min}/x_{\rm max})^{|k-4|}}{1
 - (x_{\rm min}/x_{\rm max})^{k-3}}.
\end{equation}
We use \mbox{$\rho_{\rm dust} = 0.4$} g cm$^{-3}$ (e.g., Richardson et al.\
2007), which is probably a good estimate for larger grains, but must be an
underestimate for microscopic dust.  As for $x_{\rm min}$, particles smaller
than 0.1 $\mu$m in diameter were detected in comets, e.g., in the atmosphere
of 1P/Halley with the mass spectrometers on board the Giotto and VEGA
spacecraft, but their contribution to the total mass of dust was found to
be low (Utterback \& Kissel 1990, 1995; Sagdeev et al.\ 1989), so that
adopting a lower limit of \mbox{$x_{\rm min} = 0.1$} $\mu$m seems to be
justified.

The exponent $k$ of the power law and the upper limit $x_{\rm max}$ were in
this exercise varied.  In conformity with the findings for a large
number of comets, $k$ has been constrained to a range of 3.5 to 4.1 (e.g.,
Sitko et al.\ 2011), while $x_{\rm max}$ has been selected to range from
10 cm to 10 m (e.g., Harmon et al.\ 2011).

We can now compare the approach based on the light curve (providing an upper
limit) with that based on the tail's brightness (offering a lower limit).
Turning to Fig.\ 6, we find that at the time of cataclysmic fragmentation on
December 17.6 UT, at a heliocentric distance of \mbox{$r_{\rm rel} = 0.144$}
AU, the corrected magnitude \mbox{$H_\Delta^\prime(r_{\rm rel}) = -1.6$},
which implies that \mbox{$(H_{\rm eff})_{\rm max} = +2.6$} and an upper limit
\mbox{$(X_{\rm dust})_{\rm max} = 3.5 \times 10^6$} km$^2$.  The estimated
uncertainty is at least $\pm$30 percent.  For nine different combinations of
the parameters $k$ and $x_{\rm max}$, the resulting upper limits on the mass
$({\cal M}_{\rm dust})_{\rm max}$ and the effective diameter $\cal D_{\rm max}$
are listed in Table 5.
\begin{table*}[ht]
\begin{center}
{\footnotesize {\bf Table 6}\\[0.1cm]
{\sc Total Brightness of Dust Tail of Comet C/2011 W3 on 2011 December 24.69
UT (Estimates by D.\ Seargent)}\\[0.1cm]
\begin{tabular}{c@{\hspace{0.15cm}}c@{\hspace{0.1cm}}c@{\hspace{0.3cm}}c@{\hspace{0.35cm}}c@{\hspace{0.5cm}}c@{\hspace{0.4cm}}c@{\hspace{0.75cm}}c@{\hspace{0.7cm}}c@{\hspace{0.57cm}}c@{\hspace{0.18cm}}c}
\hline\hline\\[-0.16cm]
Angular & Visual & Observed & \multicolumn{2}{@{\hspace{-0.3cm}}c}{Parameter
 $\beta$ for $(t_{\rm rel})^{\rm a}$}
 & \multicolumn{2}{@{\hspace{-0.4cm}}c}{Position angle for
 $(t_{\rm rel})^{\rm a}$} & \multicolumn{2}{@{\hspace{-0.3cm}}c}{Distance
 (AU) from} & & Effective \\[-0.04cm]
distance$\:\ell$ & magnitude & tail
 & \multicolumn{2}{@{\hspace{-0.3cm}}c}{\rule[0.7ex]{3.05cm}{0.4pt}}
 & \multicolumn{2}{@{\hspace{-0.4cm}}c}{\rule[0.7ex]{3.2cm}{0.4pt}}
 & \multicolumn{2}{@{\hspace{-0.3cm}}c}{\rule[0.7ex]{2.65cm}{0.4pt}}
 & Phase & magnitude \\[-0.04cm]
from head & estimate & width & +0.4\,day & +1.6\,days & +0.4\,day &
 +1.6\,days & Earth & Sun & angle & $H_{\rm eff}(\ell)$ \\[0.06cm]
\hline \\[-0.15cm]
4\rlap{$^\circ$} & 4 & 0$^\circ\!$.5 & 0.08 & 0.36 & 237$^\circ\!$.9
 & 231$^\circ\!$.7 & 0.64 & 0.51 & 117\rlap{$^\circ$} & 6.8 \\
8       & 4.5 & 1 & 0.17 & 0.71 & 238.4 & 231.9 & 0.62 & 0.56
 & 113 & 7.0 \\
\llap{2}3            & 5 & 2 & 0.50 & 2.09 & 239.8 & 232.4 & 0.58 & 0.72
 & $\;\:$97 & 6.6 \\
\llap{2}8\rlap{$^{\rm b}$} & \ldots{\hspace{-0.3cm}}$^{\rm c}\;$ & \ldots
 & 0.62 & 2.58 & 240.1 & 232.6 & 0.58 & 0.77 & $\;\:$93 & \ldots \\[0.1cm]
\hline \\[-0.25cm]
\multicolumn{11}{l}{\parbox{16.1cm}{$^{\rm a}$\,{\scriptsize Assumed time
of dust release reckoned from the time of the comet's perihelion
passage.}}} \\[-0.07cm]
\multicolumn{11}{l}{\parbox{16.1cm}{$^{\rm b}$\,{\scriptsize Length of the
tail seen by Seargent with the naked eye.}}} \\[0.015cm]
\multicolumn{11}{l}{\parbox{16.1cm}{$^{\rm c}$\,{\scriptsize According to
Seargent's report, over its last 5$^\circ$ the tail was very faint and
difficult{\vspace*{-0.06cm}} with the naked eye and could not be traced in
the 25$\times$100 binocular telescope.}}}\\[0.05cm]
\end{tabular}}
\end{center}
\end{table*}

The source of the best information that we have been able to find for the
brightness of the dust tail of C/2011 W3 at this early phase of investigation
is an attempt undertaken by Seargent (2011) on 2011 December 24.69 UT.  Using
an out-of-focus method of brightness estimation for comparison stars, he
determined the total brightness of the dust tail in the field of view (of an
estimated diameter of $\sim$3$^\circ$) of his 25$\times$100 binocular telescope
at three angular distances $\ell$ from the head that he estimated at
magnitude 4.8.  The results of Seargent's observation are presented in columns
1 to 3 of Table 6.  The four subsequent columns provide information on the
particles and the expected tail orientation at each distance from the head
for two assumed times of release.  The geocentric and heliocentric distances,
the phase angle, and the resulting effective visual magnitude $H_{\rm
eff}(\ell)$, all of which are --- to the given precision --- independent of
the choice for the time of release, are listed in columns 8 to 11.  The
fundamental conclusion from the tabulated values of $H_{\rm eff}(\ell)$
is that the {\it total\/} brightness in each band of the tail was
virtually independent of the distance from the head and equal to \mbox{$6.8
\pm 0.2$}:\ the {\it surface\/} brightness was decreasing with distance, but
the tail width was increasing.  And since Seargent could detect no tail with
his binocular telescope beyond $\sim$23$^\circ$ from the head, we assume in
our conservative estimate that it ended there.  The whole tail was then
brighter by a factor of $\frac{23}{3}$, or 2.2 magnitudes, than the average
in each 3$^\circ$ band, which leads to a lower limit of \mbox{$(H_{\rm
eff})_{\rm min} = 4.6$}, or 2 magnitudes fainter than indicated by $(H_{\rm
eff})_{\rm max}$ from the light curve.  Consequently, the lower limit on the
total cross sectional area of the comet's dust released after perihelion is
\mbox{$(X_{\rm dust})_{\rm min} = 0.6 \times 10^6$} km$^2$.  To obtain lower
limits on the mass and diameter estimates, the values in Table 5 should be
multiplied by 0.16 and 0.54, respectively.

It appears that the derived lower limit is relatively secure.  Seargent's value
for the brightness of the head is about 0.9 magnitude below the light curve
in Fig.\ 6 (suggesting a personal magnitude scale difference) and our
truncation of the tail's length also adds a few tenths of a magnitude, because
it implies a loss of light in the binocular magnitudes relative to those
obtained with the unaided eye.  A fairly narrow range resulting from the
two entirely independent approaches applied is encouraging and may indicate
that upon the terminal breakup all atomic sodium escaped very rapidly, leaving
hardly any lasting signature in the comet's light curve, given that at 0.144
AU from the Sun the Na photoionization lifetime is less than 1 hour (Huebner,
Keady, \& Lyon 1992; also:\ Cremonese et al.\ 1997, Combi, DiSanti, \& Fink
1997).

Because of the unacceptably large mass estimates, greatly exceeding 10$^{13}$
grams, we suggest that the exponent $k$ of the size-distribution power law in
Table 5 could not possibly be near or below 3.5.  However, there are fairly
good odds that the nucleus could have been as large as 150--200 meters in
diameter and that it could have had a residual mass as much as $\sim$10$^{12}$
grams shortly after passing through perihelion.  

\section{Properties of Dust in the Tail Before and After Perihelion}

Examining eleven of the brighter SOHO Kreutz-system comets with
prominent tails, all on their way to perihelion between 1996 and 1998,
Seka\-nina (2000) noticed that the narrow and nearly straight tails deviated
strikingly from the antisolar direction.  The highest solar radiation-pressure
accelerations, to which the dust in the tails was subjected, was found to have
been $\beta = 0.6$, suggesting the presence of a population of submicron-sized
grains dielectric in nature, most probably silicates (e.g., Sekanina et al.\
2001 and the references listed there in Sec.\ II.A.2).  The sampled comets
also showed rather consistently that the peak production of dust occurred some
20--30 {\Rsun} from the Sun, about 1 day before perihelion, and that activity
essentially terminated shortly afterwards.  Detailed modeling of one of
these objects showed that its tail was a syndyname of $\beta = 0.6$ far from
the head, but a synchrone referring to a release time of 0.8 day before
perihelion (19 {\Rsun} from the Sun), near the head.  A sharp bend or knee,
but no gap, separated the two parts of the tail. The author suggested that
the time of the abrupt termination was possibly nuclear-size dependent
(Sekanina 2000).

\begin{figure*}[ht]
 \vspace*{-5.03cm}
 \hspace*{-0.6cm}
 \centerline{
 \scalebox{1.05}{
 \includegraphics{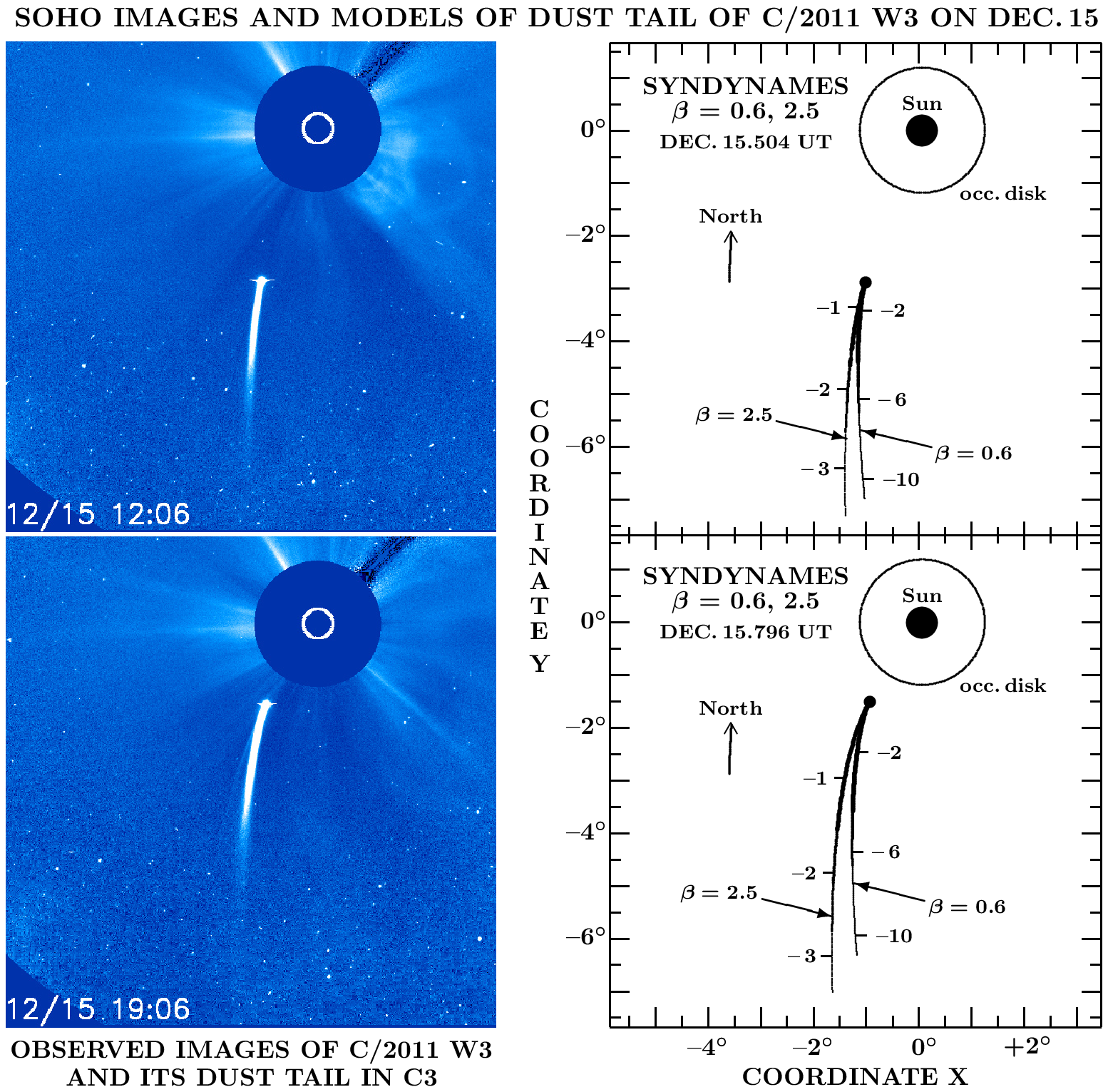}}}
 \vspace*{-9.75cm}
 \caption{Comet C/2011 W3 with its dust tail shortly before reaching
perihelion, as imaged by the C3 coronagraph on board the SOHO spacecraft.  The
upper image was taken on December 15.504 UT, or 0.508 day before perihelion,
the lower image on December 15.796 UT, or 0.216 day before perihelion.  In the
panels next to the images the observations are compared with two syndynames;
the one for \mbox{$\beta = 2.5$} is slightly to the left of the other, which
refers to \mbox{$\beta = 0.6$}.  The syndynames are calibrated by the times of
dust release, reckoned in days from the time of the comet's perihelion passage
(the negative numbers indicate days before the comet's perihelion).  The panels
also provide the scale and orientation of the images.  We note that in both
images the tail displays a slightly lesser curvature than the syndynames, but
lies generally between them, which implies that the comet released
submicron-sized dust of both dielectric and strongly absorbing nature. (Image
credit:\ ESA/NASA/LASCO consortium.)}
\vspace*{0.55cm}
\end{figure*}

To learn more about this possible relation and about the dust-emission
pattern of comet C/2011 W3, we have examined the properties of its tail
both before and after perihelion.  Two images taken by the C3 coronagraph
on board the SOHO spacecraft shortly before perihelion, on December 15.504
and 15.796 UT, are reproduced in Fig.\ 7.  Cursory inspection reveals a tail
whose curvature is slight but increases with time, with no obvious knee.  Two
syndynames are plotted for each image in the right-hand side panel:\ the one
on the left corresponds to $\beta = 2.5$, the other to $\beta = 0.6$.  The
two radiation-pressure accelerations have been chosen because they are of
particular significance for cometary dust (e.g., Sekanina et al.\ 2001):\ the
first is the highest acceleration known to affect dust in comets and is typical
of strongly absorbing grains (such as carbon-rich, organic material), whereas
the second, as mentioned above, is characteristic of the peak acceleration on
dielectric grains (such as silicates).  In both cases the particles involved
are in the submicron-size range.  Superposition of the syndynames on the
images in Fig.\ 7 shows for both observation times that the tail lies largely
between the two syndynames and that it is slightly less curved than either of
them.  A conclusion from these images alone is that C/2011 W3 was releasing
both dielectric and absorbing dust on its way to perihelion.  The main
difference between the two syndynames in Fig.\ 7 is that at a given distance
from the nucleus, the dust on the syndyname of \mbox{$\beta = 0.6$} was
released much earlier than that on the syndyname of \mbox{$\beta = 2.5$}.

It is instructive to compare the comet's appearance in the two preperihelion
images with very different views offered shortly after perihelion by the SOHO
C2 coronagraph in combination with the COR1 and COR2 coronagraphs on board the
STEREO-A and STEREO-B spacecraft.  The three selected images are of course
merely snapshots of continuous changes in the comet's figure that are fully
revealed only by the dynamic, time-lapse imaging that involves the entire data
set.  Nevertheless, given the spatial distribution of the three spacecraft,
the stereoscopic quality of the gained information makes even snapshot views
extremely valuable.

\begin{figure*}[ht]
 \vspace*{-5.1cm}
 \hspace*{0.05cm}
 \centerline{
 \scalebox{1.05}{
 \includegraphics{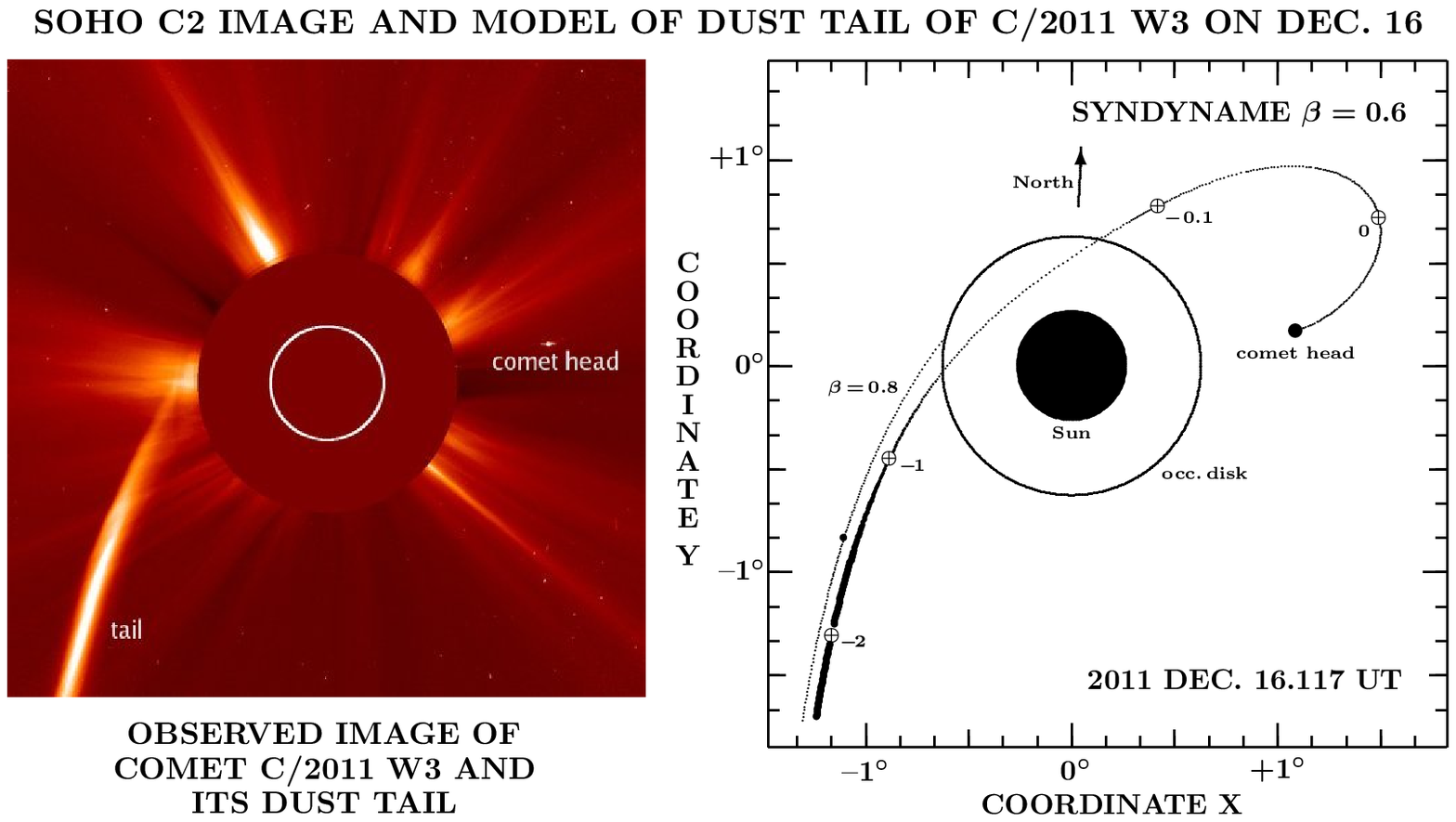}}}
 \vspace*{-16.6cm}
 \caption{Appearance of comet C/2011 W3 and its dust tail in an
image taken with the C2 coronagraph on board the SOHO spacecraft on December
16.117 UT, or 0.105 day after perihelion.  The tail, to the south-southeast of
the Sun, is seen to be completely disconnected from the comet's head, to the
west of the Sun.  The head's image is saturated (with some blooming being
apparent) but with no tail extension.  The tail lies entirely between the
syndynames \mbox{$\beta = 0.6$} and 0.8, as shown in the panel to the right.
The first of the two syndynames is plotted thicker and is calibrated by the
times of dust release, reckoned in days from the time of the comet's perihelion
passage, pinpointing the locations of particles released at perihelion and
0.1, 1, and 2 days before perihelion.  The dot on the segment of the syndyname
\mbox{$\beta = 0.8$} identifies the location of dust released 1 day before
perihelion.  We note that the entire northwestern branch of the tail is
missing.  The tail's brightness is largely determined by forward scattering
of sunlight by dust, as the phase angle increases toward the bottom of the
image.  The scale and orientation of the image are the same as those of the
panel. (Image credit:\ ESA/NASA/LASCO consortium.)}
\vspace*{0.5cm}
\end{figure*}

Figure 8 shows the comet's head and a part of its tail in a frame taken with
the C2 coronagraph on December 16.117 UT, or 0.105 day after perihelion.
The bright tail, to the south-southeast of the Sun, is entirely disconnected
from the head.  The panel to the right of the image, showing the complete
syndyname \mbox{$\beta = 0.6$}, suggests that, contrary to the observation,
the tail should have reappeared to the northwest of the occulting disk.  The
comet's head, although clearly saturated and displaying some ``blooming'',
is essentially stellar in appearance.  This means that no detectable
postperihelion tail developed by the time the image was taken, about 0.1 day
after perihelion.

An example of the peculiar appearance of C/2011 W3 is presented in Fig.\ 9,
which shows the comet imaged with the COR2 coronagraph on board the STEREO-A
spacecraft on December 16.246 UT, or 0.234 day after perihelion.  Unlike in
the SOHO's C2 coronagraph, the comet now consists of three components: the
head --- with a short wisp of a new tail that must have begun to develop before
0.2 day after perihelion --- to the east-northeast of the Sun and two separate
branches of the old tail, one to the southeast and the second to the northwest
of the Sun.  The southeastern branch is relatively sharp but faint, a far cry
from its luster in Fig.\ 8.  It can be matched with the syndyname \mbox{$\beta
= 0.6$}, but the resolution is poor as the syndynames along this tail are
``crowded''.  The northwestern branch is blob shaped.  Its first view in this
COR2 coronagraph coincided approximately with the re-appearance of the
comet's head from behind the occulting disk, nearly 3 hours after perihelion.
In projection onto the plane of the sky, this branch may unfortunately have
been superposed on top of a weak but broad coronal mass ejection.  Because of
this interference, the tail's exact contours are hard to establish, but the
syndyname \mbox{$\beta = 0.6$} seems to be again involved with the feature.
The blob terminates just before crossing the synchrone for a release time of
0.1 day before perihelion.  Again, there is no tail in the areas corresponding
to release times near perihelion.  Neither of the two branches becomes obvious
in any postperihelion image taken with the COR1 coronagraph of the STEREO-A
spacecraft.

\begin{figure*}[ht]
 \vspace*{-5.1cm}
 \hspace*{-0.05cm}
 \centerline{
 \scalebox{1.05}{
 \includegraphics{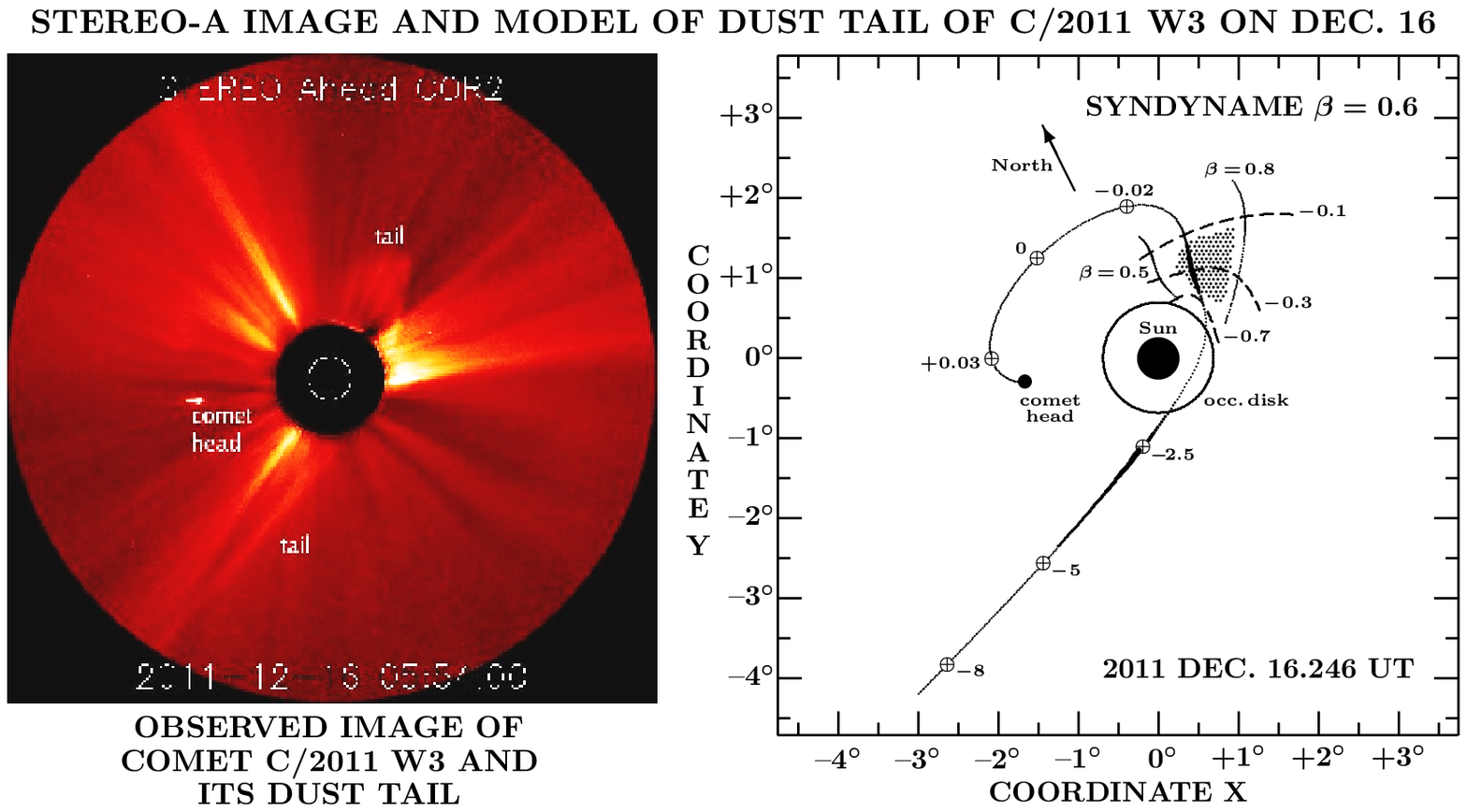}}}
 \vspace*{-16.6cm}
 \caption{Peculiar appearance of comet C/2011 W3 and its dust tail
in an image taken with the COR2 coronagraph on board the STEREO-A spacecraft
on December 16.246 UT, or 0.234 day after perihelion.  The tail has two
branches located, respectively, to the southeast and northwest of the Sun,
while the comet head's site is nearly to the east of the Sun.  The tail's
southeastern branch is faint because of backscatter of sunlight, while the
blob-shaped northwestern branch appears to be at least in part contaminated
by a relatively weak, but broad coronal mass ejection.  The panel to the
right identifies the locations on the syndyname \mbox{$\beta = 0.6$}, which
are populated by dust released from the nucleus at six different times
between 8 days before perihelion and 0.03 day after perihelion.  Three
additional release times --- 0.1, 0.3, and 0.7 day before perihelion --- are
identified by the synchrones drawn by the broken curves.  The panel also
shows the area of the tail's blob to the northwest of the Sun and short
segments of the syndynames for $\beta$ equaling 0.5 and 0.8.  No tail is
detected for release times between 0.1 day before perihelion and about
0.1 day after perihelion, but the head's image is elongated in the direction
away from the Sun, suggesting that a postperihelion tail already began to
develop during a couple of hours before the frame was taken. (Image credit:\
NASA/SECCHI consortium.)}
\vspace*{0.5cm}
\end{figure*}

In a long series of images taken with the COR1 and COR2 coronagraphs on board
the STEREO-B spacecraft during much of the first day after perihelion, the look
of the comet with its tail is downright bizarre.  In fact, hours before the
comet's head emerged from behind the occulting disk to the west-southwest,
a second branch of the tail began to show up as a steadily growing sharp
spike to the northeast of the Sun, joining the southwestern branch that had
thrived since preperihelion times.  In addition, the comet's head, after
it emerged, was, just as in Fig.\ 9, disconnected from either of the two
branches of the tail.  Overall, therefore, the comet again consisted of three
discrete components, as seen in Fig.\ 10.  The displayed image was taken with
the COR2 coronagraph on December 16.517 UT, or 0.505 day after perihelion.
The syndyname \mbox{$\beta = 0.6$} provides an excellent fit to both branches
of the tail simultaneously, although the syndynames are again fairly
``crowded'' along much of the tail.  This image shows that the northeastern
branch extends to a point that is populated by dust released $\sim$0.1 day
before perihelion, in agreement with the result from Fig.\ 9.  Careful
inspection of the comet's head reveals its elongation similar to that
detected in the STEREO-A image.

\begin{figure*}[ht]
 \vspace*{-5.1cm}
 \hspace*{-0.05cm}
 \centerline{
 \scalebox{1.05}{
 \includegraphics{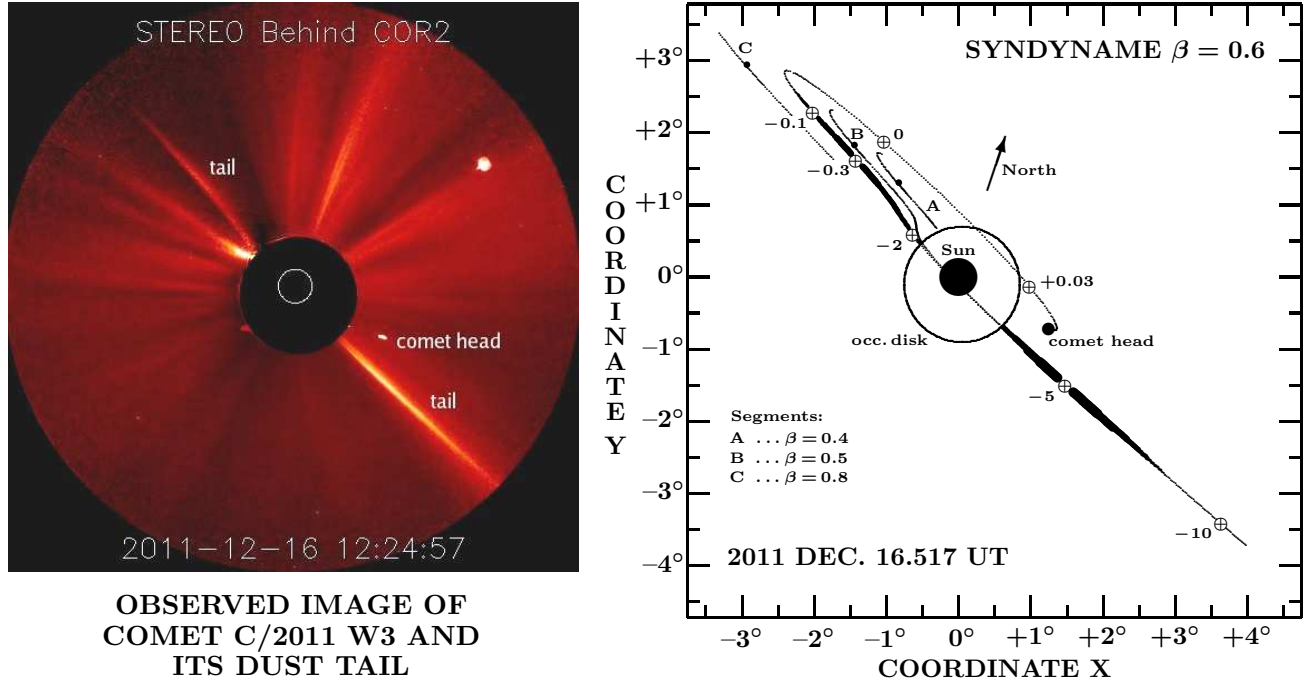}}}
 \vspace*{-16.5cm}
 \caption{Bizarre appearance of comet C/2011 W3 and its dust
tail in an image taken with the COR2 coronagraph on board the STEREO-B
spacecraft on December 16.517 UT, or 0.505 day after perihelion.  The tail
has two branches located, respectively, to the southwest and northeast of
the Sun, while the comet's head is to the west-southwest of the Sun and north
of the brightest part of the tail's southwestern branch.  Careful inspection
shows that the head's image is elongated along the line with the Sun, evidence
that a new, postperihelion tail is being developed.  The entire tail is fitted
with a syndyname of \mbox{$\beta = 0.6$}, as shown in the panel that also
provides the scale and orientation of the image.  Only the heavily-drawn
segments of the syndyname are seen as the tail in the image.  The syndyname
is calibrated by the times of dust release, reckoned in days from the time
of the comet's perihelion passage (a negative number means before perihelion,
and vice versa).  Brightness-wise, the southwestern branch dominates because
of forward scattering of sunlight.  A broad secondary maximum is noted on the
northeastern branch at a location corresponding to a release time of about
0.3 day before perihelion (December 15.7 UT).  The tail then rapidly fades and
disappears for a release time near 0.1 day before perihelion, suggesting an
attentuation of activity hours before perihelion.  Short segments of the
syndynames \mbox{$\beta = 0.4$}, 0.5, and 0.8 are also plotted, marked as A, B,
and C, respectively, to illustrate a general deficit of dust with accelerations
different from 0.6.  The dots on these syndyname segments show the locations
of particles released 0.1 day before perihelion.  The northeastern tail's
boundary is sharper to the south than to the north, indicating that dust with
\mbox{$\beta \gg 0.6$} is more scarce than that with \mbox{$\beta < 0.6$}.  No
tail is apparent along the wide swath of space referring to release times
during the first hours after perihelion.  (Image credit:\ NASA/SECCHI
consortium.)}
\vspace*{0.5cm}
\end{figure*}

To extract information on the dust particles that populate the tail, we have
examined more closely the orbital properties of dust particles along the
syndyname \mbox{$\beta = 0.6$} with the help of Table 7, which lists their
perihelion distance, the eccentricity, and the perihelion time as a function
of $\beta$ and the time of release from the comet.  The numbers along this
syndyname in Figs.\ 8--10 are the times of release in days from the comet's
perihelion time.  While it is rather clear that the bright branch of the
tail to the south-southeast in Fig.\ 8 is identical with the fairly pale
southeastern branch in Fig.\ 9 and with the gleaming southwestern branch in
Fig.\ 10, the reason for the enormous differences in brightness is not obvious.
The orbital elements of the dust particles in Table 7 allow one to calculate
the geometry for the times of the three images and to find out that the culprit
is forward scattering of sunlight.  For example, the particles in the bright
portion of the tail in Fig.\ 8, located on the syndyname \mbox{$\beta =
0.6$} and released 2 days before perihelion, were at the time of observation
nearly 7 {\Rsun} from the Sun and their phase angle was 113$^\circ$, while
for the particles that left the nucleus 1 day before perihelion, although
they were only 3.6 {\Rsun} from the Sun, the phase angle was 90$^\circ$ and
for those released 0.7 day before perihelion and located near the edge of
the occulting disk, the phase angle was only 70$^\circ$, the reason why this
part of the tail looks faint.  For the same reason, the brightness of the
tail's southeastern branch in Fig.\ 9 cannot compare with that of the
southwestern branch in Fig.\ 10 even though it is the same feature:\ at the
location of particles on the
syndyname \mbox{$\beta = 0.6$} released 5 days before perihelion, in the
middle of this tail, the phase angle is 55$^\circ$ in Fig.\ 9, but
109$^\circ$ in Fig.\ 10.

It is further noted from Table 7 that particles on that same syndyname that
were released earlier than 2.5 days before perihelion did not reach their
perihelion points until after December 16.5 UT and, at the times the images
in Figs.\ 8--10 were taken, were still on their way toward the Sun.  By
contrast, particles released later than 0.5 day before perihelion reached
their perihelion points no later than about December 16.1 UT and were
therefore in all three images already moving away from the Sun.  This is
important because the relevant perihelion distances are very small.  Table
7 shows that on the syndyname \mbox{$\beta = 0.6$} this minimum distance
drops from 2.37 {\Rsun} for dust that left the nucleus 0.5 day before
perihelion to 2.21 {\Rsun} for dust that left 0.3 day before perihelion
and to 1.81 {\Rsun} for dust that left 0.1 day before perihelion.  And, as
expected, it is equal to the perihelion distance of the comet, 1.19 {\Rsun},
for dust that left at perihelion.  Since the phase angle for particles on the
syndyname \mbox{$\beta = 0.6$} released at perihelion is in Fig.\ 8 equal to
88$^\circ$, nearly identical with the phase angle for particles released 0.7
day before perihelion (see above), the utter absence of dust from emissions
at perihelion has nothing to do with particle sunlight scattering effects.
And since Figs.\ 8-10 consistently show the absence of a tail for release
times between approximately 0.1 day before perihelion and 0.1 day after
perihelion, that is, for particles whose perihelion distances were always
smaller than $\sim$1.8 {\Rsun}, the most likely explanation for this
``activity attenuation'' is the effect of sublimation of microscopic dust
near the Sun, as discussed in detail in the following section.

Probably the most diagnostic evidence for this effect is offered by the
northeastern branch of the tail in Fig.\ 10, which is made up of the dust
that left the nucleus later than 2.2 days before perihelion.  At the time
of observation, these particles were more than 3 {\Rsun} from the Sun;
the phase angle reached 110$^\circ$ for those at the edge of the occulting
disk but dropped rapidly along the tail to 70$^\circ$ about 1$^\circ$ from
the Sun and was nearly constant around 50$^\circ$ along much of the rest
of it.  The apparent brightening at the location of particles released
$\sim$0.3 day before perihelion is thus likely to reflect their increased
production around that time.  Table 7 indicates that the dust at the edge
of the occulting disk reached perihelion, at 2.7 {\Rsun} from the Sun, only
about 0.45 day after the nucleus, while the dust at the far end of this
tail's branch was nearest the Sun, at merely $\sim$1.8 {\Rsun}, practically
simultaneously with the nucleus.  This means that at the time of observation
all particles along the northeastern tail were already moving away from the
Sun.  The fading and disappearance of this tail at a location populated
by submicron-sized particles released $\sim$0.08--0.1 day before perihelion
and moving in highly hyperbolic orbits (eccentricity $\sim$2.3 or higher)
strongly indicate that their perihelion distance of $\sim$1.8 {\Rsun} is
indeed the {\it limit at which dust begins to sublimate profusely.}
\begin{table*}
\noindent
\begin{center}
{\footnotesize {\bf Table 7}\\[0.1cm]
{\sc Orbital Elements of Dust Particles in the Tail of Comet C/2011 W3}\\[0.1cm]
\begin{tabular}{c@{\hspace{0.2cm}}c@{\hspace{0.2cm}}c@{\hspace{0.25cm}}c@{\hspace{0.25cm}}c@{\hspace{0.25cm}}c@{\hspace{0.25cm}}c@{\hspace{0.4cm}}c@{\hspace{0.17cm}}c@{\hspace{0.17cm}}c@{\hspace{0.17cm}}c@{\hspace{0.4cm}}c@{\hspace{0.17cm}}c@{\hspace{0.17cm}}c@{\hspace{0.17cm}}c@{\hspace{0.17cm}}c}
\hline\hline\\[-0.16cm]
 & & \multicolumn{14}{@{\hspace{0.02cm}}c}{Orbital elements of released dust
particles$^{\rm a}$ as function of parameter $\beta$} \\[-0.04cm]
Time & Distance
 & \multicolumn{14}{@{\hspace{0.02cm}}c}{\rule[0.7ex]{13.55cm}{0.4pt}} \\
 [-0.04cm]
of & from & \multicolumn{5}{@{\hspace{-0.15cm}}c}{Perihelion distance
 ({\Rsun})} & \multicolumn{4}{@{\hspace{-0.19cm}}c}{Orbit eccentricity$^{\rm
 c}$} & \multicolumn{5}{@{\hspace{0.02cm}}c}{Perihelion time (days$^{\rm
 b}$)} \\[-0.04cm]
release & Sun
 & \multicolumn{5}{@{\hspace{-0.15cm}}c}{\rule[0.7ex]{3.68cm}{0.4pt}}
 & \multicolumn{4}{@{\hspace{-0.19cm}}c}{\rule[0.7ex]{3.85cm}{0.4pt}}
 & \multicolumn{5}{@{\hspace{0.02cm}}c}{\rule[0.7ex]{5.25cm}{0.4pt}} \\[-0.04cm]
(days$^{\rm b}$) & ({\Rsun}) & 0.3 & 0.6 & 1.0 & 1.5 & 2.5 & 0.3 & 0.6
 & 1.5 & 2.5 & 0.3 & 0.6 & 1.0 & 1.5 & 2.5 \\[0.06cm]
\hline\\[-0.15cm]
\llap{$-$1}0 & \llap{10}8.51 & 1.69 & 2.87 & \llap{1}1.40 & \llap{3}8.50
 & \llap{6}5.98 & 1.0132 & 1.0790 & 1.1240 & 1.0241 & +0.748 & +1.862 & +4.686
 & +4.148 & $-$0.038 \\
\llap{$-$}5  & \llap{6}7.94  & 1.69 & 2.81 & 9.01 & \llap{2}4.86 & \llap{4}1.58
 & 1.0212 & 1.1237 & 1.1921 & 1.0383 & +0.387 & +0.967 & +2.248 & +1.931
 & $-$0.095 \\
\llap{$-$}2  & \llap{3}6.37  & 1.67 & 2.69 & 6.59 & \llap{1}4.18 & \llap{2}2.60
 & 1.0393 & 1.2212 & 1.3368 & 1.0704 & +0.164 & +0.407 & +0.816 & +0.653
 & $-$0.106 \\
\llap{$-$}1  & \llap{2}2.51  & 1.65 & 2.55 & 5.19 & 9.41 & \llap{1}4.26
 & 1.0628 & 1.3396 & 1.5074 & 1.1116 & +0.087 & +0.206 & +0.356 & +0.259
 & $-$0.095 \\
\llap{$-$}0\rlap{.5} & \llap{1}3.80 & 1.62 & 2.37 & 4.06 & 6.34 & 9.01
 & 1.1007 & 1.5154 & 1.7536 & 1.1766 & +0.044 & +0.098 & +0.139 & +0.082
 & $-$0.078 \\
\llap{$-$}0\rlap{.2} & 7.09 & 1.56  & 2.07 & 2.91 & 3.83 & 4.94
 & 1.1884 & 1.8778 & 2.2456 & 1.3224 & +0.015 & +0.027 & +0.024 & $-$0.002
 & $-$0.056 \\
\llap{$-$}0\rlap{.1} & 4.22 & 1.48  & 1.81 & 2.24 & 2.67 & 3.17
 & 1.3009 & 2.2894 & 2.7905 & 1.5023 & +0.004 & +0.003 & $-$0.004 & $-$0.018
 & $-$0.040 \\
0 & 1.19 & 1.19 & 1.19 & 1.19 & 1.19 & 1.19 & 1.8570 & 3.9998 & 4.9999 & 2.3333
 & $\;\;\:$0.000 & $\;\;\:$0.000 & $\;\;\:$0.000 & $\;\;\:$0.000
 & $\;\;\:$0.000 \\
\llap{+}0\rlap{.1$^{\rm d}$} & 4.22 & 1.48 & 1.81 & 2.24 & 2.67 & 3.17
 & 1.3009 & 2.2894 & 2.7905 & 1.5023 & $-$0.004 & $-$0.003 & +0.004 & +0.018
 & +0.040 \\
\llap{+}0\rlap{.3} & 9.55 & 1.59 & 2.21 & 3.38 & 4.77 & 6.44
 & 1.1428 & 1.6955 & 2.0001 & 1.2473 & $-$0.025 & $-$0.051 & $-$0.060
 & $-$0.022 & +0.066 \\[0.1cm]
\hline\\[-0.25cm]
\multicolumn{16}{l}{\parbox{16cm}{$^{\rm a\,}${\scriptsize Release (ejection)
 velocity assumed to be zero.}}} \\[-0.11cm]
\multicolumn{16}{l}{\parbox{16cm}{$^{\rm b\,}${\scriptsize Reckoned from the
 time of perihelion passage of the comet; minus sign means before perihelion,
 and vice versa.}}} \\[-0.05cm]
\multicolumn{16}{l}{\parbox{16cm}{$^{\rm c\;}${\scriptsize For $\beta = 1$
(motion in a straight line), the eccentricity is by definition infinitely
large regardless of the release time.}}} \\[-0.11cm]
\multicolumn{16}{l}{\parbox{16cm}{$^{\rm d\,}${\scriptsize This entry is
included to illustrate, by comparison with the entry $-$0.1 day, the symmetry
with respect to perihelion.}}}
\end{tabular}}
\end{center}
\end{table*}

The syndyname of $\beta = 0.6$ continues to extend in Fig.\ 10 a little farther
from the Sun than the disappearing northeastern tail and then turns sharply
back, almost 180$^\circ$, running nearly parallel to the tail, in part behind
the occulting disk, all the way to the comet's head.  No trace of dust debris,
all of which moved in orbits with perihelion distances smaller than $\sim$1.8
{\Rsun}, is --- just as in Figs.\ 8 and 9 --- detected along this arc of
the syndyname.  Thus, no particles released between about 0.1 day before
perihelion and at least 0.1 day after perihelion appear to have survived,
again pointing to $\sim$1.8 {\Rsun} as the sublimation cutoff.

The panel in Fig.\ 10 also displays segments of the syndynames \mbox{$\beta =
0.4$}, 0.5, and 0.8 for a part of the northeastern tail, where they do no
overlap with the syndyname \mbox{$\beta = 0.6$}.  A clearly apparent property
of this tail branch is its more gradual fading to the north of its sharp edge
than to the south, where the drop is abrupt.  Dust with \mbox{$\beta < 0.6$},
located to the north of the edge, has the perihelion distances systematically
smaller, in agreement with the more limited presence of surviving dust than
at \mbox{$\beta = 0.6$}.  The tail no longer appears to contain particles
with the acceleration parameter $\beta$ significantly exceeding 0.6, which
should be located to the south.

On the other hand, the tail's bright southwestern branch in Fig.\ 10 was made
up of dust that left the nucleus between 8 and 3.6 days before perihelion.  At
the time of observation these particles were between 16 {\Rsun} and 5 {\Rsun}
from the Sun.  Table 7 shows that they passed through perihelion between 1.5
and 0.7 days after the comet, so they were all still approaching the Sun in
Fig.\ 10.  And, as already alluded to above, this southwestern branch of the
tail looks bright because of effects of forward scattering of sunlight, the
phase angle varying from 93$^\circ$ for dust released 8 days before perihelion
to 134$^\circ$ for dust released 3.6 days before perihelion.  At perihelion,
the distance from the Sun was from 2.77 {\Rsun} to 2.86 {\Rsun}.

The comet head's elongation in Fig.\ 10, already mentioned above, indicates
that dust particles were released from the nucleus starting about 4 hours, or
slightly less than 0.2 day, after perihelion, and were therefore $\sim$8 hours
old at the time of observation.  This new, postperihelion tail should have
indeed extended from the nucleus in the antisolar direction.  If consisting
of particles with \mbox{$\beta = 0.6$}, its length at the time the image was
taken should have been between 4$^\prime$ and 6$^\prime$.

There is no evidence in Figs.\ 8--10 for dust with \mbox{$\beta \mbox{\gapeq}
1$}.  The question of what happened to it is again answered with the help
of Table 7, which conveys two important facts on the whereabouts of particles
in convex hyperbolic orbits:\ (i) those released from the comet up to about
2 days before perihelion moved in orbits with perihelion distances always
exceeding (often considerably) 9~{\Rsun}, whereas (ii) more recent ones
passed through perihelion nearly simultaneously with the comet, so that at
the times the images in Figs.\ 8--10 were taken this dust already was at
heliocentric distances much larger than the perihelion distance.  In summary,
dust particles with \mbox{$\beta \mbox{\gapeq} 1$} did not contribute to the
tail's postperihelion brightness because of their dispersal over a large
volume of space and reduced light scattering efficiency far from the Sun.

To address the issue of what kind of dust was released by comet C/2011 W3
after perihelion, we collected information on the tail length in the
period of time from late December 2011 to mid-March 2012.  A selection
of 54 reported photographic and visual observations is listed in Table 8.
Nearly all tail lengths from times after December 21 can be explained by
submicron-sized debris from the event of December 17.6 UT (Sec.\ 5), if
the radiation-pressure accelerations $\beta$ of up to 2.5 are allowed.
On the other hand, constraining $\beta$ to 0.6 requires that in most cases
the tail (or at least its far reaches) derive from the activity prior to
the event of December 17.  To distinguish between the two scenarios, it will
be necessary to measure accurately the position angles of the tail, which
after December 26 became almost perfectly straight.  At present we are unaware
of any such measurements.

\begin{table*}
\begin{center}
{\footnotesize {\bf Table 8} \\[0.1cm]
{\sc Selection of Reported Postperihelion Tail Lengths of C/2011 W3 and
Lengths Derived for Two Parameters $\beta_0$}\\[0.1cm]
\begin{tabular}{l@{\hspace{0.08cm}}l@{\hspace{0.08cm}}r@{\hspace{0.55cm}}c@{\hspace{0.35cm}}c@{\hspace{0.28cm}}c@{\hspace{0.65cm}}c@{\hspace{0.78cm}}c@{\hspace{0.95cm}}l@{\hspace{0.8cm}}l}
\hline\hline\\[-0.15cm]
\multicolumn{3}{@{\hspace{-0.15cm}}c}{Observation} & Reported
  & \multicolumn{2}{@{\hspace{-0.4cm}}c}{Predicted length$^{\rm a}$}
  & \multicolumn{2}{@{\hspace{-0.55cm}}c}{Required release time$^{\rm b}$\,(UT)}
  & & \\[-0.03cm]
\multicolumn{3}{@{\hspace{-0.15cm}}c}{time} & length
  & \multicolumn{2}{@{\hspace{-0.4cm}}c}{\rule[0.7ex]{2.55cm}{0.4pt}}
  & \multicolumn{2}{@{\hspace{-0.55cm}}c}{\rule[0.7ex]{3.75cm}{0.4pt}} &
  & \\[-0.01cm]
\multicolumn{3}{@{\hspace{-0.1cm}}c}{(UT)} & of tail
  & $\beta_0 = 2.5$ & $\beta_0 = 0.6$
  & $\beta_0 = 2.5$ & $\beta_0 = 0.6$
  & Observer$^{\rm c}$ & Reference$^{\rm d}$ \\[0.15cm]
\hline \\[-0.15cm]
 & & & $\;\;\;\;\;{^\circ}$ & $\;\;{^\circ}$ & $\;\;{^\circ}$ &
 & & & \\[-0.26cm]
2011 & Dec. & 21.07 &   10   &  $\;\:$7.2 & $\;\:$1.8 & Dec.\,17.3
 & Dec.\,16.4 & N.\,Wakefield & Mailing list \\
     &      & 21.30 & $\!\!>$5 & $\;\:$8.1 & $\;\:$2.0 & \,\,\ldots\ldots\ldots
 & \llap{$<$}Dec.\,16.8 & \llap{$^\ast$}W.\,Souza & Green (2012a) \\
     &      & 21.7$\;\:$ & 13 & $\;\:$9.8 & $\;\:$2.4 & Dec.\,17.3
 & Dec.\,16.4 & L.\,Barnes & Mailing list \\
     &      & 21.73 & 14\rlap{.3} & 10.0 & $\;\:$2.4 & Dec.\,17.2\rlap{5}
 & Dec.\,16.3\rlap{6} & R.\,McNaught & $\;\;\;\;$'' \\
     &      & 22.30 & 10 & 12.7 & $\;\:$3.1 & \,\,\ldots\ldots\ldots
 & Dec.\,16.6 & \llap{$^\ast$}W.\,Souza & Green (2012a) \\
     &      & 22.30 & 13 & 12.7 & $\;\:$3.1 & Dec.\,17.5\rlap{7} & Dec.\,16.5
 & \llap{$^\ast$}A.\,Amorim & $\;\;\;\;$'' \\[0.1cm]
     &      & 22.32 & 20 & 12.8 & $\;\:$3.1 & Dec.\,17.1\rlap{5} & Dec.\,16.3
 & \llap{$^\ast$}M.\,Goiato & $\;\;\;\;$'' \\
     &      & 22.7$\;\:$ & 16 & 14.8 & $\;\:$3.6 & Dec.\,17.5 & Dec.\,16.4
 & L.\,Barnes & Mailing list \\
     &      & 22.7$\;\:$ & 15 & 14.8 & $\;\:$3.6 & Dec.\,17.5\rlap{8}
 & Dec.\,16.5 & \llap{$^\ast$}D.\,Seargent & $\;\;\;\;$'' \\
     &      & 23.31 & 15 & 18.3 & $\;\:$4.5 & \,\,\ldots\ldots\ldots
 & Dec.\,16.5\rlap{5} & \llap{$^\ast$}M.\,Goiato & Green (2012a) \\
     &      & 23.7$\;\:$ & 22 & 20.7 & $\;\:$5.1 & Dec.\,17.5 & Dec.\,16.4
 & J.\,Tilbrook & Mailing list \\
     &      & 23.72 & 21\rlap{.7} & 20.8 & $\;\:$5.1 & Dec.\,17.5\rlap{5}
 & Dec.\,16.4 & R.\,McNaught & $\;\;\;\;$'' \\[0.1cm]
%
%
     &      & 24.32 & 20 & 24.8 & $\;\:$6.1 & \,\,\ldots\ldots\ldots
 & Dec.\,16.5 & \llap{$^\ast$}M.\,Goiato & Green (2012a) \\
     &      & 24.69 & 28 & 27.4 & $\;\:$6.8 & Dec.\,17.5\rlap{7} & Dec.\,16.4
 & \llap{$^\ast$}D.\,Seargent & Mailing list \\
     &      & 24.7$\;\:$ & 27 & 27.5 & $\;\:$6.8 & \,\,\ldots\ldots\ldots
 & Dec.\,16.4 & T.\,Barry & $\;\;\;\;$'' \\
     &      & 25.7$\;\:$ & 30 & 34.7 & $\;\:$8.8 & \,\,\ldots\ldots\ldots
 & Dec.\,16.5 & J.\,Dunphy & $\;\;\;\;$'' \\
%
%
%
     &      & 26.60 & 33 & 41.2 & 10.8 & \,\,\ldots\ldots\ldots & Dec.\,16.5
  & \llap{$^\ast$}R.\,Kaufman & $\;\;\;\;$'' \\
     &      & 26.63 & 38 & 41.4 & 10.9 & \,\,\ldots\ldots\ldots & Dec.\,16.4
 & J.\,Drummond & $\;\;\;\;$''\\[0.1cm]
     &      & 26.68 & 37 & 41.8 & 11.0 & \,\,\ldots\ldots\ldots
 & Dec.\,16.4\rlap{5} & \llap{$^\ast$}D.\,Seargent & $\;\;\;\;$'' \\
     &      & 26.71 & 38 & 42.0 & 11.1 & \,\,\ldots\ldots\ldots & Dec.\,16.4
 & M.\,Mattiazzo & $\;\;\;\;$'' \\
     &      & 28.28 & 30 & 52.3 & 15.1 & \,\,\ldots\ldots\ldots & Dec.\,16.8
 & \llap{$^\ast$}W.\,Souza & $\;\;\;\;$'' \\
     &      & 28.74 & 30 & 55.0 & 16.4 & \,\,\ldots\ldots\ldots
 & Dec.\,16.8\rlap{5} & \llap{$^\ast$}A.\,Pearce & $\;\;\;\;$'' \\
     &      & 29.59 & 32 & 59.2 & 18.8 & \,\,\ldots\ldots\ldots & Dec.\,16.9
 & R.\,Kaufman & $\;\;\;\;$'' \\
     &      & 29.73 & 32\rlap{.0} & 59.9 & 19.2 & \,\,\ldots\ldots\ldots
 & Dec.\,16.9 & \llap{$^\ast$}A.\,Pearce & $\;\;\;\;$'' \\[0.1cm]
     &      & 30.71 & 25 & 63.7 & 22.1 & \,\,\ldots\ldots\ldots & Dec.\,17.4
 & M.\,Mattiazzo & $\;\;\;\;$'' \\
     &      & 31.58 & 30 & 66.3 & 24.7 & \,\,\ldots\ldots\ldots & Dec.\,17.3
 & R.\,Kaufman & $\;\;\;\;$'' \\
     &      & 31.67 & 45 & 66.6 & 24.9 & \,\,\ldots\ldots\ldots
 & Dec.\,16.7\rlap{5} & \llap{$^\ast$}D.\,Seargent & $\;\;\;\;$'' \\
     &      & 31.72 & 25 & 66.7 & 25.1 & \,\,\ldots\ldots\ldots
 & \,\,\ldots\ldots\ldots & \llap{$^\ast$}R.\,McNaught & $\;\;\;\;$'' \\
2012 & Jan. &  1.73 & 22 & 68.8 & 28.0 & \,\,\ldots\ldots\ldots
 & \,\,\ldots\ldots\ldots & \llap{$^\ast$}R.\,McNaught & $\;\;\;\;$'' \\
     &      &  2.27 & 10 & 69.6 & 29.4 & \,\,\ldots\ldots\ldots
 & \,\,\ldots\ldots\ldots & \llap{$^\ast$}M.\,Goiato & Green (2012a) \\[0.1cm]
     &      &  2.6$\;\:$ & \llap{$>$}40 & 69.9 & 30.3 & \,\,\ldots\ldots\ldots
 & \llap{$<$}Dec.\,17.1 & L.\,Barnes & Mailing list \\
     &      &  2.73 & 36 & 70.1 & 30.6 & \,\,\ldots\ldots\ldots & Dec.\,17.3
 & \llap{$^\ast$}R.\,McNaught & $\;\;\;\;$'' \\
     &      &  3.73 & 30 & 70.7 & 33.0 & \,\,\ldots\ldots\ldots
 & \,\,\ldots\ldots\ldots & \llap{$^\ast$}R.\,McNaught & $\;\;\;\;$'' \\
     &      &  4.63 & 33 & 70.9 & 34.8 & \,\,\ldots\ldots\ldots
 & \,\,\ldots\ldots\ldots & R.\,Kaufman & $\;\;\;\;$'' \\
     &      &  4.69 & 30 & 70.9 & 35.0 & \,\,\ldots\ldots\ldots
 & \,\,\ldots\ldots\ldots & M.\,Mattiazzo & $\;\;\;\;$'' \\
     &      &  4.70 & 38 & 70.9 & 35.0 & \,\,\ldots\ldots\ldots
 & Dec.\,17.4 & J.\,Tilbrook & $\;\;\;\;$'' \\[0.1cm]
     &      &  5.70 & 35\rlap{.5} & 70.6 & 36.7 & \,\,\ldots\ldots\ldots
 & \,\,\ldots\ldots\ldots & J.\,Tilbrook & $\;\;\;\;$'' \\
     &      &  6.74 & 21 & 69.9 & 38.0 & \,\,\ldots\ldots\ldots
 & \,\,\ldots\ldots\ldots & R.\,McNaught & $\;\;\;\;$'' \\
     &      & 13.48 & \llap{$>$}18 & 61.1 & 39.2 & \,\,\ldots\ldots\ldots
 & \,\,\ldots\ldots\ldots & R.\,Kaufman & $\;\;\;\;$'' \\
     &      & 14.48 & 46 & 59.7 & 38.7 & \,\,\ldots\ldots\ldots & Dec.\,17.0
 & R.\,Kaufman & $\;\;\;\;$'' \\
%
%
     &      & 15.48 & \llap{$>$}45 & 58.2 & 38.1 & \,\,\ldots\ldots\ldots
 & \llap{$<$}Dec.\,17.0 & R.\,Kaufman & $\;\;\;\;$'' \\
     &      & 15.5$\;\:$ & 47 & 58.2 & 38.1 & \,\,\ldots\ldots\ldots
 & Dec.\,16.9 & L.\,Barnes & $\;\;\;\;$'' \\[0.1cm]
     &      & 16.48 & 45 & 56.8 & 37.4 & \,\,\ldots\ldots\ldots & Dec.\,17.0
 & L.\,Barnes & $\;\;\;\;$'' \\
%
%
%
     &      & 18.50 & 37 & 54.1 & 36.0 & \,\,\ldots\ldots\ldots & Dec.\,17.5
 & L.\,Barnes & $\;\;\;\;$'' \\
     &      & 20.48 & 39 & 51.6 & 34.7 & \,\,\ldots\ldots\ldots & Dec.\,17.2
 & R.\,Kaufman & $\;\;\;\;$'' \\
     &      & 22.5$\;\:$ & 29 & 49.4 & 33.4 & \,\,\ldots\ldots\ldots
 & \,\,\ldots\ldots\ldots & L.\,Barnes & $\;\;\;\;$'' \\
     &      & 23.5$\;\:$ & 39 & 48.4 & 32.7 & \,\,\ldots\ldots\ldots
 & Dec.\,17.0 & L.\,Barnes & $\;\;\;\;$'' \\
     &      & 24.5$\;\:$ & 30 & 47.5 & 32.1 & \,\,\ldots\ldots\ldots
 & \,\,\ldots\ldots\ldots & L.\,Barnes & $\;\;\;\;$'' \\[0.1cm]
     &      & 25.55 & 26 & 46.6 & 31.5 & \,\,\ldots\ldots\ldots 
 & \,\,\ldots\ldots\ldots & R.\,Kaufman & $\;\;\;\;$'' \\
     &      & 26.52 & 31 & 45.7 & 31.0 & \,\,\ldots\ldots\ldots
 & \,\,\ldots\ldots\ldots & L.\,Barnes & $\;\;\;\;$'' \\
     &      & 28.52 & 17\rlap{.5} & 44.2 & 30.0 & \,\,\ldots\ldots\ldots
 & \,\,\ldots\ldots\ldots & L.\,Barnes & $\;\;\;\;$'' \\
     &      & 30.59 & 26 & 42.8 & 29.0 & \,\,\ldots\ldots\ldots
 & \,\,\ldots\ldots\ldots & R.\,Kaufman & $\;\;\;\;$'' \\
     & Feb. & 12.46 & $\;\:$7 & 36.7 & 24.8 & \,\,\ldots\ldots\ldots
 & \,\,\ldots\ldots\ldots & R.\,Kaufman & $\;\;\;\;$'' \\
     & Mar. & 16$\;\;\;\;\,$ & $\;\:$1\rlap{.2} & 28.3 & 18.9
 & \,\,\ldots\ldots\ldots & \,\,\ldots\ldots\ldots & L.\,Barnes
 & Barnes (2012)\\[0.1cm]
\hline \\[-0.2cm]
\multicolumn{10}{l}{\parbox{15.8cm}{$^{\rm a}\;${\scriptsize Assuming the
tail consists only of dust released during the outburst that peaked
on Dec.\ 17.6 UT, about 1.6 days after perihelion.}}}\\[-0.07cm]
\multicolumn{10}{l}{\parbox{15.8cm}{$^{\rm b}\:${\scriptsize For reference,
the passage through perihelion occurred on Dec.\ 16.01 UT.}}}\\[-0.01cm]
\multicolumn{10}{l}{\parbox{15.8cm}{$^{\rm c}\:${\scriptsize Asterisk
preceding the observer's name indicates a visual detection.}}}\\[-0.06cm]
\multicolumn{10}{l}{\parbox{15.8cm}{$^{\rm d}\,${\scriptsize Mailing list =
http://tech.groups.yahoo.com/group/comets-ml; messages
18968--19383.}}}
\end{tabular}}
\end{center}
\end{table*}

At the other end of the particle-size spectrum, a question arises as to how
large the dust debris from the nucleus disintegration event of December 17
must be in order that it be released into elliptical orbits and not be lost
to interstellar space.  This can readily be determined by equating the
orbital velocities of the comet and the debris on December 17.6 UT.  For
the debris to move below the escape limit, one requires that its $\beta_{\rm
ell}$ be less than $\beta_{\rm par}$ for the parabolic limit, which follows
from a condition
\begin{equation}
\frac{2}{r_{\rm rel}} - \left( \frac{1}{a} \right)_{\rm comet} \!\! = (1 -
\beta_{\rm par}) \: \frac{2}{r_{\rm rel}},
\end{equation}
where $(1/a)_{\rm comet}$ is the inverse semi-major axis of the comet's orbit
and $r_{\rm rel}$ is the heliocentric distance at the time of release,
\mbox{$r_{\rm rel} = 0.144$} AU (Sec.\ 5).  Thus, for the debris in elliptical
orbits
\begin{equation}
\beta_{\rm ell} < \frac{1}{2} r_{\rm rel} \left( \frac{1}{a} \right)_{\rm
 comet} \!\! = 0.000915,
\end{equation}
and the particle diameters $x_{\rm ell}$ at an assumed bulk density of 0.4 g
cm$^{-3}$ are $>$0.3 cm, if released during the December 17 event.  The
correction to the barycenter of the solar system would change this result
imperceptibly.  The same condition for particles released at perihelion would
yield \mbox{$\beta_{\rm ell} < 0.000035$} and \mbox{$x_{\rm ell} > 8.2$} cm.
We conclude that only coarse-grain and larger debris continues to orbit the
Sun in elliptical orbits.  This is a way to generate SOHO-like minicomets
that move in orbits very similar to that of the comet from which they
originate, except for the orbital period.

\section{Sublimation of Microscopic Dust Near the Sun}

The existence of the so-called dust-free zone around the Sun, a result of
profuse sublimation of interplanetary dust, has been extensively studied.
The problem is well reviewed by, for example, Mann et al.\ (2004), based in
part on earlier works by Krivov, Kimura, \& Mann (1998), Kimura, Ishimoto, \&
Mukai (1997), and others.  More recently this topic was also discussed by Kama,
Min, \& Dominik (2009) in connection with the problem of an inner boundary of
protoplanetary discs.  Mann et al.\ (2004) list the radius of the sublimation
zone for different materials, as computed by various authors.  The numbers of
particular interest are: 1.5--4 {\Rsun} for quartz (compact), 4 {\Rsun}
for glassy carbon (both compact and fluffy), 5--6.5 {\Rsun} for Mg-rich
pyroxene (both modes), 10 {\Rsun} and 13.5--15.5 {\Rsun} for Mg-rich olivine
(compact and fluffy, respectively), 14 {\Rsun} for astronomical silicate
(compact), and 11--24 {\Rsun} for a compact iron sphere.  In most cases,
however, the particles were assumed to orbit the Sun in a nearly-circular
path until complete attrition, while for cometary particles moving along
strongly hyperbolic orbits the critical perihelion distance is expected to
be even slightly less.

The mineralogical and morphological properties of silicate grains in
sungrazing comets were studied extensively by Kimura et al.\ (2002), who
used a large set of light curves of the SOHO Kreutz minicomets as the data
base.  The authors concluded that the dust tails contained aggregates of
submicron-sized crystalline grains, but not amorphous grains.  The
sublimation of fluffy olivine aggregates was proposed by them to explain the
downturn on the light curves near 11--12 {\Rsun}.  An increasing outflow
of pyroxene fluffy aggregates was suggested as the source for a secondary
maximum on the light curves near 7 {\Rsun}.  Their hypothesis does not appear
to assign any significance to the range of heliocentric distances between 1.5
and 2 {\Rsun}.

To examine the sublimation properties of microscopic dust surviving in
hyperbolic orbits with these small perihelion distances, consider a
spherical dust particle of a bulk density $\rho$ and radius $a_{\rm rel}$
released from the nucleus at time $t_{\rm rel}$.  Let the particle, which
moves in a field of reduced effective gravity, sublimate at a temporally
variable rate $\dot{a}_{\rm subl}(t)$, such that at the time of observation,
$t_{\rm obs}$, the radius is reduced to \mbox{$a_{\rm fin} \ll a_{\rm rel}$},
so that
\begin{equation}
a_{\rm fin} = a_{\rm rel} - \! \int_{t_{\rm rel}}^{t_{\rm obs}} \!\!
 \dot{a}_{\rm subl}(t) \, dt,
\end{equation}
where the sublimation rate is defined as a positive quantity.  The rate
$\dot{a}_{\rm subl}$ depends strongly on the particle's equilibrium
temperature $T(t)$ and can be expressed in terms of the mass sublimation
rate per unit area, $\dot{\cal Z}_{\rm subl}$ (in g cm$^{-2}$ s$^{-1}$):
\begin{equation}
\dot{a}_{\rm subl}(T) = \frac{\dot{\cal Z}_{\rm subl}(T)}{\rho},
\end{equation}
where $\dot{\cal Z}_{\rm subl}$ is a function of the sublimation pressure,
$\wp_{\rm subl}(T)$,
\begin{equation}
\dot{\cal Z}_{\rm subl}(T) = \left( \frac{\mu}{2 \pi \Re T}
 \right)^{\frac{1}{2}} \wp_{\rm subl}(T),
\end{equation}
$\mu$ is the molar weight (in g mol$^{-1}$), $\Re$ is the gas constant (in
erg K$^{-1}$ mol$^{-1}$), and
\begin{equation}
\wp_{\rm subl}(T) = \Lambda \exp \! \left[ - \frac{L}{\Re T} \right],
\end{equation}
with $\Lambda$ and $L$, the latent heat of sublimation, being constants.

The position at time $t$ of a particle, moving in a concave hyperbolic orbit
of the perihelion distance $q$ and eccentricity $e$, is given by a hyperbolic
eccentric anomaly $f$, which is related to $t$ by
\begin{equation}
e \sinh f - f = \frac{k_{\rm grav} (1 - \beta)^{\frac{1}{2}}
 (e-1)^{\frac{3}{2}}}{q^{\frac{3}{2}}} \, (t - t_\pi^\ast),
\end{equation}
where $k_{\rm grav}$ is the Gaussian gravitational constant, $\beta < 1$ is
again the radiation-pressure parameter from Eq.\ (1), and $t_\pi^\ast$ is the
time of the particle's passage through perihelion.  The anomaly $f$ can for
any $t - t_\pi^\ast$ be computed by successive iterations of Eq.\ (24).  If
$t$ is reckoned from $t_\pi$, the time of the comet's passage through
perihelion, then in Eq.\ (24) one substitutes \mbox{$t - t_\pi^\ast = t -
t_\pi - (t_\pi^\ast - t_\pi)$}.  Since the heliocentric distance at time $t$
is equal to
\begin{equation}
r(t) = \frac{q}{e-1} (e \sinh f - 1),
\end{equation}
we can insert quantities from Eqs.\ (21) to (25) into Eq.\ (20) to obtain
the integral in the form
\begin{equation}
\int_{t_{\rm rel}}^{t_{\rm obs}} \!\! \dot{a}_{\rm subl} \, dt = \Psi \!\!
  \int_{f_{\rm rel}}^{f_{\rm obs}} \! T^{-\frac{1}{2}} \exp \! \left[
  -\frac{L}{\Re T} \right] r \, df,
\end{equation}
where $f_{\rm rel}$ and $f_{\rm obs}$ are the hyperbolic anomalies at times
$t_{\rm rel}$ and $t_{\rm obs}$, and
\begin{equation}
\Psi = \frac{\Lambda \mu^{\frac{1}{2}}}{\rho} \,
\frac{q^{\frac{1}{2}}}{k_{\rm grav}} \,
[2 \pi \Re (1 - \beta) (e - 1)]^{-\frac{1}{2}}.
\end{equation}

For a given $t_{\rm obs}$, the time of release $t_{\rm rel}$ and the
radiation-pressure parameter $\beta$ are determined by fitting the imaged
tail's morphology with a set of syndynames that describe the distribution
of dust particles in the tail.  These computations automatically provide
the particles' perihelion distance $q$ and eccentricity $e$.  Now the
magnitude of the integrated sublimation effect depends on four quantities
--- the sublimation-pressure constant $\Lambda$ (related to the entropy of
the particle material), the latent heat (or enthalpy) of sublimation $L$,
the molar weight $\mu$, and the bulk density $\rho$.  The effect is also a
function of the variations in the equilibrium temperature $T$ along the
particles' hyperbolic orbits.  In practice, the computations are difficult
because this temperature is a complicated function of the optical,
thermophysical, and morphological properties of the particle material, all
of which vary with time and heliocentric distance.

We limit ourselves to a brief examination of the likelihood that the dust of
C/2011 W3 that begins to sublimate profusely at $\sim$1.8 {\Rsun} consists of
olivine-dominated silicate particles released from the nucleus, respectively,
0.08 and 0.12 day before perihelion and subjected to solar radiation pressure
of \mbox{$\beta = 0.6$}.  The first, more conservative case implies that
\mbox{$q = 1.7294$} {\Rsun}, \mbox{$e = 2.4513$}, and \mbox{$t_\pi^\ast -
t_\pi = -0.0008$} day, whereas the other case leads to \mbox{$q = 1.8841$}
{\Rsun}, \mbox{$e = 2.1678$}, and \mbox{$t_\pi^\ast - t_\pi = +0.0078$} day.
For the temperature we use only a fairly crude approximation based on a
black-body temperature and an assumed thermal radiative regime,
\begin{equation}
T(r) = \frac{280\,{\rm K}}{r^{\frac{1}{2}}}.
\end{equation}

The sublimation rate and other properties of olivine depend on the ratio of
magnesium to iron in its formula Mg$_{_{\scriptstyle z}}$Fe$_{2-z}^{2+}$SiO$_4$,
where \mbox{$0 \leq z \leq 2$}.  Following Barthelmy (2010), we adopt \mbox{$z
= 1.6$} for the empirical formula, \mbox{$\mu = 153.31$} g mol$^{-1}$ for the
molar weight, and \mbox{$\rho = 3.32$} g cm$^{-3}$, an average mineralogical
density, for the bulk density of submicron-sized olivine-dominated grains.  

We now require that the integrated sublimation effect computed from Eq.\ (26)
amounts to 0.2 $\mu$m in the particle radius, the average size of silicate
grains near \mbox{$\beta = 0.6$} (e.g., Kimura, Ishimoto, \& Mukai 1997;
Kimura et al.\ 2002) and search for the relationship between the two remaining
unknowns --- the sublimation-pressure constant $\Lambda$ and the heat of
sublimation $L$ --- that satisfy this total loss in the particle radius between
$t_{\rm rel}$ and $t_{\rm obs}$.  The two values of $t_{\rm rel}$ thus lead to
two sets of solutions, which are insensitive to the observation time, as long
as the observation takes place after the particle's passage through perihelion.
The solutions are presented in Fig.\ 11 in a range of 100 to 150 kcal mol$^{-1}$
for the heat of sublimation and compared with several sources of information
on sublimation of silicates and other potentially relevant materials.  Because
Eqs.\ (26) and (27) show that the integrated sublimation effect depends on
$\Lambda$ only through a function $\Lambda \mu^{\frac{1}{2}} \rho^{-1}$, it is
this function that is plotted in Fig.\ 11 against the heat of sublimation $L$.

Nagahara, Mysen, \& Kushiro (1994) published the sublimation-pressure constants
for the two endmembers of the olivine solid solution system.  Including the
molar weight and density, the data are as follows:\ \mbox{$\Lambda = 6.72 \times
10^{13}$} Pa, \mbox{$L = 129.74$} kcal mol$^{-1}$, \mbox{$\mu = 140.69$} g
mol$^{-1}$, and \mbox{$\rho = 3.27$} g cm$^{-3}$ for forsterite (Mg-endmember,
for which \mbox{$z = 2$}); and \mbox{$\Lambda = 2.48 \times 10^{15}$} Pa,
\mbox{$L = 119.94$} kcal mol$^{-1}$, \mbox{$\mu = 203.78$} g mol$^{-1}$,
and \mbox{$\rho = 4.39$} g cm$^{-3}$ for fayalite (Fe-endmember, for which
\mbox{$z = 0$}).  Because the sublimation properties vary smoothly between
the two endmembers (e.g., Gail \& Sedlmayr 1999), we plot in Fig.\ 11 the
dependence of $L$ on $\Lambda \mu^{\frac{1}{2}} \rho^{-1}$ for the entire set
of olivine members.  In their study, Kimura et al.\ (2002) used Nagahara,
Mysen, \& Kushiro's (1994) sublimation parameters for forsterite, but the molar
weight for olivine with \mbox{$z = 1.1$}, and the density of 3.3 g cm$^{-3}$,
so that in Fig.\ 11 their point would be located just 0.04 in log$(\Lambda
\mu^{\frac{1}{2}} \rho^{-1})$ to the right of forsterite.  For reference and
orientation we also plot in the figure:\ (i) the data point for ``silicates'',
as used by Kimura, Ishimoto, \& Mukai (1997) --- \mbox{$\Lambda = 1.07 \times
10^{13}$} Pa, \mbox{$L = 113.98$} kcal mol$^{-1}$, \mbox{$ \mu = 67.0$} g
mol$^{-1}$, and \mbox{$\rho = 2.37$} g cm$^{-3}$; (ii) the numbers that the
most recent CRC Handbook for Chemistry and Physics (Haynes 2011) provides for
quartz (SiO$_2$):\ \mbox{$\Lambda = 1.38 \times 10^{13}$} Pa, \mbox{$L =
134.55$} kcal mol$^{-1}$, \mbox{$\mu = 60.08$} g mol$^{-1}$, and \mbox{$\rho =
2.65$} g cm$^{-3}$; and (iii) an extrapolated relationship between $\Lambda$
and $L$, as used by Sekanina (2003) for some metallic elements in his study of
the light curves of SOHO sungrazers, with an average value of 4 for the ratio
$\sqrt{\mu}/\rho$, which varied roughly between 1 and 7.

To the extent that we can --- despite the approximations --- express some
confidence from the meaning of the plot in Fig.\ 11, we conclude that silicates
dominated by Mg-rich olivine could represent a plausible candidate for the
material that made up dust particles that began to sublimate profusely in the
tail of comet C/2011 W3 near a heliocentric distance of 1.8 {\Rsun}.

An effect that has not been accounted for in this paper is the deviation of
the radiation-pressure acceleration from the $r^{-2}$ dependence very close
to the Sun, as briefly mentioned in Sec.\ 6.  The reason for this deviation
is twofold:\ (i) very close to the Sun, the particle is subjected to the
radiation coming only from that part of the photosphere that is above the
particle's horizon, and (ii) the photosphere is closer to the particle by up
to 1 {\Rsun} than the Sun's center.  The first fact decreases the magnitude of
the radiation pressure relative to that from the standard inverse-square power
law, while the latter one increases it.  Combined, the second influence
prevails, and the integration over the relevant region shows that at a distance
$r$ from the Sun's center the actual acceleration, $\beta^\ast$, exceeds the
standard value, $\beta$, by
\begin{equation}
\frac{\beta^\ast}{\beta} = 2 \left( \! \frac{r}{\mbox{\Rsun}} \! \right)^2
 \! \left[ 1 - \sqrt{1 - \left( \! \frac{\mbox{\Rsun}}{r} \! \right) ^2}
 \, \right].
\end{equation}
It is clear that $\beta^\ast = \beta$ for large values of the ratio
$r/\mbox{\Rsun}$, while $\beta^\ast/\beta$ is equal to 1.01 for \mbox{$r =
5$}~{\Rsun}, 1.15 for \mbox{$r = 1.5$}~{\Rsun}, and 1.29 for \mbox{$r =
1.2$}~{\Rsun}.  For the borderline heliocentric distances involved in the
sublimation problem this effect does not exceed 10 percent and is not
significant enough to invalidate the results.

\begin{figure*}[ht]
 \vspace*{-4.5cm}
 \hspace*{-0.6cm}
 \centerline{
 \scalebox{0.95}{
 \includegraphics{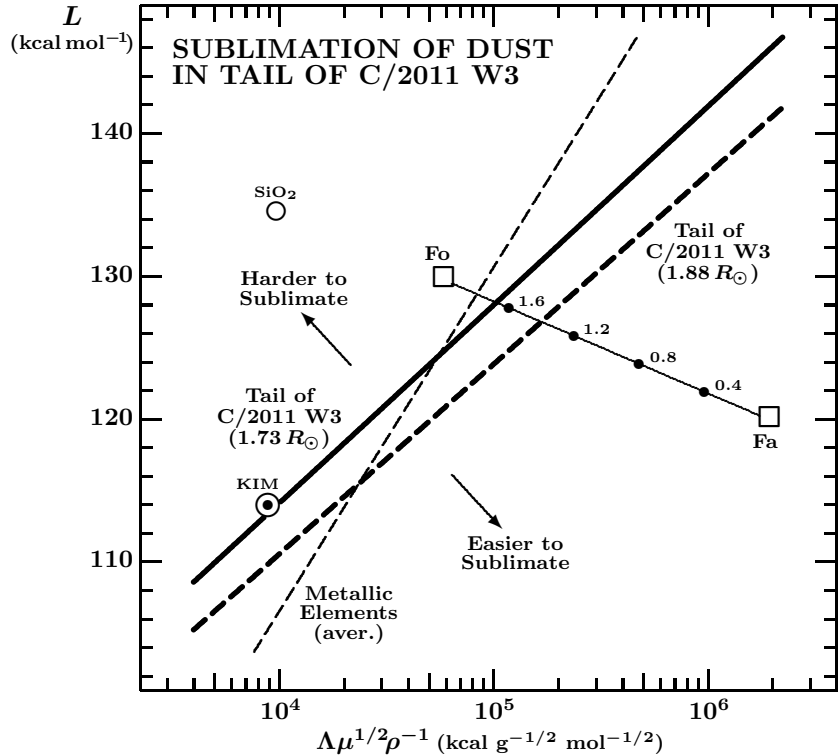}}}
 \vspace*{-13.65cm}
 \caption{The relationship between the parameter $\Lambda \mu^{ \frac{1}{2}}
\rho^{-1}$ --- a function of the sublimation-pressure constant, molar weight,
and bulk density --- and the latent heat of sublimation $L$.  The uninterrupted
thick curve is the set of solutions that refer to the onset of profuse
sublimation of dust in the tail of comet C/2011 W3 at 1.73 {\Rsun} from the
Sun.  The dashed thick curve is the set of solutions that refer to the onset
of profuse sublimation at 1.88 {\Rsun}.  The two squares are the locations in
the plot of the two endmembers of the olivine solid solution system, based on
the sublimation-pressure constants by Nagahara, Mysen, \& Kushiro (1994):\
forsterite (Fo; Mg-endmember) and fayalite (Fa; Fe-endmember).  The dots on
the Fo-Fa connecting line refer, from left to right, to the members of the
olivine system, whose formula is, respectively, Mg$_{1.6}$Fe$_{0.4}$SiO$_4$,
Mg$_{1.2}$Fe$_{0.8}$SiO$_4$, Mg$_{0.8}$Fe$_{1.2}$SiO$_4$, and
Mg$_{0.4}$Fe$_{1.6}$SiO$_4$.  For reference and orientation, we also plot the
data points for ``silicates,'' as used by Kimura, Ishimoto, \& Mukai (1997)
(KIM), and for quartz, SiO$_2$ (Haynes 2011), and an extrapolated relationship
between $\Lambda$ and $L$, as used by Sekanina (2003) for some metallic
elements in his study of the light curves of SOHO sungrazers, with an average
value of 4 for the ratio $\sqrt{\mu}/\rho$.  The constants for olivine used
by Kimura et al.\ (2002) are not shown, as they nearly coincide with those for
forsterite.}
\vspace*{0.5cm}
\end{figure*}

\section{Rapid Buildup of Thermal Stress in Nucleus' Interior and Process
of Cata\-clysmic Fragmentation}

From the information presented, the process of disintegration of comet C/2011
W3 appears to have been followed for at least four days immediately after
perihelion.  A drop in activity around December 15.7 UT, or 0.3 day before
perihelion, may have been related to sublimation of
dust (Sec.\ 11) and was not necessarily an early signature of the process of
disintegration.  On the other hand, the sudden and permanent loss of the
nuclear condensation observed first between December 19.4 and 20.3 UT (Fig.\
1) was clear evidence of terminal collapse.  There has been no trace of new
activity detected from December 20 on, indicating that the comet's progressive
fading after this date only reflected the rate of dispersal in space of the
dust ejecta released during the December 16--20 period of activity.

It is not absolutely clear whether in this critical period of time the activity
proceeded more or less continuously or was dominated by discrete outbursts
other than the terminal event(s).  Either way, this activity must account for
the formation of the quasi-parabolic envelope, encircling the spine tail, and
also visible in its developing phase in the images from December 17 through 19.
Although the particle-ejection velocities in the spine tail were much lower
than in the envelope (Sec.\ 8), the outburst that triggered the formation of
the spine tail was obviously the most devastating for the comet in that it
was associated with the cataclysmic fragmentation of the entire residual mass
of the nucleus.  The low velocities of mostly large-sized debris may explain
why to an earth-based observer the process appeared to have taken at least
another two days to complete.  Simple calculations support this scenario.
Indeed, released on December 17.6 UT and expanding at $\sim$20 m s$^{-1}$,
large dust in the coma would occupy a volume of about 6000 km, or
10$^{\prime\prime}$, in diameter on December 19.4 UT, when the comet still
displayed a nuclear condensation (Fig.\ 1), and nearly 10,000 km, or
17$^{\prime\prime}$, in diameter on December 20.3 UT, when the condensation
was already gone.  The diameter of the
bright condensation in the Malargue image from December 19
is in fact about 30$^{\prime\prime}$, which suggests a major contribution
of dust from the previous activity, while the breadth of the nascent spine
tail in the Malargue image from December 20 is just about the expected
17$^{\prime\prime}$.  The bright streamer in the December 19 image --- the
early appearance of the spine tail --- can be traced to a distance of at least
6$^\prime$ from the nucleus; its end point indicates the radiation-pressure
acceleration parameter of \mbox{$\beta \simeq 0.12$}, that is, the presence
of dust particles larger than 10 $\mu$m in diameter, when released during the
December 17.6 outburst.

It is important to realize that the timing of this event offers unequivocal
evidence of a {\it gradual deterioration\/} of the comet's health and thereby
demonstrates its appreciable {\it resistance\/} to the hostile environment in
close proximity of the Sun.  This conclusion clearly shows that the nucleus
could not possibly have been held together merely by self-gravity.  The comet's
significantly delayed response does not favor models that describe the cometary
nucleus as a strengthless or very poorly cemented rubble-pile structure, which
--- by virtue of being unconsolidated debris --- reacts to tidal forces
promptly (e.g., Asphaug \& Benz 1994, 1996; Solem 1994).  Evidence from comet
C/2011 W3 thus leads to the same conclusion as evidence from the close-up
imaging of the nuclei of 81P/Wild and 9P/Tempel, which were independently
found incompatible with the rubble pile model by, respectively, Brownlee et
al.\ (2004) and Thomas et al.\ (2007).

In a very recent study, Gundlach et al.\ (2012) explain the survival of
sungrazing comets like C/2011 W3 within the Roche limit of the Sun by the
counterpressure due to virtually isotropic outgassing from the nucleus, which
more than compensates for the tidal disruption force.  For C/2011 W3 such a
scenario is, however, unnecessary, because its nuclear diameter at perihelion
was almost certainly less than 1 km and the tidal stress therefore less than
$\sim$10 Pa (as seen from Fig.\ 2 in Gundlach et al.'s paper), a benign effect
for any comet except an essentially strengthless one.  A major problem with
Gundlach et al.'s scenario is that it is inconsistent with the extensively
observed duplicity of the sungrazing comet C/1965 S1 (e.g., Pohn 1965,
Thackeray 1965, Iannini 1966, Lourens 1966), whose estimated diameter of
less than 9 km (Knight et al.\ 2010) and perihelion distance of 1.67 {\Rsun}
place the comet deep inside Gundlach et al.'s ``safe'' zone.

What property was it then that made C/2011 W3 survive perihelion on the one
hand but suddenly disintegrate nearly 2 days later on the other hand?  Always
suspected in any substantially delayed response are the thermal effects in the
comet's nuclear interior, as transport of heat takes time.  The resulting
temperature changes, dramatically enhanced in the sungrazers near perihelion,
impose a severe burden in the form of thermal stress, which may not
only dwarf the tidal stress but could last for extended periods of time.

To investigate heat transport effects in the nucleus of C/2011 W3, we have
followed the procedure outlined and applied by Sekanina (2009) to study the
giant explosions of 17P/Holmes.  The approach assumes that the solar radiation
impinging on the nucleus is spent --- in rather extensive bare areas of the
surface, devoid of ice --- only on thermal reradiation and conduction of heat
into the interior.  This case is therefore, in principle, contrary to that
considered by Gundlach et al.\ (2012).  Being interested merely in basic,
order-of-magnitude information on thermal stress, we have applied only the
standard (isothermal) version of the one-dimensional heat-transfer problem
in a spherical object.  We have assumed that an initial central temperature
of the nucleus is 60\,K and a coefficient of effective thermal conductivity
\mbox{$K_{\rm eff} = 0.2$} W m$^{-1}$ K$^{-1}$, which was found the most
likely value to fit the recurrence of the giant explosions of comet 17P once
every 115 years or so (Sekanina 2009) and which is also well within the
interval compatible with the range of the thermal inertia that Davidsson,
Guti\'errez, \& Rickman (2009) derived in their analysis of near-infrared
thermal emission spectra of features on the nucleus of comet 9P/Tempel.
With the numbers used for the specific heat and the bulk density, this
value of $K_{\rm eff}$ is equivalent to an effective thermal diffusivity of
\mbox{$\kappa_{\rm eff} \simeq 0.006$} cm$^2$ s$^{-1}$.

The results of applying the heat-transfer equation to the nucleus of
comet C/2011 W3 are presented in Figs.\ 12 and 13.  Figure 12, a plot of
temperature against depth beneath the surface, exhibits an enormous perihelion
asymmetry.  The plot shows for example that the same temperature reached 0.6
day before perihelion at a depth of less than 10 meters is reached 1.6 days
after perihelion at a depth of more than 40 meters.  A significant conclusion
from Fig.\ 13, a plot of the postperihelion temperature with heliocentric
distance, is that except in the topmost 15-meter-deep layer, the temperature
keeps increasing as the comet recedes from the Sun; it is still increasing at
large heliocentric distances.  In this regime, the effects of thermal stress
must keep increasing and must further be enhanced by sudden activity from
scattered reservoirs or ``pockets'' of highly volatile substances, such as
amorphous water ice, reached by the rapidly penetrating thermal wave deep
inside the nucleus.  Sooner or later, the combination of thermal stress and
such explosive events is bound to have catastrophic consequences for the
nucleus.  We thereby find an {\it entirely plausible mechanism for cascading
fragmentation of sungrazers, large or small, well away from the Sun\/}.

\begin{figure}[ht]
 \vspace*{-4.6cm} 
 \hspace*{0.9cm} 
 \centerline{
 \scalebox{1.05}{
 \includegraphics{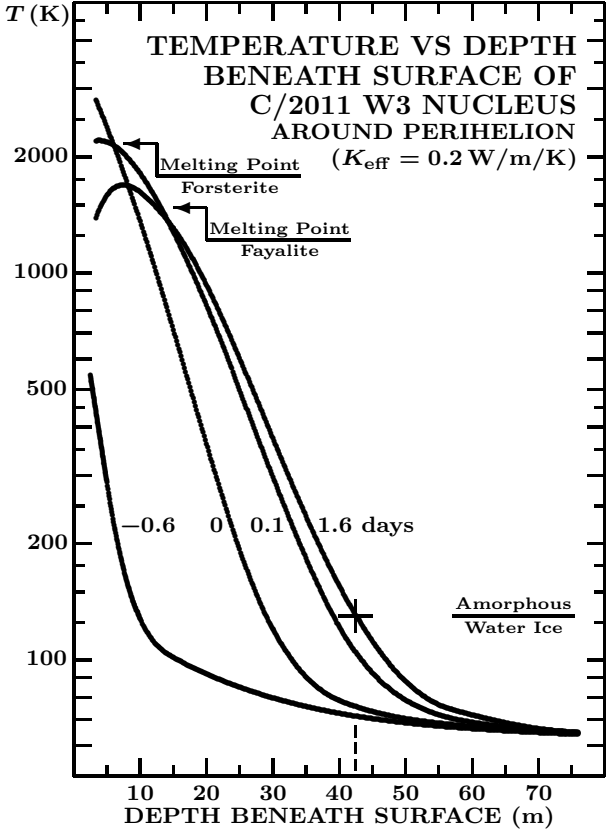}}}
 \vspace*{-15.6cm} 
 \caption{Temperature variations with depth beneath the surface derived for
the nucleus of comet C/2011 W3 with a maximum diameter of 600 meters.  The
one-dimensional heat-transfer equation has been solved, averaging the incident
solar energy over the surface, assuming that the energy is spent only on the
conduction into the nucleus and thermal reradiation (bare surface, devoid of
ices or other substances that would sublimate), and adopting a Bond albedo of
0.03, an emissivity of unity, a coefficient of effective thermal conductivity
of \mbox{$2 \times 10^4$} erg cm$^{-1}$ s$^{-1}$ K$^{-1}$, a specific heat
capacity of \mbox{$8 \times 10^6$} erg g$^{-1}$ K$^{-1}$, and a bulk density
of 0.4 g cm$^{-3}$.  The individual curves are identified by the time from
perihelion:\ 0.6 day before perihelion, at perihelion, and 0.1 and 1.6 days
after perihelion, when comet C/2011 W3 was at heliocentric distances of,
respectively, 0.0731, 0.00555, 0.0196, and 0.144 AU.  Also marked are the
crystallization temperature of amorphous water ice and the melting points
of forsterite and fayalite.}
\vspace*{0.25cm} 
\end{figure}

\begin{figure}
 \vspace*{-4.75cm} 
 \hspace*{0.85cm}
 \centerline{
 \scalebox{1.1}{ 
 \includegraphics{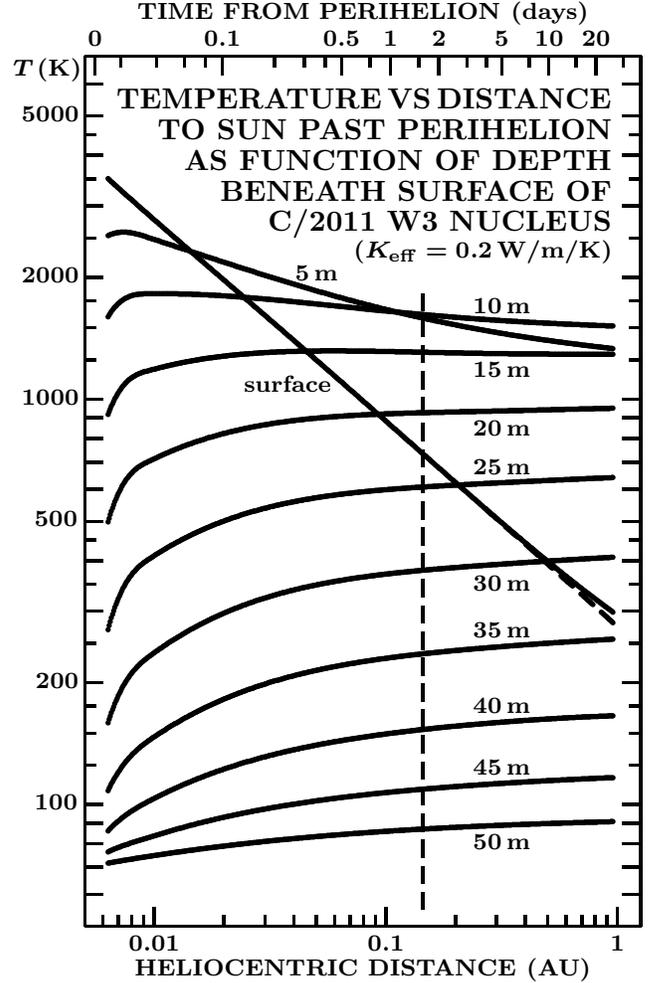}}}
 \vspace*{-14.95cm} 
 \caption{Temperature variations with heliocentric distance at the surface
and at 10 depths beneath the surface derived for the nucleus of comet
C/2011 W3.  The computations are based on the model and assumptions
described in the caption to Fig.\ 12.  The surface-temperature variations
are shown for preperihelion (broken) and postperihelion branches of the
orbit, all other curves are postperihelion.  The broken line parallel to
the axis of ordinates shows the heliocentric distance 0.144 AU, at which
the cataclysmic fragmentation event occurred on December 17.6 UT.}
\mbox{}\\[-0.2cm]
\end{figure}

The general equations for the radial and tangential components of thermal
stress, $\sigma_\parallel$, $\sigma_\perp$, at distance $h$ from the center
of a sphere of radius $h_{\displaystyle \star}$ are given by (Timoshenko \&
Goodier 1970)
\begin{eqnarray}
\sigma_\parallel(h) &  = & 2 \chi \left[
 \frac{\Phi(h_{\displaystyle \star})}{h_{\displaystyle \star}^3}
 - \frac{\Phi(h)}{h^3} \right], \nonumber \\[-0.2cm]
 & & \\[-0.2cm]
\sigma_\perp(h) &  = & \chi \left[ \frac{2
 \Phi(h_{\displaystyle \star})}{h_{\displaystyle \star}^3}
 + \frac{\Phi(h)}{h^3} - T(h) \right], \nonumber
\end{eqnarray}
where
\begin{equation}
\chi = \frac{\alpha_{\rm L} E_{\rm Y}}{1 - \nu}
\end{equation}
is the thermal stress parameter and
\begin{equation}
\Phi(h) = \int_{0}^{h} \!\! T(s) \, s^2 ds,
\end{equation}
with $T(s)$ being the temperature at distance $s$ from the center.  The
object's thermal and mechanical properties determine the coefficient of linear
thermal expansion $\alpha_{\rm L}$, Young's modulus $E_{\rm Y}$, and Poisson's
ratio $\nu$.  From the first of Eqs.\ (30) we note that at the surface
$(h = h_{\displaystyle \star})$ the radial component $\sigma_\parallel$ of
thermal stress is always nil.  On the other hand, at the center the radial
and tangential components have the same magnitude equal to
\begin{equation}
\lim_{h \rightarrow 0} \sigma_\parallel = \lim_{h \rightarrow 0} \sigma_\perp =
 2 \chi \left[ \frac{\Phi(h_{\displaystyle \star})}{h_{\displaystyle \star}^3}
 - {\textstyle \frac{1}{3}}
 T_0 \right],
\end{equation}
where $T_0 = T(0)$ is the central temperature.  In addition, there exists a
mean interior temperature $\langle T \rangle$ such that
\begin{equation}
\int_{0}^{h_{\scriptstyle \star}} \!\! T(s) \, s^2 ds = \langle T \rangle \!\!
 \int_{0}^{h_{\scriptstyle \star}} \!\! s^2 ds = {\textstyle \frac{1}{3}}
 \langle T \rangle \, h_{\displaystyle \star}^3,
\end{equation}
so that
\begin{equation}
\lim_{h \rightarrow 0} \sigma_\parallel = \lim_{h \rightarrow 0} \sigma_\perp
 = {\textstyle \frac{2}{3}} \chi \left[ \langle T \rangle - T_0 \right].
\end{equation}
As the distribution of temperature in the nucleus' interior keeps steadily
increasing with time and heliocentric distance following the perihelion
passage, so do both $\langle T \rangle$ (much more so than $T_0$) and the
thermal stress in the center.  The tendency toward a sungrazer's breakup far
from the Sun due to thermal forces is confirmed even without the existence
of subsurface reservoirs of highly volatile ices.

In Eqs.\ (30) we have divided the entire range of distances from the center,
\mbox{$0 \leq h \leq h_{\displaystyle \star}$}, into $n$ separate intervals
\mbox{$h_{i-1} \leq h \leq h_i \,(i=1,\ldots,n)$}, where $h_0 = 0$ and
$h_n = h_{\displaystyle \star}$, and fitted the temperature $T$ from Fig.\ 12
inside each interval by a polynomial of power $m$,
\begin{equation}
T(h) = \sum_{k=0}^{m} a_{k,i} h^k, \;\;\; i = 1, \ldots, n.
\end{equation}
The integral $\Phi(h_{\displaystyle \star})$ is then given by
\begin{equation}
\Phi(h_{\displaystyle \star}) = \sum_{i=1}^{n} \sum_{k=0}^{m}
 \frac{a_{k,i}}{k+3} \left( h_i^{k+3} - h_{i-1}^{k+3} \right)
\end{equation}
and for distance $h$ from the center in a \mbox{$(j \! + \! 1)$-st} interval,
\mbox{$h_j \leq h \leq h_{j+1}$} \mbox{($0 \le j \le n \! - \! 1$)}, the
integral $\Phi(h)$ is equal to
\begin{eqnarray}
\Phi(h) &  = & \sum_{i=1}^{j} \sum_{k=0}^{m} \frac{a_{k,i}}{k+3}
 \left( h_i^{k+3} - h_{i-1}^{k+3} \right) \nonumber \\
 &  & \!\! + \sum_{k=0}^{m} \frac{a_{k,j+1}}{k+3} \left( h^{k+3} - h_j^{k+3}
 \right).
\end{eqnarray}
Because Fig.\ 12 shows that the temperature changes very little at depths
exceeding 60--70 meters, the described approach can also be used to approximate
the magnitude and distribution of thermal stress in the interior of the nucleus
of smaller dimensions, whose center is assumed to be at depth $h_{\displaystyle
\star} - \hbar$ rather than at $h_{\displaystyle \star}$.  Reckoning distance
$h$ (but not the interval boundaries $h_{i-1}$ and $h_i$) from the new center
at this depth, we can write Eq.\ (36) in the form
\begin{equation}
T(h) = \sum_{k=0}^{m} a_{k,i} (h + \hbar)^k = \sum_{k=0}^{m} b_{k,i} h^k,
 \;\;\; i = j, \ldots, n,
\end{equation}
when $\hbar$ is in an interval \mbox{$h_{j - 1} \leq \hbar \leq h_j$}
\mbox{$(1 \leq j \leq n)$}, and for \mbox{$0 \leq k \leq m$}
\begin{equation}
b_{k,i} = \sum_{l=k}^{m} \left( \!\! \begin{array}{c}l \\ k \end{array}
 \!\! \right) a_{l,i} \hbar^{l-k}, \;\;\; i = j, \ldots, n.
\end{equation}
Coefficients $b_{k,i}$ now replace $a_{k,i}$ in Eqs.\ (37) and (38), in which
the interval boundaries have to be modified accordingly.

This set of formulas can also be used when the size of the nucleus is truncated
to \mbox{$h_{\displaystyle \bullet} < h_{\displaystyle \star}$} on the outside,
that is, when it is desirable to eliminate the contribution to thermal stress
from a surface layer once it has been removed by erosion or otherwise
destroyed.  In this case the integral $\Phi(h_{\displaystyle \bullet})$ is
computed from Eq.\ (38) rather than (37), either with the coefficients
$a_{k,i}$ or $b_{k,i}$, depending on whether $\hbar$ is or is not nil.

\begin{figure}[h]
 \vspace*{-4.6cm}
 \hspace*{-0.25cm}
 \centerline{
 \scalebox{0.94}{
 \includegraphics{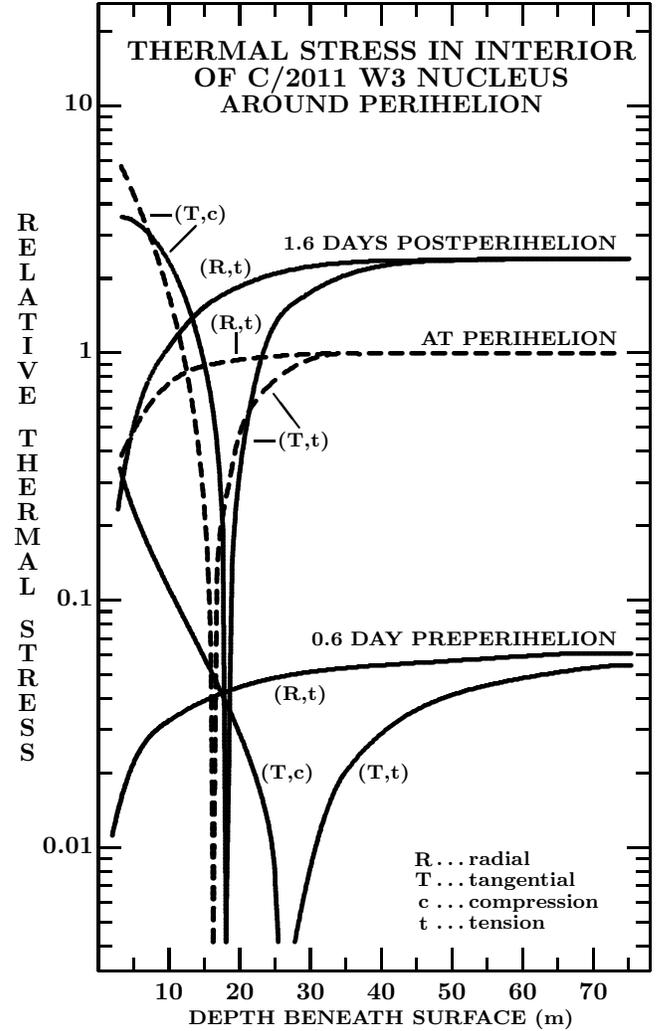}}}
 \vspace*{-9.4cm}
 \caption{Relative thermal stress in the interior of comet C/2011 W3
as a function of depth beneath the surface.  Derived from the temperature
variations 0.6 day before perihelion for the nucleus 400 meters in diameter
(solid), at perihelion for 280 meters in diameter (broken), and 1.6 days
after perihelion for 150 meters in diameter (solid), the stress curves are
normalized to thermal stress at the nucleus' center at perihelion.  The
segments of the curves are marked to distinguish between the radial and
tangential stress components and whether the stress was a compression or
tension.}
\vspace*{0.3cm}
\end{figure}

To allow for a gradual erosion of the surface layer in the proximity of the
Sun, the internal-temperature distribution presented in Figs.\ 12 and 13 has
been used to compute the thermal-stress profiles for the nucleus 400 meters
in diameter at 0.6 day before perihelion, 280 meters in diameter at perihelion,
and 150 meters in diameter 1.6 days after perihelion.  The results are
presented in two steps.  In the first step, the relative variations in the
radial and tangential components of thermal stress, normalized to its magnitude
at the center of the nucleus at perihelion, are displayed against depth beneath
the surface in Fig.\ 14.  The major properties of the stress curves are as
follows:

(1) Just as with temperature, thermal stress continues to climb even as the
comet recedes from the Sun; measured by its magnitude at the nucleus' center,
the thermal stress at perihelion is about 16 times higher than 0.6 day before
perihelion, but 2.4 times {\it lower\/} than 1.6 days after perihelion.

(2) At a given time, the radial component of thermal stress is always tension,
increasing from zero at the surface, as already noted; the tangential component
reaches its maximum at the surface as compression, but changes to tension a few
tens of meters beneath the surface, increasing toward the center.

(3) Both components reach maximum tension at the center of the nucleus, where
they have the same magnitude, as implied by Eq.\ (33).

(4) Cracks, whose formation is triggered by increasing thermal stress at large
depths due to the thermal wave's gradual penetration into the interior, are
bound to open preexisting subsurface reservoirs of highly volatile ices (such
as amorphous water ice), initiating their explosion when heated up to the
activation temperature.

(5) The steady increase in thermal stress at the center of the nucleus over long
periods of time after perihelion is likely to be a driving force behind the
episodes of seemingly spontaneous cascading fragmentation at large heliocentric
distances, which account for the SOHO sungrazers' observed scatter in
elements other than orbital period.

Measured in absolute units, the magnitude of thermal stress depends on the
stress parameter $\chi$, given by Eq.\ (31) and consisting of three physical
quantities:\ Young's modulus $E_{\rm Y}$, the coefficient of linear thermal
expansion $\alpha_{\rm L}$, and Poisson's ratio $\nu$.  Since their values
for cometary material are unknown, we make in the following an effort to
constrain each of them to our best ability by using meteoritic, lunar, and
terrestrial analogues.  Given our conclusion in Sec.\ 11 on the nature of
dust sublimating from comet C/2011 W3 near the Sun, a plausible candidate
among the terrestrial samples appears to be an olivine-based silicate material.

{\small \bf Young's modulus.}  There is a large number of papers with
information on Young's modulus, but they mostly refer to samples of zero or
near-zero porosity.  For the Moon, Pritchard \& Stevenson (2000) estimated
that the effective Young's modulus in the outer layers of the lunar lithosphere
is between 5 and 10 GPa.  For ordinary chondrites the topic was recently
updated by Kimberley \& Ramesh (2011).  They also provided the results of their
experiments aimed at determining the elastic and mechanical properties of a
particular L5 ordinary chondrite found in Antarctica, showed that the outcome
depends on whether the data were obtained under quasistatic or dynamic
conditions, compared their results to the findings from the numerous studies
of the atmospheric breakup of meteors, and emphasized that the propagation of
cracks is gradual because the crack speeds are much lower than the limit
defined by elastodynamic fracture theory.  Although Kimberley \& Ramesh did not
consider thermal stress, their conclusions are relevant, because they found
that the value of the Young's modulus for a particular sample with a 13 percent
porosity depended on a strain rate and range of strain; at slow rates and
strains of up to 0.01, \mbox{$E_{\rm Y} = 3.2$} GPa, at strains ranging from
0.01 to 0.03 and much higher strain rates (appropriate in hypervelocity impact
phenomena), \mbox{$E_{\rm Y} = 8.5$} GPa; this behavior has also been observed
in some brittle terrestrial materials, e.g., ceramics (Wang \& Ramesh 2004).
Pertinent to our problem is the lower value of $E_{\rm Y}$.

From analysis of limited samples of meteorites, rarely including carbonaceous
chondrites (e.g., Flynn, Kl\"{o}ck, \& Krompholz 1999), most values found for
Young's modulus were one order of magnitude higher than the quoted result by
Kimberley \& Ramesh, but their porosities were mostly lower than 13 percent.
A key work was published by Yomogida \& Matsui (1983) on the mechanical,
thermal, and elastic properties (measured quasistatically) of 20 ordinary
chondrites of H and L types.  For some of these properties the authors
compared the meteorites with lunar samples studied by others.  One of Yomogida
\& Matsui's findings was the strong dependence of the elastic properties on
porosity.  They showed that even a low porosity reduces the values of the
elastic parameters, including Young's modulus, much more for the meteorites
and the lunar samples than for terrestrial rocks.  While there are fairly large
differences between the individual samples, the steep drop in Young's modulus
$E_{\rm Y}(\psi)$ with porosity $\psi$ is for 19 out of the 20 chondrites
fitted by
\begin{equation}
E_{\rm Y}(\psi) = {\rm const} \cdot \exp \! \left[ -U \! \left( \!
\frac{\psi}{1 \! - \! \psi} \! \right)^{\!\frac{1}{4}} \right] \!
\;\; ({\rm for} \;\,\psi \geq \psi_0),
\end{equation}
where \mbox{$U = 13.8 \pm 1.2$} and \mbox{${\rm const} = 85_{-42}^{+82} \times
10^{12}$\,Pa} \mbox{$(\psi_0 = 0.05)$} for 13 samples (high $E_{\rm Y}$ group)
and \mbox{$31_{-10}^{+15} \times 10^{12}$\,Pa} \mbox{$(\psi_0 = 0.02)$} for 6
samples (low $E_{\rm Y}$ group).  This expression provides a substantially
better fit than Warren's (1969) formula used by Yomogida \& Matsui (1983) in
their Fig.\ 8 to match a normalized elastic modulus of both the meteorites
and the lunar samples.  Although by 2001 the porosity (with values of up to
30 percent) had already been known for more than 450 stony meteorites (Britt
\& Consolmagno 2003), information on their elastic moduli seems to be lagging
behind.

A variety of studies of terrestrial samples provides constraints on Young's
modulus of olivine.   However, many measurements refer to temperatures in
excess of 1000\,K.  For example, Tait (1992) used a value of 197 GPa at
1470\,K, while Hiraga, Anderson, \& Kohlstedt (1993) gave values of 159, 155,
and 154 GPa at, respectively, 1373\,K, 1473\,K, and 1523\,K.  Chung (1970)
measured the bulk modulus and Poisson's ratio of olivine as a function of
the Mg/Fe ratio at room temperature (296\,K) and zero pressure.  In terms of
Young's modulus, his results show a systematic trend from a maximum of 198 GPa
for forsterite to a minimum of 141 GPa for fayalite.  An extensive study of
the dependence on the temperature of the elastic properties of a single
crystal of forsterite was published by Suzuki, Anderson, \& Sumino (1983), who
found that the elastic moduli increase with decreasing temperature at a rate
of about 1.3 percent per 100\,K between 0\,K and 1200\,K.  From their values
of the bulk modulus and Poisson's ratio, Young's modulus of forsterite comes
out to be 175 GPa at 1500\,K and 207 GPa at 100\,K, an increase of 18 percent.
All these numbers are fairly consistent with the highest values of Young's
modulus near 140 GPa measured by Yomogida \& Matsui (1983) for the meteorites
of low porosity.  It therefore appears that the temperature effect is
insignificant relative to the porosity effect.  A detailed investigation of
Young's modulus variations with porosity for a variety of terrestrial materials
was published by Phani \& Sanyal (2005), who determined that for colloidal gel
derived silica the relative modulus varied as
\begin{equation}
\frac{E_{\rm Y}(\psi)}{(E_{\rm Y})_0} = \frac{(1 -\psi)^2}{1 + {\textstyle
\frac{3}{2}} \psi}
\end{equation}
for porosities smaller than 0.85.  The slope of the curve is somewhat
similar to that shown by Yomogida \& Matsui (1983) for a sintered material
and tuff in their Fig.\ 8; it is considerably less steep than the slope for
the meteorites and the lunar samples.

{\small \bf Coefficient of linear thermal expansion.}  This coefficient is on
a recently compiled ``wish'' list of 12 physical quantities of meteorites,
the data on which are desired most by the scientific community (Consolmagno,
Britt, \& Opeil 2010), a fact that cogently illustrates the urgent need for
this practically nonexistent information.

It has been known for nearly a century that the thermal expansion coefficient
is generally temperature dependent and that its rate of variation in crystallic
solids is, at least at relatively low temperatures, approximately proportional
to that of the specific heat capacity at constant volume (e.g., Austin 1952).
Given that our primary interest is the temperature range from $\sim$70\,K
to $\sim$300\,K, we have focused on a search for relevant sets of data on
olivine-based silicates that should be as consistent with this temperature
interval as possible.  Once available, the data on the coefficient of linear
thermal expansion have been fitted to satisfy a properly scaled formula for
the Einstein crystal model (e.g., Rogers 2005),
\begin{equation}
\alpha_{\rm L}(T) = \lambda \left( \! \frac{\Omega}{T} \! \right)^{\!2} \!
\left[ \sinh \! \left( \! {\frac{\Omega}{T}} \! \right) \! \right]^{-2},
\end{equation}
where $\lambda$ and $\Omega$ are constants to be determined from the data. (The
known fact that this equation does not {\it predict\/} correct values of heat
capacity at low temperatures is irrelevant to our using it to {\it fit\/} a set
of $\alpha_{\rm L}$ values.)

Samples of a hot-pressed olivine aggregate were used by Aizawa, Ito, \&
Tatsumi (2001) to measure its volumetric coefficient of thermal expansion,
$\alpha_{\rm V}$, but the temperature range was from 300\,K up, which makes
the data, requiring much extrapolation, only marginally useful.  After
conversion to \mbox{$\alpha_{\rm L} = {\textstyle \frac{1}{3}} \alpha_{\rm
V}$}, we found that fitting the data between 300\,K and 600\,K by Eq.\ (43)
requires \mbox{$\lambda = 12.0 \times 10^{-6}$\,K$^{-1}$} and \mbox{$\Omega
= 304$\,K}.  With these constants $\alpha_{\rm L}$ equals 1.02, 5.85, and
\mbox{$8.62 \times 10^{-6}$\,K$^{-1}$} at the temperatures of, respectively,
100\,K, 200\,K, and 300\,K.

Another set of volumetric thermal expansion coefficients is found in the paper
by Suzuki, Anderson, \& Sumino (1983), already mentioned in the subsection on
Young's modulus.  Their data refer to the crystal of forsterite and cover the
desired range of temperatures.  A fit by Eq.\ (43) requires that \mbox{$\lambda
= 9.14 \times 10^{-6}$\,K$^{-1}$} and \mbox{$\Omega = 251.4$\,K}.  The fit
gives $\alpha_{\rm L}$ equal to 1.53, 5.54, and \mbox{$7.27 \times
10^{-6}$\,K$^{-1}$} at 100\,K, 200\,K, and 300\,K, respectively.  The fit to
the measured values is perfect for the first two temperatures, but too low by
12 percent for the third.

Unfortunately, most information available on the thermal expansion coefficient
refers to geology of the earth's crust, volcanology, and related scientific
fields (e.g., Afonso, Ranalli, \& Fern\`andez 2005).  As a result, one seldom
finds data pertaining to low temperatures and low pressures.  In addition, 
the effect of porosity on thermal expansion is seldom addressed, and when it
is, the results and implications depend on the type of material and physical
conditions (e.g., Faivre, Bellet, \& Dolino 2000; Moretti et al.\ 2005; Hunter
\& Brownell 2006; Ghabezloo \& Sulem 2009; Ghabezloo 2010), although generally
there appears to be a tendency for the thermal expansion coefficient to
increase with increasing porosity.

{\small \bf Poisson's ratio.} Besides the elastic moduli, Yomogida \& Matsui
(1983) also measured Poisson's ratio $\nu$ for the already mentioned 20
meteorites.  With the exception of a single anomalous sample (the same one
that did not fit the dependence of Young's modulus on porosity), all 19
remaining had their values of $\nu$ between 0.12 and 0.29.  In this data set
there is a slight tendency for $\nu$ to decrease with increasing porosity,
but the scatter is too large for a functional fit; on the average,
\mbox{$\langle \nu \rangle = 0.21 \pm 0.05$}.  

For olivine, Chung (1970) found that Poisson's ratio was increasing
systematically with the Fe/Mg ratio from 0.242 for forsterite to 0.308 for
fayalite.  Just as with the thermal expansion coefficient, the dependence of
Poisson's ratio on porosity for terrestrial materials appears to vary
from case to case.  Gel derived silica, for example, has Poisson's ratio
nearly constant between 0.15 and 0.19 at porosities smaller than 0.6,
but it then rapidly climbs to 0.26 at porosity 0.82 (Ashkin, Haber, \&
Wachtman 1990; Phani \& Sanyal 2005).  On the other hand, Poisson's ratio
of sintered iron decreases from 0.30 at zero porosity to 0.20 at porosity
0.2 (Kov\'a\v{c}ik 2005).

The temperature dependence of Poisson's ratio for forsterite is shown by
Suzuki, Anderson, \& Sumino's (1983) data.  The rate amounts to $<$0.001 per
100\,K and can be neglected.  Compared to Young's modulus and the thermal
expansion coefficient, Poisson's ratio has a relatively minor effect in
computing thermal stress.

{\small \bf Adopting stress parameter {\boldmath $\chi$} for C/2011 W3.}
Whereas the computed relative variations in thermal stress in the interior of
comet C/2011 W3 in Fig.\ 14 suggest an enormous growth near and after
perihelion, it remains unclear whether the effect is sufficient to crumble
and collapse the nucleus.  Investigations of this kind can only be conducted
in a framework of a particular comet model, with realistic estimates for the
quantities that enter the stress parameter $\chi$.  The extensive discussion
of Young's modulus, the thermal expansion coefficient, and Poisson's ratio
was intended to facilitate this effort.

To anchor our efforts in the framework of a numerical-simulation construct
for comet formation, evolution, and morphological makeup, we focus on Lasue
et al.'s (2009, 2011) work on aggregation and collisional interaction of
cometesimals in the primordial solar nebula and the internal structure of
cometary nuclei.  Their developed model is based on a homogeneity exponent
determining an aggregation regime, accounts for disruptive, sticking,
compaction, and sintering processes, allows comparison with observational
constraints and future in situ observations, and provides predictions for
the cohesive strength and the radial porosity profile of the nucleus'
interior.  In our choice of parameters we concur with Lasue et al., who
clearly prefer the layered structure of the nucleus, which is determined
by the range of 0.4 to 0.6 for their homogeneity exponent and which offers
the characteristic values of the porosity and cohesive strength that we use
in this study.

Keeping in mind the complete disintegration of C/2011 W3 after perihelion, our
primary interest is the temporal variation in thermal stress {\it deep inside
the nucleus\/}, which we refer to as the {\it central\/} thermal stress.  By
this term we understand stress in the volume of the nucleus that does not
include the surface layer, where the thermal stress profile is very different
from that at greater depths.  In Fig.\ 14 the boundary between the surface
layer and the deep interior approximately coincides with the apparent minimum
on the stress curve, associated with the transition of the tangential component
$\sigma_\perp$ from compression to tension.

The extensive areas of the surface and the adjacent, relatively shallow
subsurface layer of the nucleus that are presumed to be devoid of ices, must
in the proximity of perihelion suffer from even higher levels of thermal
stress than the rest of the nucleus (Fig.\ 14).  They are primary candidates
for thermal-stress damage, being probably riddled with crisscross cracks and
the debris removed soon after perihelion.  These areas may in fact provide for
the comet's earliest postperihelion activity.  We will not deal with them in
this paper.

In Fig.\ 14, a rather interesting property of the thermal stress distribution
in deeper layers, say, from a depth of $\sim$30 meters on, is that the
magnitude of the total effect,
\begin{equation}
\sigma = \sqrt{ \sigma_\parallel^2 + \sigma_\perp^2 },
\end{equation}
is almost independent of depth, especially at perihelion and afterwards.
Because $\sigma$ fairly rapidly converges to \mbox{$\sigma_\parallel \sqrt{2}
= \sigma_\perp \sqrt{2}$} at the nucleus' center, it is both convenient and
appropriate to deal with this limit as a measure of thermal stress in our
consideration of its temporal variations in absolute units --- hence the
reason for the term {\it central\/}.

The discussion of the quantities that make up the stress parameter $\chi$
suggests that in an effort to find their most appropriate values one will
encounter considerable difficulties and uncertainties.  The two most severe
problems are the strong dependence of Young's modulus on porosity and the
thermal expansion coefficient's variation with temperature.  In the former
case, there is the need to extrapolate the data way outside the covered range
of porosity.  In the latter instance, it is troubling to find a parameter,
which the thermal stress theory handles as a constant, to be so strongly
temperature dependent.  Only the selection of a value for Poisson's ratio
appears to be somewhat less controversial, thanks to this quantity's relative
invariability.  Yet, for all three quantities it is necessary to resort to
other classes of objects and/or substances to approximate cometary material.
Also, we have no choice but to rely on a comet model to estimate the porosity
in the interior.  With these caveats in mind we proceed to the next step.

We note that Lasue et al.'s (2011) aggregate model predicts a porosity close
to 0.65 at large depths and near the center of a cometary aggregate regardless
of whether the size distribution function of the cometesimals is Gaussian or a
power law.  Equation (41) then yields \mbox{$E_{\rm Y} = 8.6$} MPa from the
high $E_{\rm Y}$ group of meteorites and \mbox{$E_{\rm Y} = 3.1$} MPa from the
low $E_{\rm Y}$ group.  For the coefficient of linear thermal expansion
$\alpha_{\rm L}$ we have to adopt a mean value from a range of less than 1 to
at least 6 or \mbox{$7 \times 10^{-6}$\,K$^{-1}$}, based on Suzuki, Anderson,
\& Sumino's (1983) data for a single crystal of forsterite.  Because of the
possibility that an explosion from a reservoir of amorphous water ice at depths
in excess of 40 meters was a trigger of, or a contributor to triggering, the
terminal outburst (Fig.\ 12), we have chosen \mbox{$\alpha_{\rm L} = 3 \times
10^{-6}$\,K$^{-1}$}, which is near the mean and is also attained at 130\,K, the
temperature of activation of the ice's exothermic process of crystallization
(e.g., Schmitt et al.\ 1989).  Because of the dearth of relevant information,
we have not included any porosity correction.  For Poisson's ratio we have
adopted \mbox{$\nu = 0.1$}, which is based primarily on the sample of
meteorites and allows for an assumed modest decrease with porosity.  Its value
has the role of a minor correction to the stress parameter, which comes out to
be $\chi = 29$ Pa and 10 Pa with the two values of $E_{\rm Y}$.

To address the issue of whether the magnitude of thermal stress determined in
this way for C/2011 W3 could lead to the crumbling and collapse of this comet's
nucleus, we have compared our results with the expected cohesive strength of
a cometary aggregate provided by Lasue et al.'s model.  Specifically, we have
taken approximate upper and lower limits near the homogeneity exponent of 0.5,
as presented in Fig.\ 7b of Lasue et al.\ (2009), because they alluded to this
point only in general terms in their second paper (Lasue et al.\ 2011); the
understanding the reader gets is that the introduction of a size distribution
function of cometesimals had no significant effect on the cohesive strength.

The resulting thermal stress curves, referring to the two adopted values of
Young's modulus, are plotted against time from perihelion in Fig.\ 15.  In
broad terms, we conclude that thermal stress developing in the interior of
the nucleus is:\ (1) insufficient to disrupt the object before perihelion; (2)
comparable to the cohesive strength in the immediate proximity (within hours)
of perihelion, thus weakening the nucleus' structure; and (3) greater than
the strength of the comet after perihelion to the extent that it can no longer
hold together.

Because of the uncertainties involved, one needs to exercise a caution and
emphasize that this conclusion is model and parameter dependent.  Nevertheless,
in the case of the thermal expansion coefficient, its chosen constant value
has a tendency to suppress the degree of asymmetry of thermal stress relative
to perihelion, that is, it overestimates thermal stress before perihelion and
underestimates it after perihelion.  As a matter of fact, this relation can be
quantified:\ from Eq.\ (43) it follows that a change $\Delta T$ in the
temperature leads to the following change $\Delta \alpha_{\rm L}$ in the
coefficient of linear thermal expansion:
\begin{equation}
\frac{\Delta \alpha_{\rm L}}{\alpha_{\rm L}} = 2 \left[ \frac{\Omega}{T} \,
 \coth \! \left( \! \frac{\Omega}{T} \! \right) - 1 \right] \!
 \frac{\Delta T}{T}.
\end{equation}
This formula indicates that, for example, a modest temperature increase of
5\,K entails an increase in the relative rate of the $\alpha_{\rm L}$
coefficient --- and therefore in thermal stress --- of 2.4 percent at 200\,K,
5.3 percent at 150\,K, 15.5 percent at 100\,K, and an enormous 37 percent at
70\,K.  This effect comes on top of the thermal stress increase that we have
discussed below Eq.\ (35), and enhances the role of thermal stress as the
driving force behind the events of cascading fragmentation at large
heliocentric distances, as mentioned in point (5) between Eqs.\ (40) and
(41); it also aids the role of thermal stress in unlocking potential
reservoirs of highly volatile ices in the nucleus' interior.

In summary, our suspicion that the delayed response of the nucleus of C/2011 W3
to the extremely high temperatures in the proximity of perihelion was due to
heating its interior and to the resulting effect of thermal stress finds
support in this comprehensive analysis, which is limited only by incomplete
information on cometary material --- the problem that no hypothesis can avoid.
If our scenario is valid at least with an order-of-magnitude accuracy, it
strengthens the notion that C/2011 W3 was able to withstand thermal stresses
on the order of several kPa and therefore possessed far more cohesion than
necessary to avoid collapse and disintegration right at perihelion, in the
inner solar corona.
 
\section{Standing of C/2011 W3 in the Kreutz System and Its Possible Past
Evolutionary Path}

The place of the now defunct comet C/2011 W3 in the hierarchy of the Kreutz
system of sungrazers cannot be pinpointed with certainty.  However, within
a given model of the system, the comet's evolutionary path can be traced
in some detail.  The exercise involves modeling and interpretation of
long-term changes in (a) the orbital period and (b) the angular elements
and perihelion distance.  As amply explained in Sec.\ 1, the first issue is
addressed in terms of the tidal-assisted fragmentation of the early parent
objects of C/2011 W3 in the close proximity of perihelion, while the second
issue involves a sequence of fragmentation episodes of its more recent
precursors far from the Sun.

\begin{figure*}[ht]
 \vspace*{-5.65cm}
 \hspace*{-1.3cm}
 \centerline{
 \scalebox{0.9}{
 \includegraphics{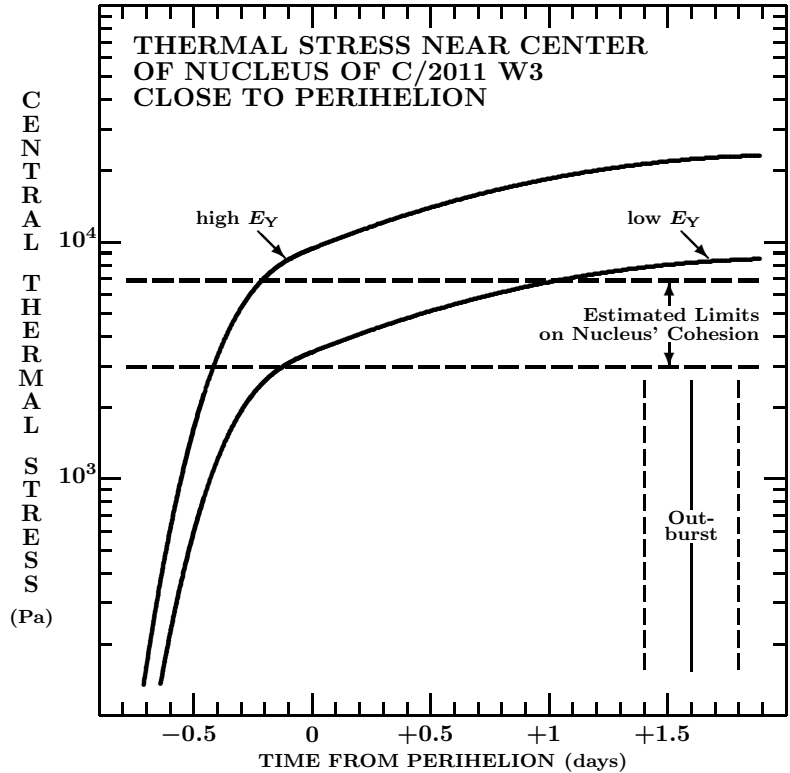}}}
 \vspace*{-10.65cm}
 \caption{Computed thermal stress near the center of the nucleus of C/2011 W3.
 The solid curves show the stress growth through perihelion for the two
adopted values of Young's modulus, the horizontal dashed lines mark the
estimated limits on the cohesive strength of the nucleus (from Lasue et al.\
2009, 2011).  The stress curves represent a conservative estimate; because
the thermal expansion coefficient increases with the temperature, the actual
postperihelion growth of thermal stress in the comet's interior is even
steeper.  The time of the terminal outburst of C/2011 W3, \mbox{$1.6 \pm 0.2$}
days after perihelion, is shown by the solid vertical line, its uncertainty
by the parallel broken lines.}
\vspace*{0.4cm} 
\end{figure*}

As an example, we elaborate on the evolutionary path of C/2011 W3 in the
context of the hierarchical scenario B of the Kreutz system, introduced in
our recent paper (Sekanina \&  Chodas 2007).  This scenario is based on
the postulated identity of the spectacular sungrazer X/1106 C1 with a comet
of A.D.\ 467.  The true orbital period of this progenitor (or superfragment,
in the terminology used in Sekanina \& Chodas 2004) was obviously 639
years.  The integration of the orbit of C/2011 W3 back in time indicates
(Table 4) that the previous return to perihelion occurred in January 1329
(with a formal uncertainty of $\pm$2 years, which for this exercise is not
critical).  It should be emphasized that the 1329 orbital set is valid only
in the unlikely absence of any fragmentation events involving the comet's
precursors during the entire revolution about the Sun.  Only in this
hypothetical case does the true orbital period come out to be almost exactly
683 years.  In order for a fragment --- an early parent of C/2011 W3 --- of
comet 467 to arrive at perihelion in 1329, its true orbital period must have
been 862 years, or 223 years longer than that of the principal fragment, comet
X/1106 C1.  This {\it increase\/} in the orbital period requires that the
parent fragment of C/2011 W3 split off from the main body of comet 467 near
perihelion (say, at a heliocentric distance of 0.01 AU) with a separation
velocity of $\sim$2.2 m s$^{-1}$ essentially in the direction of the orbital
motion (Table 8 of Sekanina 2002a).  When this parent fragment arrived at
perihelion in 1329, its orbital period was (with some uncertainty due to the
indirect planetary perturbations) still 862 years.  Since the true orbital
period of C/2011 W3 is 683 years, the necessary {\it decrease\/} of 179 years
in the orbital period requires that a new fragment broke off from the parent
fragment in 1329, again in the immediate proximity of perihelion, with a
separation velocity of $\sim$3.6 m s$^{-1}$ essentially in the direction
opposite the direction of orbital motion.  Separation rates of fragments
corresponding to relative velocities of up to 5 or 6 m s$^{-1}$ are known to
be necessary for explaining the tidal-assisted splitting of sungrazing comets
(Sekanina \& Chodas 2007), so the required velocities of less than 4 m s$^{-1}$
provide a plausible evolutionary model for comet C/2011 W3.  Similar numbers
would result if the adopted scenario B should be replaced with another one.

\begin{figure*}[ht]
 \vspace*{-4.35cm}
 \hspace*{-0.5cm}
 \centerline{
 \scalebox{0.98}{
 \includegraphics{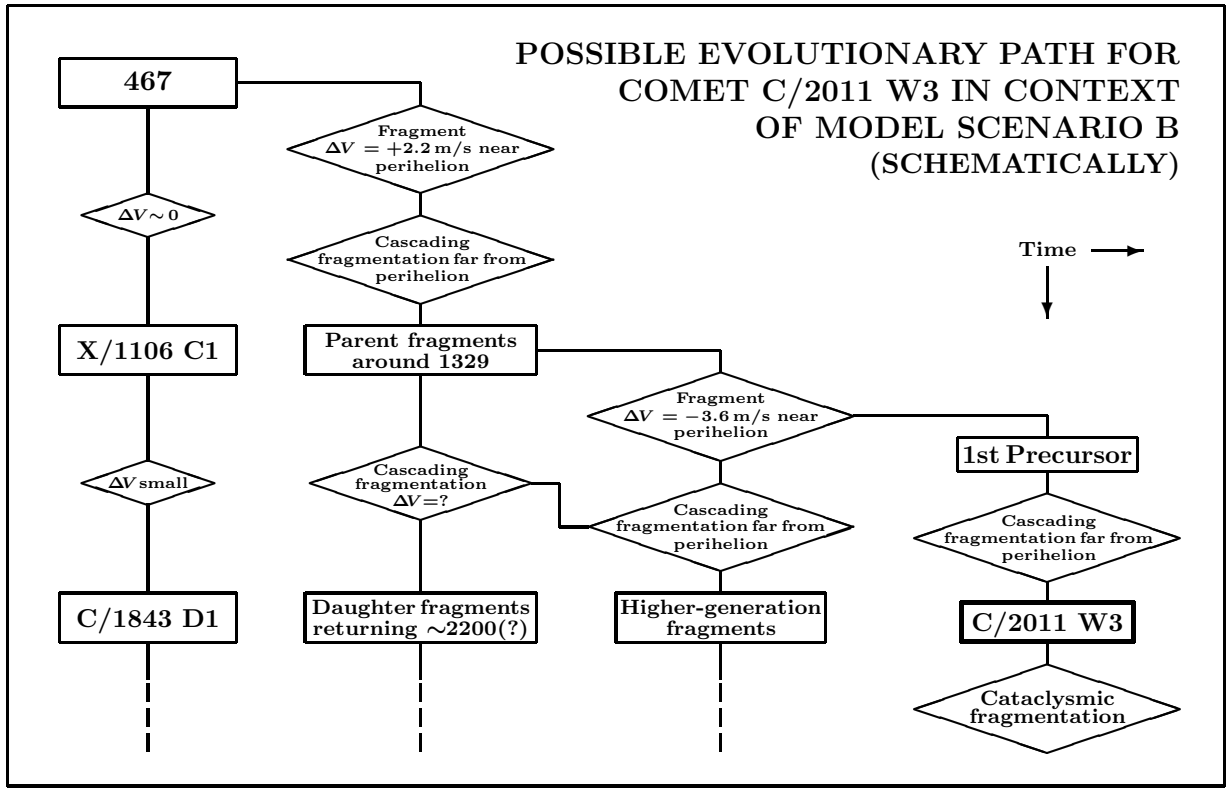}}}
 \vspace*{-13.8cm}
 \caption{Possible evolutionary path for comet C/2011 W3 in a broader
context of the model scenario B (Sekanina \& Chodas 2007), which assumes that
comet C/1843 D1, rather than C/1882 R1, is the most massive fragment of X/1106
C1.  The parallel fragmentation branch involving C/1882 R1 is not shown in this
schematic representation.  The diamond-shaped boxes describe events of the
tidal-assisted and cascading nontidal fragmentation processes, the rectangular
boxes are their products.  The separation velocities involved are given as
$\Delta V$; the sign refers to the direction along the orbital-velocity
vector:\ positive means ahead, negative behind.  The fragmentation hierarchy
is presented schematically, with only some paths among the great multitude of
possible chain events being depicted.}
\vspace*{0.5cm}
\end{figure*}

Because of potential differences in the angular elements and perihelion
distance between the orbits of C/2011 W3 and the new fragment referred to
above, it is highly unlikely that the two objects are identical.  Much
more probable is that the new fragment was a precursor to C/2011 W3, which
between 1329 and 2011 underwent additional, nontidal fragmentation at large
heliocentric distances, and that during these episodes one of the fragments
acquired the necessary orbital orientation and perihelion distance to become
C/2011 W3.  If the first precursor's angular elements in the 14th century
were in the range of most SOHO sungrazing comets, that is, close to those of
C/1963 R1, the differences relative to C/2011 W3 may have been substantial,
about 35$^\circ$ to 40$^\circ$ in the argument of perihelion and the longitude
of the ascending node, some 10$^\circ$ in the inclination, and up to 10 percent
of the Sun's radius in the perihelion distance.  To bridge the gaps of this
magnitude requires several nontidal fragmentation events in the general
proximity of aphelion (at heliocentric distances of, typically, 100 AU or so)
at an assumed average separation velocity of 2--3 m s$^{-1}$, a requirement
by no means excessive.  A simulation model for a sequence of nontidal
fragmentation events scaled for the Kreutz system from the case of D/1993 F2
(Shoemaker-Levy 9) shows that the number of these episodes can easily exceed
a dozen during one revolution about the Sun (Sekanina 2002a, 2002b).  Their
high frequency is also obvious from the arrival rate of the SOHO minicomets,
the high-generation products of the cascading fragmentation process, the age
of each of them being manifestly shorter than one revolution about the Sun
(e.g., Sekanina \& Chodas 2004).

In the broader context of the fragmentation hierarchy of the branch of the
Kreutz system, which in our Scenario B also contains the sungrazer C/1843 D1,
the described evolutionary path of C/2011 W3 is depicted in Fig.\ 16.  In this
more general scheme, the parent fragment may have reached perihelion years or
even decades before or after 1329, because fragmentation at large heliocentric
distances entails modest changes in the orbital period of up to a few years
per event (Table 8 of Sekanina 2002a).

Regardless of the details of the evolution of C/2011 W3, its origin appears to
be linked to the expected new, 21st century cluster of bright sungrazers.  What
remains unclear is its subgroup membership.  While already Kreutz (1901) was
thinking in terms of two subgroups when investigating historical sungrazers,
this division for the bright members of the Kreutz system became a hit after
Marsden's (1967) publication of his classical paper, which listed C/1843 D1,
C/1880 C1, and C/1963 R1 as the definite members of subgroup I, while
C/1882 R1, C/1945 X1, and C/1965 S1 belonged to subgroup II.  In addition,
Marsden tabulated four and two additional comets, respectively, as possible
members of the two categories.  A dent in this classification scheme was
made by C/1970 K1, whose orbit did not fit either subgroup.  Because its orbit
was closer to subgroup II, Marsden (1989) classified it subsequently as
subgroup IIa.  Now comes C/2011 W3, whose perihelion distance fits subgroup I,
but the angular elements are incompatible with any of the subgroups I, II, or
IIa.  This is worrisome, because Marsden (1967) regarded the longitude of the
nodal line, one of the orbital angles, as the prime classifier for the
subgroups.  If one considers, by extension, C/2011 W3 to be a representative
of a new subgroup III, one ends up with a total of eight bright sungrazers
distributed into four categories, or two members per subgroup on the average,
not to mention that the orbit of C/1945 X1, one of the eight, is not really
all that well fixed (see Marsden 1989 and our comment about Cunningham's
computations in Sec.\ 7).

\begin{figure*}[ht]
 \vspace*{-4.25cm}
 \hspace*{0.5cm}
 \centerline{
 \scalebox{1}{ 
 \includegraphics{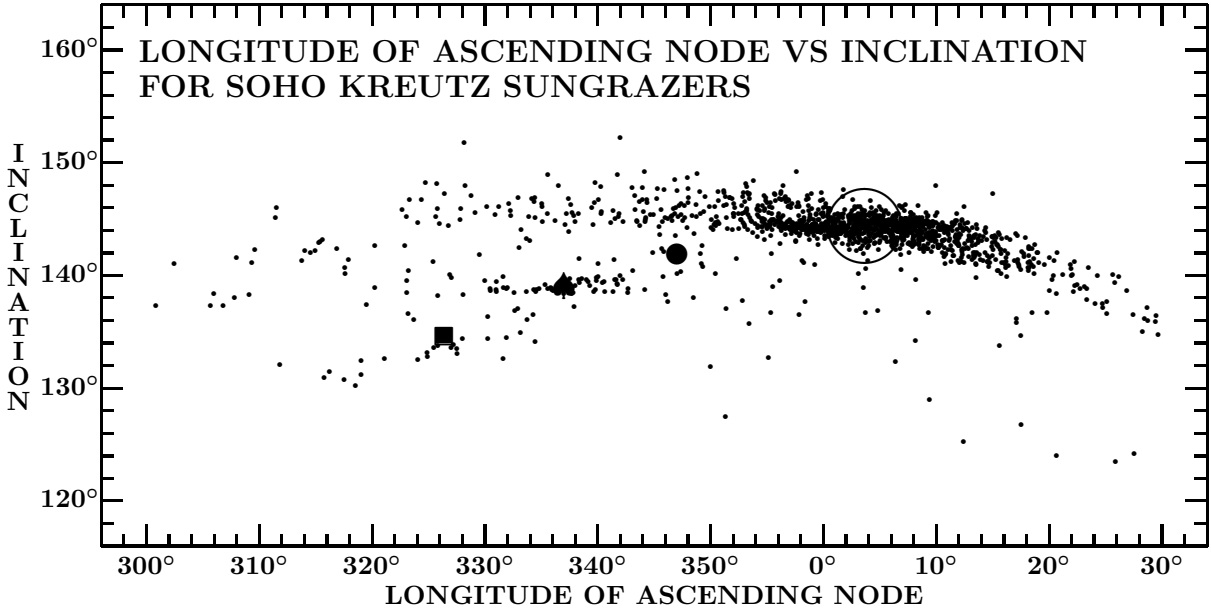}}}
 \vspace*{-17.4cm} 
 \caption{Comets C/2011 W3, C/1970 K1, C/1965 S1, and C/1843 D1 in a
plot of the longitude of the ascending node against the orbit inclination
for 1565 SOHO Kreutz sungrazers from the period January 1996 to June 2010.
C/2011 W3 is identified by the square, C/1970 K1 by the spadesuit mark,
C/1965 S1 by the filled circle, while the location of C/1843 D1 coincides
with the dense concentration of the SOHO sungrazers and is in the center of
the open oversized circle.  Comet C/2011 W3 appears to be orbitally related
to a cluster of up to nearly 30 SOHO sungrazers with the ascending nodes
between 310$^\circ$ and 335$^\circ$ and with the inclinations between
130$^\circ$ and 137$^\circ$.}
\vspace*{0.75cm}
\end{figure*}

An overwhelming majority of SOHO, SMM, and SOLWIND sungrazers belongs to
subgroup I.  This was shown to be the case by Marsden (1989) for the SMM and
SOLWIND comets and is illustrated for 1565 SOHO sungrazers in a plot of the
longitude of the ascending node against the orbit inclination in Fig.\ 17.
However, nearly 100 SOHO sungrazers, or about 6 percent of the total, make
up a second branch, which in the range of ascending node longitudes from
$\sim$320$^\circ$ to at least $\sim$345$^\circ$ lies about 6$^\circ$ below
an extended arm of the subgroup I set.  And some 4$^\circ$ to 5$^\circ$ below
this second branch there is yet another cluster of nearly 30 SOHO sungrazers
with their nodal longitudes between 310$^\circ$ and 335$^\circ$.

For the sake of comparison, we have also included four bright sungrazers in
Fig.\ 17.  Comet C/1843 D1, of subgroup I, is right in the dense core of the
SOHO sungrazers, while C/1965 S1 (of subgroup II), C/1970 K1 (of subgroup
IIa), and C/2011 W3 are all located off the main branch.  A surprising finding
is that the second, much less populated but still well defined branch of the
SOHO sungrazers belongs to subgroup IIa (not II, as has generally been
assumed).  There is practically no concentration at all of the SOHO objects
near the location of C/1965 S1.  On the other hand, the agreement between the
positions of C/2011 W3 and a third, very sparsely populated branch is rather
obvious.  Thus, when it comes to the SOHO sungrazers, the orientation of the
orbit of C/2011 W3 in space is certainly not unique.  And while the three
branches essentially merge into a common relation in the plot of the longitude
of the ascending node against the argument of perihelion in Fig.\ 18, C/2011
W3 is again surrounded by a number of SOHO sungrazers.

On the whole, the classification of the bright sungrazers into the subgroups
is somewhat questionable.  A suggestion for its eventual abandonment is
supported by our finding (Sekanina \& Chodas 2007) that even following a
single fragmentation event at large heliocentric distance, two fragments of
the same parent can end up in orbits formally belonging to different subgroups.
In fact, the progenitor's splitting far from the Sun into two superfragments
with widely different orbits is a prerequisite behind the idea of introducing
the concept of subgroups in the first place (Sekanina \& Chodas 2004).

The point of contention with C/2011 W3 is thus not whether it could derive
from a precursor belonging to another subgroup, but, rather, that no
plausible candidate for a sungrazing comet was ever recorded in the first
half of the 14th century.  Hasegawa \& Nakano (2001) considered only one
{\it possible\/} sungrazer, X/1381 V1, during the entire 14th century.  All
three comets between 1282 and 1368 in England's (2002) list have a relatively
low sungrazer ranking, even though the author does not entirely exclude the
possibility for each of them, including one seen in Europe in October 1314,
to be a member of the Kreutz system.

On the other hand, only if the cumulative effect in the orbital period from a
number of fragmentation episodes at large heliocentric distances could, in an
extreme and quite unlikely run of events, amount to as much as $\sim$50 years,
could X/1381 V1 have been a parent to C/2011 W3 in the sense that one of the
products of its tidal-assisted breakup in 1381 would have become the first
precursor to C/2011 W3.

Of course, a large number of comets recorded in the late 13th century and
during the 14th century were cataloged by Ho (1962), by Hasegawa (1980), and
by Kronk (1999).  However, most of them fail to fit a Kreutz sungrazer because
of the time of the year, the location in the sky, the apparent motion, or the
appearance, while for the rest of them the reported information is too vague
for the identification.  The future modeling of the Kreutz-system evolution
should proceed with these circumstances taken into account.

\begin{figure*}[ht]
 \vspace*{-5.3cm} 
 \hspace*{0.5cm}
 \centerline{
 \scalebox{1}{ 
 \includegraphics{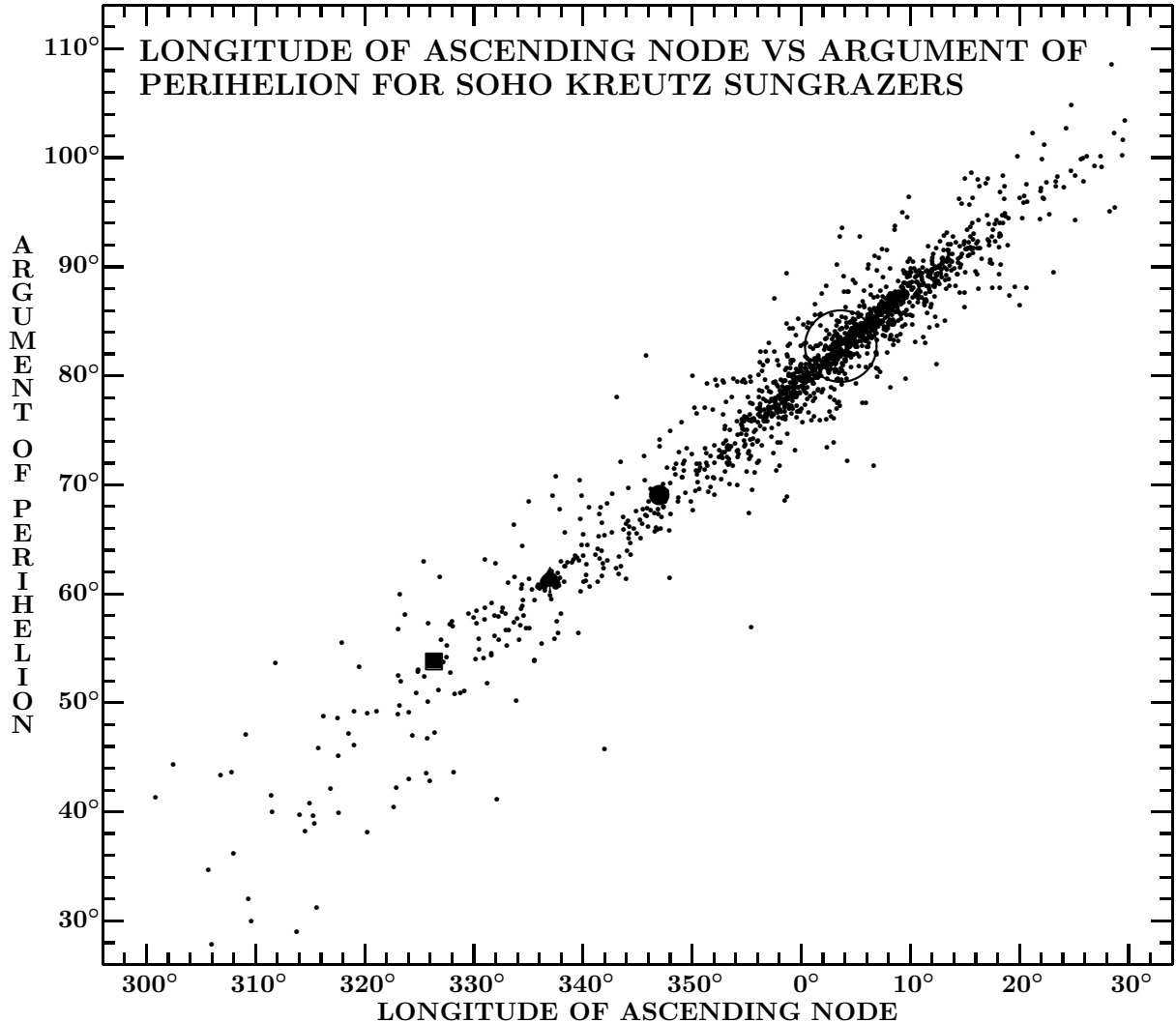}}}
 \vspace*{-10.4cm} 
 \caption{Comets C/2011 W3, C/1970 K1, C/1965 S1, and C/1843 D1 in a plot
of the longitude of the ascending node against the argument of perihelion
for 1565 SOHO Kreutz sungrazers from the period January 1996 to June 2010.
The four bright comets are identified the same way as in Fig.\ 17.  Although
the three branches now essentially merge into one relation, the position of
C/2011 W3 in the plot appears to be again closely matched by a modest number
of SOHO sungrazers.}
\vspace*{0.5cm}
\end{figure*}

\section{Conclusions}

The issues and results presented in this investigation are summarized into
the following conclusions:

(1) Having survived the perihelion passage, comet C/2011 W3 was observed to
undergo major morphological changes, culminating in the permanent loss of the
nuclear condensation between 2011 December 19.4 and 20.3 UT, some 4 days after
perihelion.

(2) This transformation was accompanied by a parallel development of a narrow,
rectilinear dust streamer, extending essentially along the tail's axis in the
images of December 19 and rapidly evolving, in less than 24 hours, into a
ribbon-like spine tail.

(3) From the temporal variations in the projected orientation of its axis
between December 19 and January 18, we find that this spine tail was a
synchronic feature, whose brightest part consisted of submillimeter-sized dust
grains released at velocities not exceeding 30 m s$^{-1}$ and whose origin was
an outburst or a terminal (cataclysmic) fragmentation event (possibly even a
rapid sequence of such very brief episodes), which peaked on December
\mbox{$17.6 \pm 0.2$} UT.

(4) Accordingly, the comet's postperihelion activity, which lasted less
than 48 hours and resulted in releasing up to an estimated 10$^{12}$ grams of
dust (equivalent to a sphere 150-200 meters across), was the sole source of
the spectacular dust tail, which was observed for 3 months, until mid-March.

(5) The delayed response of C/2011 W3 to the extremely adverse environment in
the close proximity of the Sun, which the nucleus' disintegration must have
signified, is of utmost importance.  This overdue reaction speaks volumes
about the cohesion of the nucleus and provides a clue to the probable nature
of the breakup mechanism --- as a product of a gradual heat pulse penetration
to great depths of the nucleus and an entailing steady buildup of thermal stress
throughout the interior.

(6) Modeling of the heat-transfer process and the thermal stress distribution
inside the nucleus, whose dimensions diminish with time because of continuing
erosion at the surface, shows that the rate of propagation of the heat pulse
around perihelion was extremely rapid and highly asymmetric in that at greater
depths beneath the surface both temperature and thermal stress were much higher
at the same heliocentric distance after than before perihelion and continued
to grow far from the Sun, on the way to aphelion.

(7) Based on the elastic constants for meteoritic, lunar, and some terrestrial
analogs, the magnitude of thermal stress throughout much of the nucleus is
estimated at more than 10 kPa shortly after perihelion, overcoming the
object's expected cohesive strength (on the order of several kPa according
to Lasue et al.'s aggregate model), riddling the entire interior with a dense
network of clefts and cracks, thus causing eventually the nucleus' collapse
into a cloud of debris.

(8) The same mechanism may be responsible for the ubiquitous process of
cascading fragmentation that sungrazers are subjected to at all heliocentric
distances, including very far from the Sun.

(9) On the other hand, survival in the solar corona and the delayed response
to the adverse conditions in its environment is inconsistent with the model
of a strengthless or poorly cemented rubble-pile nucleus made up essentially
of unconsolidated debris.

(10) If there were isolated reservoirs of volatile ices deep in the interior
of the nucleus, major cracks caused by excess thermal stress would have opened
them to the propagating heat wave, with the ensuing explosion contributing
to the destruction of the nucleus; our model predicts that 1.6 days after
perihelion, pockets of amorphous water ice more than 40 meters beneath the
surface would have been exposed to a temperature of $\sim$130\,K and thereby
an exothermic reaction would be triggered, associated with amorphous ice's
crystallization; only data on temporal variations in the comet's production of
water vapor and/or other gas species could settle this issue.

(11) The disappearance of the nuclear condensation deprived observers of
obtaining postperihelion astrometry needed for an accurate orbit computation
by traditional techniques, so that the orbital period of C/2011 W3 remained
essentially indeterminate even at the time of termination of observations.

(12) We argue that the missing nucleus must have been located on the
synchrone defined by the axis of the spine tail whose orientation and sunward
tip have been measured; the missing nucleus also must have been situated on
the line of forced orbital-period variation, computed from the orbital
solutions based on high-quality preperihelion astrometry from the ground. 

(13) We succeed in deriving the astrometric positions of this missing nucleus
as the coordinates of the points of intersection of the spine tail's synchrone
and the line of orbital-period variation; a high-quality orbital solution is
obtained by linking the preperihelion observations with these derived positions
of the missing nucleus.

(14) The resulting orbit gives \mbox{$698 \pm 2$} years for the osculating
orbital period and confirms that C/2011 W3 is the first major member of
the expected new, 21st century cluster of bright Kreutz-system sungrazers,
whose existence was predicted by these authors in 2007.

(15) From the spine-tail evolution we determine that the measured
sunward tip receded antisunward from the computed position of the missing
nucleus, the distance increasing from 1$^\prime$ on December 23 to more
than 10$^\prime$ in mid-January and that this terminus was populated by
dust particles 1-2 mm in diameter.

(16) The bizarre appearance of the comet's dust tail in images taken with
the coronagraphs on board the SOHO and STEREO spacecraft only hours after
perihelion is modeled, using the standard dynamical method, in terms
of the ratio $\beta$ of the acceleration by solar radiation pressure to that
by the solar gravitational attraction; the tail brightness was affected
by forward scattering of sunlight by dust to a considerable degree.

(17) Modeling of the SOHO's C3 preperihelion images of the dust tail
shows a population of predominantly microscopic particles with a broad range
of sizes and composition, described both by \mbox{$\beta > 1$} (strongly
absorbing, such as carbon-rich, grains moving in convex hyperbolic orbits)
and by \mbox{$\beta < 1$} (essentially dielectric, perhaps mostly silicate,
grains in concave hyperbolic orbits); the ones that were still imaged after
perihelion had \mbox{$\beta \simeq 0.6$}, probably submicron-sized silicate
particles.

(18) By contrast, nearly all microscopic dust released before perihelion into
convex hyperbolic orbits moved almost immediately away from the Sun and was
not seen in the postperihelion images of the comet.

(19) The disconnection of the comet's head from the dust tail released before
perihelion, strikingly depicted in the postperihelion images taken with the
SOHO's C2 coronagraph and with the coronagraphs on board STEREO-A and STEREO-B,
is interpreted as a result of vigorous sublimation of submicron-sized dust at
heliocentric distances smaller than about 1.8 {\Rsun}, primarily particles
with \mbox{$\beta \leq 0.6$} released between 0.1 day before and 0.1 day
after perihelion.

(20) From the integrated effect of dust particle sublimation, we establish
the relationship between the sublimation-pressure constant $\Lambda$ and the
heat of sublimation $L$ and find that it is consistent with the $L(\Lambda)$
relation for Mg-rich olivine and that therefore it is likely that the dust
sublimating very close to the Sun is dominated by olivine-based silicates,
a conclusion that is also supported dynamically by the magnitude of the
acceleration parameter $\beta$.

(21) The place of C/2011 W3 in the hierarchy of the Kreutz system of sungrazing
comets is a matter of some speculation, but one possible evolutionary path is
charted, based on our orbit integration back to the comet's previous return to
perihelion, which nominally occurred in 1329.

(22) In the context of Scenario B of the Kreutz system's hierarchy (from our
2007 paper) we follow the tidal-assisted breakup of the progenitor (or
superfragment) --- the comet of 467 --- and the expected chain of nontidal
fragmentation events into a number of parent fragments (including X/1106 C1);
a subsequent tidal-assisted splitting of one such parent fragment in the 14th
century into another generation of fragments, including the first precursor
of C/2011 W3; and, finally, another bout of nontidal cascading fragmentation
with the birth of comet C/2011 W3 as a discrete object of its own.

(23) In this scenario, C/2011 W3 is indirectly related to X/1106 C1 and
C/1843 D1, and in the coming years and decades it may be followed
by more equally bright or brighter sungrazing comets in similar orbits.

(24) Comet C/2011 W3 presents a complication for the classification of the
bright sungrazers into subgroups:\ its perihelion distance is typical for
subgroup I, but its angular elements do not fit any of the subgroups I, II,
or IIa; thus, among the bright members of the Kreutz system, the orbit of
C/2011 W3 is unique, although we are not ready to call it a representative
of a new subgroup.

(25) In a plot of the longitude of the ascending node against the inclination,
C/2011 W3 is associated with the least populated of the three branches into
which the SOHO sungrazers discriminate; from this standpoint, its orbit is not
unique.

(26) Because the orbits of the sungrazers (including the bright objects) can 
be transformed by the cascading fragmentation process from one subgroup into
another subgroup during one revolution about the Sun, the classification of
the bright sungrazers into subgroups may no longer be appropriate.

(27) Future modeling of the hierarchy and evolution of the Kreutz system of
sungrazers should account for the existence of C/2011 W3 and its apparent
relationship with the minor branch of SOHO sungrazers; at a minimum, orbital
computations should verify the degree of plausibility of a fragmentation
path of the type displayed in Fig.\ 16.\\[-0.08cm]

We thank S.\ R.\ Chesley for getting us acquainted with the correspondence
he received from R.\ H.\ McNaught regarding the imaging with the Uppsala
Schmidt Telescope, Siding Spring Survey, and in part via J.\ Giorgini, from
M.\ Kusiak regarding his and R.\ Kracht's determination of the astrometric
positions of the comet from the data collected on board the SOHO and STEREO
spacecraft.  We gratefully acknowledge Kracht's comments on his work
concerning the SOHO and STEREO-B astrometry.  We further thank J. \v{C}ern\'y
and McNaught for their permission to reproduce the comet's postperihelion
images taken, respectively, at Malargue and Siding Spring.  D.\ W.\ E. Green
kindly made available to us two older publications.  This research was carried
out at the Jet Propulsion Laboratory, California Institute of Technology, under
contract with the National Aeronautics and Space Administration.\\[-0.3cm]
\begin{center}
{\footnotesize REFERENCES}
\end{center}
\vspace{-0.3cm}
\begin{description}
{\footnotesize
\item[\hspace{-0.3cm}] Afonso, J. C., Ranalli, G., \& Fern\`andez, M. 2005,
 Phys. Earth Plan. Inter., 149, 279\\[-0.55cm]
\item[\hspace{-0.3cm}] Aizawa, Y., Ito, K., \& Tatsumi, Y. 2001, Tectonophys.,
 339, 473\\[-0.55cm]
\item[\hspace{-0.3cm}] Ashkin, D., Haber, R. A., \& Wachtman, J. B. 1990, J.
 Amer. Ceram. Soc., 73, 3376\\[-0.55cm]
\item[\hspace{-0.3cm}] Asphaug, E., \& Benz, W. 1994, Nature, 370,
 120\\[-0.55cm]
\item[\hspace{-0.3cm}] Asphaug, E., \& Benz, W. 1996, Icarus, 121,
 225\\[-0.55cm]
\item[\hspace{-0.3cm}] Austin, J. B. 1952, J. Amer. Ceram. Soc., 35,xi
 243\\[-0.55cm]
\item[\hspace{-0.3cm}] Barnes, L. 2012, http:/{\hspace{-2.8pt}}/oortcloud.org/index.php?topic=17617.0\\[-0.55cm]
\item[\hspace{-0.3cm}] Barthelmy, D. 2010, http:/{\hspace{-2.8pt}}/webmineral.com/data/Olivine.shtml\\[-0.55cm]
%
%
\item[\hspace{-0.3cm}] Biesecker, D. A., Lamy, P., St. Cyr, O. C., Llebaria,
 A., \& Howard, R. A.  2002, Icarus, 157, 323\\[-0.55cm]
\item[\hspace{-0.3cm}] Britt, D. T., \& Consolmagno, G. J. 2003, Meteorit.
 Plan. Sci., 38, 1161\\[-0.55cm]
\item[\hspace{-0.3cm}] Brownlee, D.\,E., Horz, F., Newburn, R.\,L., Zolensky,
 M., Duxbury, T.\,C., Sanford, S., Sekanina, Z., Tsou, P., Hanner, M.\,S.,
 Clark, B.\,C., Green, S.\,F., \& Kissel, J. 2004, Science, 304, 1764\\[-0.55cm]
\item[\hspace{-0.3cm}] Brueckner, G. E., Howard, R. A., Koomen, M. J.,
 Korendyke, C. M., Michels, D. J., Moses, J. D., Socker, D. G., Dere, K. P.,
 Lamy, P. L., Llebaria, A., Bout, M. V., Schwenn, R., Simnett, G. M., Bedford,
 D. K., \& Eyles, C. J. 1995, Sol. Phys., 162, 357\\[-0.55cm]
\item[\hspace{-0.3cm}] \v{C}ern\'y, J., ed. 2011, http:/{\hspace{-2.8pt}}/www.kommet.cz/page.php?al=prvni\_ snimky\_komety\_lovejoy\_ze\_zeme\\[-0.55cm]
\item[\hspace{-0.3cm}] Chesley, S. R. 2012, personal communication\\[-0.55cm]
\item[\hspace{-0.3cm}] Chung, D. H. 1970, JGR, 75, 7353\\[-0.55cm]
\item[\hspace{-0.3cm}]
Combi, M. R., DiSanti, M. A., \& Fink, U. 1997, Icarus, 130, 336
\\[-0.55cm]
\item[\hspace{-0.3cm}]
Consolmagno, G. J., Britt, D. T., \& Opeil, C. P. 2010, Eur. Plan. Sci.
Congr. Abstr., 5, 215
\\[-0.55cm]
\item[\hspace{-0.3cm}]
Cremonese, G., Boehnhardt, H., Crovisier, J., Rauer, H., Fitzsimmons, A.,
Fulle, M., Licandro, J., Pollacco, D., Tozzi, G. P., \& West, R. M. 1997,
ApJ, 490, L199
\\[-0.55cm]
\item[\hspace{-0.3cm}]
Cunningham, L. E. 1946a, IAUC 1025
\\[-0.55cm]
\item[\hspace{-0.3cm}]
Cunningham, L. E. 1946b, HAC 733
\\[-0.55cm]
\item[\hspace{-0.3cm}]
Davidsson, B. J. R., Guti\'errez, P. J., \& Rickman, H. 2009, Icarus, 201, 335
\\[-0.55cm]
%
%
\item[\hspace{-0.3cm}]
England, K. J. 2002, JBAA, 112, 13
\\[-0.55cm]
\item[\hspace{-0.3cm}]
Faivre, C., Bellet, D., \& Dolino, G. 2000, J. Appl. Phys., 87, 2131
\\[-0.55cm]
\item[\hspace{-0.3cm}]
Finson, M. L., \& Probstein, R. F. 1968, ApJ, 154, 327
\\[-0.55cm]
\item[\hspace{-0.3cm}]
Flynn, G. J., Kl\"{o}ck, W., \& Krompholz, R. 1999, Lunar Plan. Sci. Conf.,
30, 1073
\\[-0.55cm]
\item[\hspace{-0.3cm}]
Gail, H.-P., \& Sedlmayr, E. 1999, A\&A, 347, 594
\\[-0.55cm]
\item[\hspace{-0.3cm}]
Ghabezloo, S. 2010, Constr. Build. Mater, 24, 1796
\\[-0.55cm]
\item[\hspace{-0.3cm}]
Ghabezloo, S., \& Sulem, J. 2009, Rock Mech. Rock Eng., 42, 1
\\[-0.55cm]
\item[\hspace{-0.3cm}]
Green, D. W. E., ed. 2011, CBET 2930
\\[-0.55cm]
\item[\hspace{-0.3cm}]
Green, D. W. E., ed. 2012a, CBET 2967
\\[-0.55cm]
\item[\hspace{-0.3cm}]
Green, D. W. E., ed. 2012b, http:/{\hspace{-2.8pt}}/www.icq.eps.harvard.edu/
 CometMags.html
\\[-0.55cm]
%
%
\item[\hspace{-0.3cm}]
Gundlach, B., Blum, J., Skorov, Yu. V., \& Keller, H. U. 2012,
arXiv:1203.1808v1 [astro-ph.EP]
\\[-0.55cm]
\item[\hspace{-0.3cm}]
Harmon, J. K., Nolan, M. C., Howell, E. S., Giorgini, J. D., \& Taylor,
P. A. 2011, ApJ, 734, L2
\\[-0.55cm]
\item[\hspace{-0.3cm}]
Hasegawa, I. 1980, Vistas Astron., 24, 59
\\[-0.55cm]
\item[\hspace{-0.3cm}]
Hasegawa, I., \& Nakano, S. 2001, PASJ, 53, 931
\\[-0.55cm]
\item[\hspace{-0.3cm}]
Haynes, W. M., ed. 2011, CRC Handbook of Chemistry and Physics, 92nd ed.,
CRC Press, Boca Raton, FL, pp. 6-92 and 15-42
\\[-0.55cm]
\item[\hspace{-0.3cm}]
Hiraga, T., Anderson, I. M., \& Kohlstedt, D. L. 2004, Nature, 427, 699
\\[-0.55cm]
\item[\hspace{-0.3cm}]
Ho, P.-Y. 1962, Vistas Astron., 5, 127
\\[-0.55cm]
\item[\hspace{-0.3cm}]
Howard, R. A., Moses, J. D., Vourlidas, A., Newmark, J. S.,
Socker, D. G., Plunkett, S. P., Korendyke, C. M., Cook, J. W.,
Hurley, A., Davila, J. M., Thompson, W. T., St Cyr, O. C.,
Mentzell, E., Mehalick, K., Lemen, J. R., Wuelser, J. P.,
Duncan, D. W., Tarbell, T. D., Wolfson, C. J., Moore, A.,
Harrison, R. A., Waltham, N. R., Lang, J., Davis, C. J.,
Eyles, C. J., Mapson-Menard, H., Simnett, G. M., Halain, J. P.,
Defise, J. M., Mazy, E., Rochus, P., Mercier, R., Ravet, M. F.,
Delmotte, F., Auchere, F., Delaboudiniere, J. P., Bothmer, V.,
Deutsch, W., Wang, D., Rich, N., Cooper, S., Stephens, V.,
Maahs, G., Baugh, R., McMullin, D., \& Carter, T. 2008,
Space Sci. Rev., 136, 67 \\[-0.55cm]
\item[\hspace{-0.3cm}]
Huebner, W. F., Keady, J. J., \& Lyon, S. P. 1992, Ap\&SS, 195, 1
\\[-0.55cm]
\item[\hspace{-0.3cm}]
Hunter, O., Jr., \& Brownell, W. E. 2006, J. Amer. Ceram. Soc., 50, 19
\\[-0.55cm]
\item[\hspace{-0.3cm}]
Iannini, G. M. 1966, IAUC 1948
\\[-0.55cm]
\item[\hspace{-0.3cm}]
Kama, M., Min, M., \& Dominik, C. 2009, A\&A, 506, 1199
\\[-0.55cm]
\item[\hspace{-0.3cm}]
Kimberley, J., \& Ramesh, K. T. 2011, Meteorit. Plan. Sci., 46, 1653
\\[-0.55cm]
\item[\hspace{-0.3cm}]
Kimura, H., Ishimoto, H., \& Mukai, T. 1997, A\&A, 326, 263
\\[-0.55cm]
\item[\hspace{-0.3cm}]
Kimura, H., Mann, I., Biesecker, D. A., \& Jessberger, E. K. 2002, Icarus,
 159, 529
\\[-0.55cm]
\item[\hspace{-0.3cm}]
Knight, M. M., A'Hearn, M. F., Biesecker, D. A., Faury, G., Hamilton, D. P.,
Lamy, P., \& Llebaria, A. 2010, AJ, 139, 926
\\[-0.55cm]
\item[\hspace{-0.3cm}]
Kov\'a\v{c}ik, J. 2005, J. Mater. Sci., 41, 1247
\\[-0.55cm]
\item[\hspace{-0.3cm}]
Kracht, R. 2011, http:/{\hspace{-2.8pt}}/tech.groups.yahoo.com/group/comets-ml/
 message/18907
\\[-0.55cm]
\item[\hspace{-0.3cm}]
Kracht, R. 2012, personal communication
\\[-0.55cm]
\item[\hspace{-0.3cm}]
Kreutz, H. 1888, Publ. Sternw. Kiel, 3
\\[-0.55cm]
\item[\hspace{-0.3cm}]
Kreutz, H. 1891, Publ. Sternw. Kiel, 6
\\[-0.55cm]
\item[\hspace{-0.3cm}]
Kreutz, H. 1901, Astron. Abh., 1, 1
\\[-0.55cm]
\item[\hspace{-0.3cm}]
Krivov, A., Kimura, H., \& Mann, I. 1998, Icarus, 134, 311
\\[-0.55cm]
\item[\hspace{-0.3cm}]
Kronk, G. W. 1999, Cometography: A Catalog of Comets (Cam\-bridge: Cambridge
Univ. Press)
\\[-0.55cm]
\item[\hspace{-0.3cm}]
Kronk, G. W., ed. 2011, http:/{\hspace{-2.8pt}}/cometography.com/lcomets/
 2011w3.html
\\[-0.55cm]
\item[\hspace{-0.3cm}]
Lamy, P. L., Toth, I., Weaver, H. A., A'Hearn, M. F., \& Jorda, L. 2009,
A\&A, 508, 1045
\\[-0.55cm]
\item[\hspace{-0.3cm}]
Lasue, J., Botet, R., Levasseur-Regourd, A. C., \& Hadamcik, E. 2009, Icarus,
203, 599
\\[-0.55cm]
\item[\hspace{-0.3cm}]
Lasue, J., Botet, R., Levasseur-Regourd, A. C., Hadamcik, E., \& Kofman, W.
2011, Icarus, 213, 369
\\[-0.55cm]
\item[\hspace{-0.3cm}]
Lourens, J. V. B. 1966, Mon. Not. Astron. Soc. South Africa, 25, 52
\\[-0.55cm]
\item[\hspace{-0.3cm}]
Mann, I., Kimura, H., Biesecker, D. A., Tsurutani, B. T., Gr\"{u}n, E.,
McKibben, R. B., Liou, J.-C., MacQueen, R. M., Mukai, T., Guhathakurta, M.,
\& Lamy, P. 2004, Space Sci. Rev., 110, 269
\\[-0.55cm]
\item[\hspace{-0.3cm}]
Marcus, J. N. 2007, Int. Comet Quart., 29, 39
\\[-0.55cm]
\item[\hspace{-0.3cm}]
Marsden, B. G. 1967, AJ, 72, 1170
\\[-0.55cm]
\item[\hspace{-0.3cm}]
Marsden, B. G. 1989, AJ, 98, 2306
\\[-0.55cm]
\item[\hspace{-0.3cm}]
Marsden, B. G. 2005, ARA\&A, 43, 75
\\[-0.55cm]
%
%
\item[\hspace{-0.3cm}]
McNaught, R. H. 2012, http:/{\hspace{-2.8pt}}/msowww.anu.edu.au/$\sim$rmn/
 C2011W3.htm
\\[-0.55cm]
\item[\hspace{-0.3cm}]
Moretti, L., De Stefano, L., Rossi, A. M., \& Rendina, I. 2005, Appl. Phys.
Lett., 86, 061107
\\[-0.55cm]
\item[\hspace{-0.3cm}]
Nagahara, H., Mysen, B. O., \& Kushiro, I. 1994, Geochim. Cosmochim. Acta, 58,
1951
\\[-0.55cm]
\item[\hspace{-0.3cm}]
Phani, K. K., \& Sanyal, D. 2005, J. Mater. Sci., 40, 5685
\\[-0.55cm]
\item[\hspace{-0.3cm}]
Pohn, H. 1965, IAUC 1937
\\[-0.55cm]
\item[\hspace{-0.3cm}]
Pritchard, M. E., \& Stevenson, D. J. 2000, in R. M. Canup \& K. Righter
(eds.) Origin of the Earth and Moon, U. Arizona, Tucson, p. 179
\\[-0.55cm]
%
%
\item[\hspace{-0.3cm}]
Richardson, J. E., Melosh, H. J., Lisse, C. M., \& Carcich, B. 2007, Icarus,
190, 357
\\[-0.55cm]
\item[\hspace{-0.3cm}]
Rogers, D. 2005, Einstein's Other Theory: The Planck-Bose-Ein\-stein Theory of
Heat Capacity, Princeton U., Princeton, NJ, p. 73
\\[-0.55cm]
\item[\hspace{-0.3cm}]
Sagdeev, R. Z., Evlanov, E. N., Fomenkova, M. N., Prilutskii, D. F., \&
Zubkov, B. V. 1989, Adv. Space Res., 9, No. 3, p. 263
\\[-0.55cm]
\item[\hspace{-0.3cm}]
Schmitt, B., Espinasse, S., Grim, R. J. A., Greenberg, J. M., \& Klinger, J.
1989, in J.Hunt \& T. D. Guyenne (eds.) Physics and Mechanics of Cometary
Materials, ESA SP-302, ESTEC, Noordwijk, Netherlands, p. 65
\\[-0.55cm]
\item[\hspace{-0.3cm}]
Seargent, D. A. J. 2011, http:/{\hspace{-2.8pt}}/tech.groups.yahoo.com/group/
 comets-ml/message/19078
\\[-0.55cm]
\item[\hspace{-0.3cm}]
Sekanina, Z. 1984, Icarus, 58, 81
\\[-0.55cm]
\item[\hspace{-0.3cm}]
Sekanina, Z. 2000, ApJ, 545, L69
\\[-0.55cm]
\item[\hspace{-0.3cm}]
Sekanina, Z. 2002a, ApJ, 566, 577
\\[-0.55cm]
\item[\hspace{-0.3cm}]
Sekanina, Z. 2002b, ApJ, 576, 1085
\\[-0.55cm]
\item[\hspace{-0.3cm}]
Sekanina, Z. 2003, ApJ, 597, 1237
\\[-0.55cm]
\item[\hspace{-0.3cm}]
Sekanina, Z. 2008, Int. Comet Quart., 30, 3
\\[-0.55cm]
\item[\hspace{-0.3cm}]
Sekanina, Z. 2009, Int. Comet Quart., 31, 99
\\[-0.55cm]
\item[\hspace{-0.3cm}]
Sekanina, Z. 2012, CBET 2967
\\[-0.55cm]
\item[\hspace{-0.3cm}]
Sekanina, Z., \& Chodas, P. W. 2004, ApJ, 607, 620
\\[-0.55cm]
\item[\hspace{-0.3cm}]
Sekanina, Z., \& Chodas, P. W. 2007, ApJ, 663, 657
\\[-0.55cm]
\item[\hspace{-0.3cm}]
Sekanina, Z., \& Chodas, P. W. 2008, ApJ, 687, 1415
\\[-0.55cm]
\item[\hspace{-0.3cm}]
Sekanina, Z., Hanner, M. S., Jessberger, E. K., \& Fomenkova, M. N.
2001, in E. Gr\"{u}n, B. {\AA}. S. Gustafson, S. Dermott, \& H.
Fechtig (eds.) Interplanetary Dust, Springer, Berlin, Germany, p. 95
\\[-0.55cm]
%
%
\item[\hspace{-0.3cm}]
Sitko, M. L., Lisse, C. M., Kelley, M. S., Polomski, E. F., Lynch, D. K.,
Russell, R. W., Kimes, R. L., Whitney, B. A., Wolff, M. J., \& Harker,
D. E. 2011, AJ, 142, 80
\\[-0.55cm]
\item[\hspace{-0.3cm}]
Spahr, T. B., Williams, G. V., Nakano, S., \& Doppler, A., eds. 2011,
MPC 77148
\\[-0.55cm]
\item[\hspace{-0.3cm}]
Spahr, T. B., Williams, G. V., Nakano, S., \& Doppler, A., eds. 2012,
MPC 77565-77566
\\[-0.55cm]
\item[\hspace{-0.3cm}]
Suzuki, I., Anderson, O. L., \& Sumino, Y. 1983, Phys. Chem. Minerals, 10, 38
\\[-0.55cm]
\item[\hspace{-0.3cm}]
Tait, S. 1992, Amer. Mineralogist, 77, 146
\\[-0.55cm]
\item[\hspace{-0.3cm}]
Thackeray, A. D. 1965, Mon. Not. Astron. Soc. South Africa, 24, 159
\\[-0.55cm]
\item[\hspace{-0.3cm}]
Thomas, P. C., Veverka, J., Belton, M. J. S., Hidy, A., A'Hearn, M. F.,
Farnham, T. L., Groussin, O., Li, J.-Y., McFadden, L. A., Sunshine, J.,
Wellnitz, D., Lisse, C., Schultz, P., Meech, K. J., \& Delamere, W. A.
2007, Icarus, 187, 4
\\[-0.55cm]
\item[\hspace{-0.3cm}]
Timoshenko, S. P., \& Goodier, J. N. 1970, Theory of Elasticity, 3rd ed.,
McGraw-Hill, New York, NY, p. 454
\\[-0.55cm]
\item[\hspace{-0.3cm}]
Utterback, N. G., \& Kissel, J. 1990, AJ, 100, 1315{\hspace*{-0.05cm}}
\\[-0.55cm]
\item[\hspace{-0.3cm}]
Utterback, N. G., \& Kissel, J. 1995, Astrophys. Space Sci., 225, 327
\\[-0.55cm]
\item[\hspace{-0.3cm}]
Wang, H., \& Ramesh, K. T. 2004, Acta Materialia, 52, 355
\\[-0.55cm]
\item[\hspace{-0.3cm}]
Warren, N. 1969, JGR, 74, 713
\\[-0.55cm]
%
%
\item[\hspace{-0.3cm}]
Williams, G. V. 2011a, MPEC 2011-X36
\\[-0.55cm]
\item[\hspace{-0.3cm}]
Williams, G. V. 2011b, MPEC 2011-Y02
\\[-0.55cm]
\item[\hspace{-0.3cm}]
Williams, G. V. 2011c, MPEC 2011-Y23
\\[-0.61cm]
\item[\hspace{-0.3cm}]
Yomogida, K., \& Matsui, T. 1983, JGR, 88, 9513}
\end{description}
\end{document}